\documentclass[11pt,a4paper]{article}
\pdfoutput=1

\usepackage{jheppub-nosort,amsthm}

\PassOptionsToPackage{svgnames}{xcolor}
\usepackage{xcolor}
\usepackage{amsmath,amssymb,amsthm,amscd}
\usepackage{physics}
\usepackage{tikz}
\usetikzlibrary{decorations.pathmorphing}
\tikzset{node distance=2cm, auto}
\tikzset{snake it/.style={decorate, decoration=snake}}
\usepackage{epsfig}
\usepackage{psfrag,comment}
\usepackage{graphicx}
\usepackage{caption}
\usepackage{subcaption}
\usepackage{mathrsfs}
\usepackage[normalem]{ulem}
\usepackage{bm}
\usepackage{lscape}
\usepackage{diagbox}
\usepackage{upgreek}

\usepackage{url}
\newcommand\doilink[1]{\href{http://dx.doi.org/#1}{#1}}
\newcommand\arxivlink[1]{\href{http://arxiv.org/abs/#1}{#1}}

\newcommand{\be}{\begin{equation}}
\newcommand{\ee}{\end{equation}}
\newcommand{\ba}{\begin{aligned}}
\newcommand{\ea}{\end{aligned}}
\newcommand{\bea}{\begin{eqnarray}}
\newcommand{\eea}{\end{eqnarray}}
\newcommand{\bean}{\begin{eqnarray*}}
\newcommand{\eean}{\end{eqnarray*}}

\def\r{\right\rangle}

\def\1{\mathbf{1}}
\def\0{|\1\r}

\newcommand{\rme}{{\mathrm{e}}}
\newcommand{\rmi}{{\mathrm{i}}}

\def\XXint#1#2#3{{\setbox0=\hbox{$#1{#2#3}{\int}$}
     \vcenter{\hbox{$#2#3$}}\kern-.5\wd0}}

\makeatletter
\newsavebox\myboxA
\newsavebox\myboxB
\newlength\mylenA

\newcommand*\widebar[2][0.75]{%
    \sbox{\myboxA}{$\m@th#2$}%
    \setbox\myboxB\null% Phantom box
    \ht\myboxB=\ht\myboxA%
    \dp\myboxB=\dp\myboxA%
    \wd\myboxB=#1\wd\myboxA% Scale phantom
    \sbox\myboxB{$\m@th\overline{\copy\myboxB}$}%  Overlined phantom
    \setlength\mylenA{\the\wd\myboxA}%   calc width diff
    \addtolength\mylenA{-\the\wd\myboxB}%
    \ifdim\wd\myboxB<\wd\myboxA%
       \rlap{\hskip 0.8\mylenA\usebox\myboxB}{\usebox\myboxA}%
    \else
        \hskip -0.5\mylenA\rlap{\usebox\myboxA}{\hskip 0.5\mylenA\usebox\myboxB}%
    \fi}
\makeatother

\definecolor{cottoncandy}{rgb}{1.0, 0.74, 0.85}
\definecolor{cornellred}{rgb}{0.7, 0.11, 0.11}
\definecolor{darktangerine}{rgb}{1.0, 0.66, 0.07}
\newcommand{\picscale}{0.8}
\definecolor{iceberg}{rgb}{0.44, 0.65, 0.82}
\newcommand{\DiscZZ}[1]{\mathord{
\begin{tikzpicture}[baseline={([yshift=-0.8ex]current bounding box.center)}, line width=1, scale=\picscale]
\draw[line width=2pt] (0,0) circle (0.8cm);
\fill[iceberg] (0,0) circle (0.8cm);
\node at (1.35,0.6) {\footnotesize $#1$};
\end{tikzpicture}}}
\newcommand{\AnnulusZZ}[2]{\mathord{
\begin{tikzpicture}[baseline={([yshift=-0.8ex]current bounding box.center)}, line width=1, scale=\picscale]
\draw[line width=2pt] (0,0) circle (0.6cm);
\draw[line width=2pt] (0,0) circle (1.2cm);
\fill[iceberg, even odd rule] (0,0) circle[radius=1.2cm] circle[radius=0.6cm];
\node at (0.05,0.1) {\footnotesize $#1$};
\node at (1.5,0.95) {\footnotesize $#2$};
\end{tikzpicture}}}
\title{Resurgence in the Virasoro Minimal String and 3d Gravity}

\author[a]{Maximilian~Schwick}
\affiliation[a]{Albert Einstein Center for Fundamental Physics, Institute for Theoretical Physics,\\ University of Bern, CH-3012 Bern, Switzerland}
\emailAdd{maximilian.schwick@unibe.ch}

\abstract{
We compute non-perturbative, resurgent contributions to the Virasoro minimal string and 3d gravity using techniques from hermitian matrix models. In particular, we construct a fully non-perturbative partition function for the Virasoro minimal string in terms of a Zak transform. In this context, negative tension D-branes appear naturally, which in the matrix model correspond to anti-eigenvalues, or instantons on the involuted sheet of the spectral curve. We further extend this analysis to resolvents and observe resurgent wall crossing phenomena between ZZ- and FZZT-branes. Using recent results that relate the Virasoro minimal string to 3d gravity with end-of-the-world branes we proceed to study the resurgent consequences of summing over the genus in 3d gravity, where we find non-perturbative contributions of doubly exponential type. These statements are then tested using resurgent large-order asymptotics. Lastly, we compute the non-perturbative eigenvalue density for generic hermitian matrix models and identify the change of asymptotic behavior at the edge of the eigenvalue distribution with a Stokes transition. This allows us to identify oscillations in the eigenvalue density with anti-Stokes behavior of FZZT-branes. In the case of 3d gravity with end-of-the-world branes we comment how this Stokes transition coincides with the onset of black hole behaviour and compute the non-perturbative primary density. Furthermore, we apply these techniques to the eigenvalue density of JT gravity to compute higher genus corrections.
}


\arxivnumber{2602.22335}

\begin{document}

%%%%%%%%%%%%%%%%%%%%%%%%%%%%%%%%%%%%%%%%%%%%%%%%%%%%%%%%%%%%%%%%%
%%%%%%%%%%%%%%%%%%%%%%%%%%%%%%%%%%%%%%%%%%%%%%%%%%%%%%%%%%%%%%%%%
\maketitle
%%%%%%%%%%%%%%%%%%%%%%%%%%%%%%%%%%%%%%%%%%%%%%%%%%%%%%%%%%%%%%%%%
%%%%%%%%%%%%%%%%%%%%%%%%%%%%%%%%%%%%%%%%%%%%%%%%%%%%%%%%%%%%%%%%%

\vfill

\eject

\allowdisplaybreaks

%%%%%%%%%%%%%%%%%%%%%%%%%%%%%%%%%%%%%%%%%%%%%%%%%%%%%%%%%%%%%%%%%
%%%%%%%%%%%%%%%%%%%%%%%%%%%%%%%%%%%%%%%%%%%%%%%%%%%%%%%%%%%%%%%%%
\section{Introduction and Summary}
\label{sec:intro}
%%%%%%%%%%%%%%%%%%%%%%%%%%%%%%%%%%%%%%%%%%%%%%%%%%%%%%%%%%%%%%%%%
%%%%%%%%%%%%%%%%%%%%%%%%%%%%%%%%%%%%%%%%%%%%%%%%%%%%%%%%%%%%%%%%%

Understanding non-perturbative contributions in string theory has been a fruitful area of study for many years. In this context the discovery of D-branes has opened many pathways \cite{p94, p95}. However explicit computations are hard to perform and therefore simple examples are of paramount importance. For this reason non-perturbative contributions in matrix models and minimal string theories have been studied extensively in recent years, e.g. \cite{d91, d92, akk03,msw07, msw08, m08, kmr10, sss19, gs21, msy21, emms22a, emms22b,  mss22, emmms22, e23, sst23, cemm24, ams24,emm24, emt24}. In minimal string theory such contributions are due to ZZ- \cite{zz01} and FZZT-branes \cite{fzz00, t00}. In the corresponding matrix model they are associated to eigenvalue tunnelling -- the movement of eigenvalues in the diagonal matrix integral to non-perturbative saddles \cite{msw07, msw08}. 

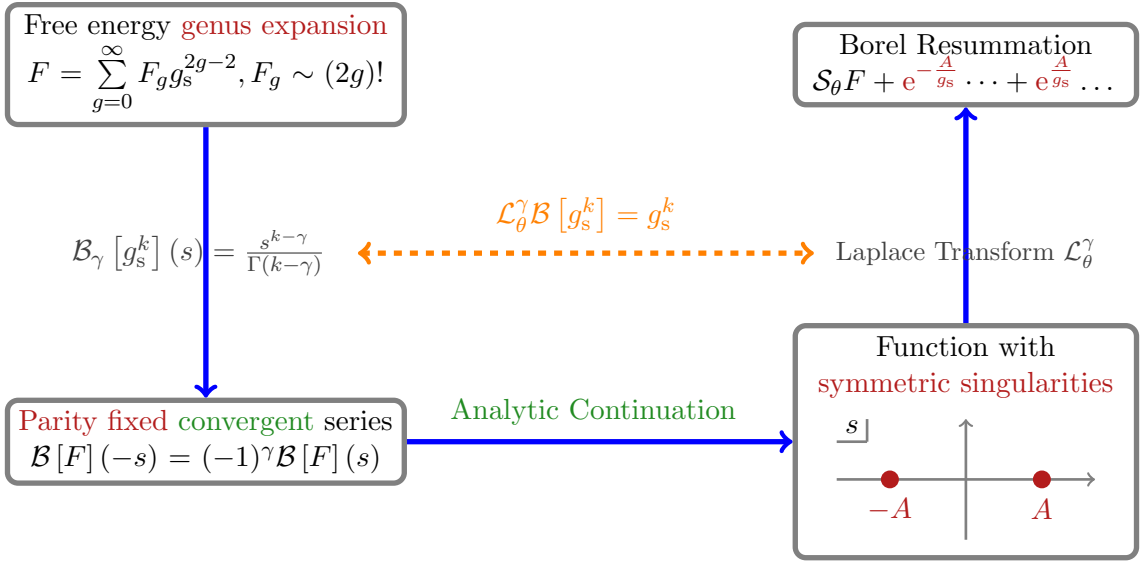
\begin{figure}
\centering
	\begin{tikzpicture}[
	grayframe/.style={
		rectangle,
		draw=gray,
		text width=11em,
		align=center,
		rounded corners,
		minimum height=2em
	},
	longgrayframe/.style={
		rectangle,
		draw=gray,
		text width=12.9em,
		align=center,
		rounded corners,
		minimum height=2em
	}, scale =1, line width=2
	]
	\node[longgrayframe] (a) at (-5,4) {Free energy {\color{cornellred}genus expansion}\\$F = \sum\limits_{g=0}^{\infty} F_g g_{\text{s}}^{2g-2}, F_g\sim(2g)!$};
	\node (ab) at (-5,1.5) {\color{darkgray}$\mathcal{B}_{\gamma}\left[g_{\text{s}}^{k}\right](s) = \frac{s^{k-\gamma}}{\Gamma\left(k-\gamma\right)}\,\,\,$};
	\node[longgrayframe] (b) at (-5,-1) {{\color{cornellred}Parity fixed} {\color{ForestGreen}convergent} series\\$\mathcal{B}\left[F\right](-s)=(-1)^{\gamma}\mathcal{B}\left[F\right](s)$};
	\draw[blue, ->] (a) -- (b);
	\node (bc) at (0.1,-0.6) {\color{ForestGreen} Analytic Continuation};
	\node[grayframe] (c) at (5,-1) {Function with\\ {\color{cornellred}symmetric singularities}\vspace{2cm}};
	\draw[blue, ->] (b) -- (c);
	\node[grayframe] (d) at (5,4) {Borel Resummation\\$\mathcal{S}_\theta F + {\color{cornellred}\rme^{-\frac{A}{g_{\text{s}}}}}\dots+ {\color{cornellred}\rme^{\frac{A}{g_{\text{s}}}}}\dots$\\};
	\draw[blue,->] (c) -- (d);
	\node (cd) at (5, 1.5) {\small \color{darkgray} Laplace Transform $\mathcal{L}_\theta^{\gamma}$};
	\draw[orange, dashed, <->] (-3, 1.5) -- (3, 1.5);
	\node[orange] at (0, 2) {$\mathcal{L}_\theta^{\gamma} \mathcal{B}\left[g_{\text{s}}^k\right]=g_{\text{s}}^k$};
	\begin{scope}[scale=1,  shift={({5},{-1.5})}]
	\draw[line width=1, gray, ->] (-1.7,0) -- (1.7,0);
	\draw[line width=1, gray, ->] (0, -0.8) -- (0,0.8);
	\draw[cornellred, fill=cornellred] (1,0) circle (.5ex);
	\draw[cornellred, fill=cornellred] (-1,0) circle (.5ex);
	\node at (-1.5, 0.7) {$s$};
	\draw[line width=1, gray] (-1.3, 0.5) -- (-1.7, 0.5);
	\draw[line width=1, gray] (-1.3, 0.5) -- (-1.3, 0.8);
	\node[cornellred] at (1, -0.4) {$A$};
	\node[cornellred] at (-1, -0.4) {$-A$};
	\end{scope}
	\end{tikzpicture}
\caption{Pictorial description of the Borel resummation of a divergent series with genus expansion. Start with the left-downward arrow by performing a Borel transform $\mathcal{B}_{\gamma}$, which yields a power-series convergent in a disk around the origin ($\gamma \in\mathbb{R}$ see \cite{m12, s14, abs18} for details). Moreover, because we started with a genus expansion the Borel transform and its analytic continuation will be parity fixed -- the resulting Borel-plane singularities will be symmetrically arranged. The final step with the right-upward arrow is resummation via Laplace transform $\mathcal{L}_{\theta}^{\gamma}$ along the angle $\theta$ ({\color{orange}inverse Borel transform}); such  transformation will turn the Borel plane singularities into exponential contributions where the location of the singularity determines the exponential weight. Therefore, because of the symmetry on the Borel plane that we inherited from the genus expansion, such exponentials appear always in pairs {\color{cornellred}$(A, -A)$} -- a phenomenon often referred to as \emph{resonance}. Notice how this argument is independent of the choice of parameter $\gamma$. This figure illustrates and builds on a graphic from \cite{mss22}.}
\label{fig:innerworkingsofresurgence}
\end{figure}

Moreover, resurgent computations for the Painlevé I equation \cite{gikm10, asv11, bssv22} as well as the Painlevé II equation  \cite{sv13, v23} -- which compute the specific heat of $(2,3)$ minimal (super-) string theory -- suggested the existence of further non-perturbative contributions that are exponentially enhanced: Each falling instanton has an exponentially growing sibling, a phenomenon that is called resonance in the resurgence literature \cite{gs21, mss22}. Such contributions usually come with a minus sign in comparison to the standard instantons which suggests that in minimal string theory they correspond to branes with negative tension \cite{sst23}. Using resurgent techniques \cite{e81} it was also argued that such instanton contributions appear in the asymptotics of generic string genus expansions \cite{s90, sst23}. As it is central to this work we quickly review this argument in figure \ref{fig:innerworkingsofresurgence}\footnote{For reviews of 1-summability and resurgence the interested reader may consult \cite{m12, s14, abs18}.}. 

In parallel to the developments described above, the playground of minimal string theories has recently been extended by the construction of the Virasoro minimal string (VMS) \cite{cemr23}. It represents a continuous family of critical worldsheet theories indexed by $c\in\mathbb{R}_{\geq 25}$ and is given via
\begin{equation}
\left({\text{Liouville CFT}\atop{c\geq 25}}\right)\quad\bigoplus\quad\left({\text{Liouville CFT}\atop{\hat{c}=26-c\leq 1}}\right)\quad\bigoplus\quad\left({{\mathfrak{b}\mathfrak{c}-\text{ghosts}}\atop{-26}}\right).
\end{equation}
\noindent
For further details we refer the reader to the original paper \cite{cemr23}. Given the natural proximity of the VMS to other minimal string theories it is natural to ask if the non-perturbative contributions mentioned above find an analogue here: we use the methods developed in \cite{msw07, mss22, sst23} to compute the exponential contributions to the free energy of the VMS. More specifically we will study the asymptotic behaviour of the free energy and the resolvent, where we find negative tension brane contributions in addition to resurgent wall crossing phenomena. In this context, we observe that not all falling non-perturbative exponentials in the VMS have their origins on the physical sheet of the spectral curve -- some standard falling D-brane contributions in fact originate in the non-physical sheet.

Putting all the contributions together we can construct the full non-perturbative partition function for the VMS -- meaning the full family of non-perturbative completions allowed by resurgence. This proposal is based on \cite{krsst25a} and we generalize it to the VMS: In this way we can write the partition function as a sum over filling fractions in the spirit of \cite{msw08, em08} and present checks on its validity by comparing to the previously computed non-perturbative effects. Consequently we discuss possible choices of non-perturbative completion -- firstly a minimal choice consistent with median summation and secondly a generic choice showing quasi-periodic asymptotics. 

It has been shown in \cite{jrw25-gen, jrsw25}, that the VMS describes a sub-sector of 3d gravity: Specifically in \cite{jrsw25} a map is constructed between the VMS and 3d gravity path integrals on manifolds that are topologically a thickened Riemann surface $\Sigma_{g,n}\times I$, where $g$ counts the genus, $n$ is the number of boundaries and $I$ is an interval with end-of-the-world (EOW) branes on both ends. We use this map to explore the resurgence non-perturbative effects that arise if one sums over the genus in 3d gravity. For our specific example we pick $n=2$ and try to understand the effects of the genus sum by using the VMS non-perturbative predictions. We find doubly exponential contributions in $c$ as expected in \cite{jrw25-gen}, which we further confirm with asymptotic checks. In addition we also find both positive and negative instantons, just as in the VMS. In this context it could be interesting to generalize the closed form VMS partition function from section \ref{sec:exactZ} to correlators ($n>0$).

%%%%%%%%%%%%%%%%%%%%%%%%%%%%%%%%%%%%%%%%%%%%%%%%%%%%%%%%%%%%%%%%%
%%%%%%%%%%%%%%%%%%%%%%%%%%%%%%%%%%%%%%%%%%%%%%%%%%%%%%%%%%%%%%%%%

Last but not least we compute the non-perturbative eigenvalue density for hermitian matrix models with resurgent methods. In this way we can understand the change of asymptotic behaviour that occurs at the edge of the eigenvalue distribution as a Stokes transition of FZZT-branes. Consequently we find a formula for the non-perturbative density that has a universal, oscillating piece -- next to additional non-perturbative effects that depend on the choice of non-perturbative completion. This formula is a generalization of a proposal for the non-perturbative eigenvalue density in \cite{sss19}. We then connect this discussion to 3d and JT gravity where the edge of the eigenvalue spectrum is identified with the onset of the Cardy growth of states and compute the associated non-perturbative densities.

This paper is organized as follows: in section \ref{sec:VMS} we will apply the matrix model formalism of \cite{msw07, mss22, sst23} to the VMS and collect results. Here in subsection \ref{subsec:instantonsVMM} we first focus on the physical sheet instantons and compare with literature results. Thereafter we compute the previously mentioned, new, non-physical sheet contributions. In subsection \ref{subsec:bcft} we reproduce negative tension brane contributions directly from boundary conformal field theory. Subsection \ref{subsec:JT} is dedicated to the Jackiw–Teitelboim two-dimensional dilaton gravity (JT) \cite{t83, j84} limit of the formulae produced in subsection \ref{subsec:instantonsVMM}. Subsequently, we will perform an asymptotic large-order study of the free energy in subsection \ref{subsec:largeorderchecks} to numerically confirm our previous results. In this context it is natural to also perform large order checks on the corresponding resolvent, which yields wall crossing phenomena between ZZ- and FZZT-branes. This is the topic of section \ref{subsec:wall-crossing}. Putting the above together, we can conjecture a fully non-perturbative partition function for the VMS in section \ref{sec:exactZ} and check it against previously computed non-perturbative contributions. Here we will construct the generic transseries -- meaning the full family of resurgent non-perturbative contributions -- in subsection \ref{subsec:generic-exactZ} and comment on specific non-perturbative completions in subsection \ref{subsec:non-pert-com}. Having established the VMS non-perturbative effects we import them to a subsector of 3d gravity using the map established in \cite{jrsw25} to find doubly exponential contributions. This is the content of section \ref{sec:3d-gravity}. Lastly, in section \ref{sec:black-hole-stokes}  we consider a generic hermitian matrix model and compute the non-perturbative density from resurgent contributions. In subsection \ref{subsec:gen-construction} we present the generic construction for hermitian matrix models, and connect this discussion to 3d and JT gravity in subsections \ref{subsec:vms-stokes} and \ref{subsec:jt-stokes} respectively.

%intro ( 1a recently been found new instantons in matrix models 1 paragraph or less, due to resurgence etc; 1b these are negative tension d-branes in minimal string duals 1 paragraph or less, again via resurgence etc; 1c recently MVS extends examples of multicritical and minimal strings 1 paragraph or less; 1d might MVS also have antiEV and negative branes?)

%%%%%%%%%%%%%%%%%%%%%%%%%%%%%%%%%%%%%%%%%%%%%%%%%%%%%%%%%%%%%%%%%
%%%%%%%%%%%%%%%%%%%%%%%%%%%%%%%%%%%%%%%%%%%%%%%%%%%%%%%%%%%%%%%%%
\section{Non-Perturbative Effects in the Virasoro Minimal String}
\label{sec:VMS}
%%%%%%%%%%%%%%%%%%%%%%%%%%%%%%%%%%%%%%%%%%%%%%%%%%%%%%%%%%%%%%%%%
%%%%%%%%%%%%%%%%%%%%%%%%%%%%%%%%%%%%%%%%%%%%%%%%%%%%%%%%%%%%%%%%%

In this section we apply recent progress in the non-perturbative study of matrix models and their double scaling limits \cite{msw07, msw08, mss22, sst23, eggls23} to the VMS. More precisely, explicit formulae have been developed that allow the direct computation of exponential corrections (resurgent instantons) purely from special geometry\footnote{Meaning that the ingredients needed to employ these formulae are: the spectral curve with a differential on it, and the Bergman kernel \cite{eo07a}.}. Here we apply those methods to the free energy of the VMS \cite{cemr23} to compute explicitly the non-perturbative effects. Let us stress that these methods come from hermitian matrix models: To make sure that they indeed apply to the VMS\footnote{This is non-trivial in the sense that the VMS does not have an explicit hermitian matrix model underlying it -- it just fulfils topological recursion, which is a more generic framework.} we perform explicit numerical large order checks\footnote{This will be important for section \ref{sec:exactZ} where we make a conjecture for the non-perturbative partition function of the VMS based on matrix model arguments.}. By doing so we will predict and numerically verify contributions from negative tension branes -- additional effects that were initially not observed in \cite{cemr23}. Lastly we will study the VMS resolvent and verify the existence of wall crossing phenomena between ZZ- and FZZT-branes. In the following we will adhere exactly to the conventions of \cite{cemr23}.

%%%%%%%%%%%%%%%%%%%%%%%%%%%%%%%%%%%%%%%%%%%%%%%%%%%%%%%%%%%%%%%%%
%%%%%%%%%%%%%%%%%%%%%%%%%%%%%%%%%%%%%%%%%%%%%%%%%%%%%%%%%%%%%%%%%
\subsection{Anti-Eigenvalues in the Virasoro Matrix Model}
\label{subsec:instantonsVMM}
%%%%%%%%%%%%%%%%%%%%%%%%%%%%%%%%%%%%%%%%%%%%%%%%%%%%%%%%%%%%%%%%%
%%%%%%%%%%%%%%%%%%%%%%%%%%%%%%%%%%%%%%%%%%%%%%%%%%%%%%%%%%%%%%%%%

%%%%%%%%%%%%%%%%%%%%%%%%%%%%%%%%%%%%%%%%%%%%%%%%%%%%%%%%%%%%%%%%%
\begin{figure}
\centering
	\begin{tikzpicture}
	\begin{scope}[scale=0.7,  shift={({-5.8},{0})}]
	\fill[fill=LightBlue,fill opacity=0.2, line width=1pt] (0,0)   to [out=90,in=95] (4,0)
	to [out=90,in=95] (8,0)
	to [out=85,in=180] (10.2, 1.1)
	to [out=70, in=290] (10.2, 2.5)
	to [out=180,in=0] (1,2.5)
    to [out=180,in=90] (-2,0)
    to [out=270, in=180] (-1, -0.5)
    to [out=0, in=270] cycle;
    \fill[fill=darktangerine,fill opacity=0.2, line width=1pt] (-2,0)
    to [out=270,in=180] (1,-2.5)
    to [out=0, in=180] (10.2, -2.5)
    to [out=70, in=290] (10.2, -1.1)
    to [out=180,in=270] (8,0)
    to [out=265,in=275] (4,0)
    to [out=265,in=270] (0,0)
    to [out=270, in=0] (-1, -0.5)
    to [out=180, in=270] cycle;
    \draw[line width=1pt] (0,0)   to [out=90,in=95] (4,0)
	to [out=90,in=95] (8,0)
	to [out=85,in=180] (9.8, 1.1);
	\draw[line width=1pt] (9.8, 2.5)
	to [out=180,in=0] (1,2.5)
	to [out=180,in=90] (-2,0);
	\draw[line width=1pt] (-2,0)
    to [out=270,in=180] (1,-2.5)
    to [out=0, in=180] (9.8,-2.5);
    \draw[line width=1pt] (9.8, -1.1)
    to [out=180,in=270] (8,0)
    to [out=265,in=275] (4,0)
    to [out=265,in=270] (0,0)
    to [out=270, in=0] (-1, -0.5);
    \draw[line width=1pt, dotted] (9.8, 1.1) -- (10.2, 1.1);
    \draw[line width=1pt, dotted] (9.8, -1.1) -- (10.2, -1.1);
    \draw[line width=1pt, dotted] (9.8, 2.5) -- (10.2, 2.5);
    \draw[line width=1pt, dotted] (9.8, -2.5) -- (10.2, -2.5);
    \draw[color=ForestGreen, line width=2pt] (-2,0) to [out=270, in=180] (-1, -0.5)
    to [out=0, in=270] (0,0);
    \draw[dashed, color=ForestGreen, line width=2pt] (-2,0) to [out=90, in=180] (-1, 0.5)
    to [out=0, in=90] (0,0);
    \draw[blue, line width=1.5pt, ->] (0,0) to [out=90, in=180] (2, 1.7);
    \draw[blue, line width=1.5pt] (2, 1.7) to [out=0, in=90] (4, 0);
    \draw[orange, line width=1.5pt, ->] (0,0) to [out=270, in=180] (2, -1.7);
    \draw[orange, line width=1.5pt] (2, -1.7) to [out=0, in=270] (4, 0);
\draw[ForestGreen, fill=ForestGreen] (-2,0) circle (.7ex);
\draw[ForestGreen, fill=ForestGreen] (0,0) circle (.7ex);
\draw[cornellred, fill=cornellred] (4,0) circle (.7ex);
\draw[cornellred, fill=cornellred] (8,0) circle (.7ex); 
\node at (0.4, 0) {$0$};
\node at (-2.5, 0) {$\infty$};
\node at (4.8, 0) {$E^{\star}_{1, +}$}; 
\node at (8.6, 0) {$E^{\star}_{\dots}$}; 
\node at (11, 0) {$\cdots$};
\end{scope}
	\end{tikzpicture}
\caption{The spectral curve of the VMS. We show the branch cut ({\color{ForestGreen}green}) starting at $0$ and ending at $\infty$. Furthermore we visualize the two sheets: the physical sheet on top ({\color{LightBlue}light blue}) and the non-physical sheet at the bottom ({\color{darktangerine}light orange}). Those two sheets meet at the cut and at the saddle points \eqref{eq:saddlesVMS} ({\color{cornellred}red}). In addition we show the integration cycle ({\color{blue}blue}) of the suppressed instanton action \eqref{eq:action} associated to the saddle point $E^{\star}_{1, +}$ in the physical sheet. On the other hand, for the same saddle-point $E^{\star}_{1, +}$, we also show the integration cycle ({\color{orange}orange}) associated to the exponentially growing instanton \eqref{eq:naiveminussignforaction} on the non-physical sheet. Both contributions are seen by the asymptotics of the free energy (see section \ref{subsec:largeorderchecks}). This plot builds on previous graphics from \cite{mss22, sst23}.}
\label{fig:MSspectralcurve}
\end{figure}
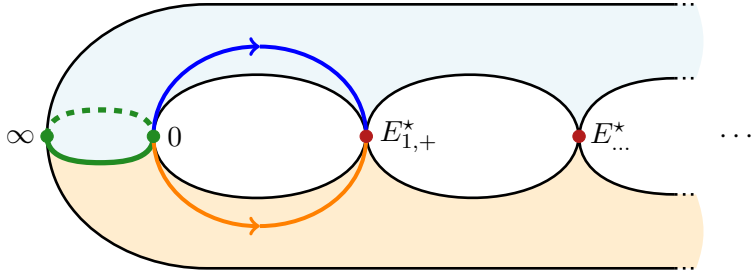
%%%%%%%%%%%%%%%%%%%%%%%%%%%%%%%%%%%%%%%%%%%%%%%%%%%%%%%%%%%%%%%%%

We are interested in non-perturbative corrections to the free energy \cite{cemr23}
\begin{equation}
\label{eq:freeEnergyVMS}
F_{\text{V}} \simeq \sum\limits_{g=0}^{\infty} F_{\text{V}, g} g_{\text{s}}^{2g-2},
\end{equation}
\noindent
where we use the subscript V to indicate that we are specialising to the VMS. This is an asymptotic series of factorial growth $F_{\text{V}, g}\sim(2g)!$ which already suggests the existence of exponential corrections. The explicit coefficients $F_{\text{V}, g}$ can be computed by running the topological recursion \cite{eo07a} on the associated spectral curve. For an energy $E$ it is given by\footnote{We follow exactly the conventions of \cite{cemr23}.}
\begin{equation}
\label{eq:spectralCurveVMS}
y_{\text{V}}(E) = -2\sqrt{2}\pi\frac{\sin\left(2\pi b \sqrt{-E}\right)\sin\left(2\pi b^{-1} \sqrt{-E}\right)}{\sqrt{-E}},
\end{equation}
\noindent
with the standard matrix model differential and Bergman kernel \cite{cemr23}. Furthermore we have the usual relation $c=1+6(b+b^{-1})^2$. This spectral curve has a square-root branch cut starting at the branch point $a=0$ and ending at positive infinity. This is shown in figure \ref{fig:MSspectralcurve}. See appendix \ref{app:data} for some explicit free energy coefficients.

\paragraph{Physical Sheet Instanton Corrections:}
We start by computing the well-known instanton corrections to the free energy \eqref{eq:freeEnergyVMS} of the form
\begin{equation}
\mathcal{F}^{(n)} \simeq \rme^{-n\frac{A}{g_{\text{s}}}} \left(\text{asymptotic series in }g_{\text{s}}\right),
\end{equation}
\noindent
by using the formulae developed in \cite{msw07, gs21, sst23}. Here non-perturbative contributions to the free energy of generic one-cut matrix models are computed through eigenvalue tunnelling, where $n$ counts the number of eigenvalues that have tunnelled to a non-perturbative saddle. Those formulae can be adapted to hold for the double scaling limit as well by moving one of the endpoints of the spectral curve to infinity \cite{gs21, sst23}. Let us quickly state the result for a general double scaled configuration and then specialize to the VMS:  focus on a saddle-point $E^{\star}$ of a generic double scaled curve of the form
\begin{equation}
\label{eq:genericCurve}
y(E) = M(E)\sqrt{-E},
\end{equation}
\noindent
with one cut running from the branch point $a=0$ to positive infinity. For the following discussion we require the moment function $M(E)$ to be analytic at the origin and furthermore $M^{\prime}(E^{\star})\neq 0$. Then the lowest contributing $g_{\text{s}}$ order to the one instanton is given by\footnote{Notice the minus sign in comparison to \cite{sst23}. This is because in \cite{sst23} the cut was assumed to run to negative infinity instead of positive infinity. The additional factor of 2 in comparison to \cite{sst23} in front of the integral \eqref{eq:action} and in front of the moment function is due to the fact that we have only considered a half cycle \cite{sst23}.} \cite{gs21, sst23}
\begin{equation}
\label{eq:generic-one-inst}
\mathcal{F}^{(1)} \simeq \rme^{-\frac{A}{g_{\text{s}}}}\left(\frac{1}{4}\sqrt{\frac{g_{\text{s}}}{4\pi M^{\prime}(E^{\star})(a-E^{\star})^{\frac{5}{2}}}}+O(g_{\text{s}}^{3/2})\right),
\end{equation}
\noindent
with the instanton action given by
\begin{equation}
\label{eq:action}
A = 2\int\limits_{a}^{E^{\star}} y(E)\text{d}E,
\end{equation}
\noindent
where we have chosen the integration path in the principal sheet of $y$ (see the blue integration contour in figure \ref{fig:MSspectralcurve}). Notice also the factor of $2$ that promotes the half cycle in \eqref{eq:action} to a full cycle. Specialising the above to the VMS we find, for each saddle point
\begin{equation}
\label{eq:saddlesVMS}
E^{\star}_{k, \pm} = -\frac{k^2 b^{\pm 2}}{4},\qquad k\in\mathbb{Z}_{\geq 1},
\end{equation}
\noindent
the corresponding free energy contribution and their instanton action
\begin{equation}
\label{eq:oneinst}
\mathcal{F}^{(1)}_{\text{V}, k,\pm} \simeq \frac{\sqrt{(-1)^k g_{\text{s}}}}{4\cdot 2^{3/4} \pi^{3/2}k\sqrt{b^{\pm 1}\sin(\pi k b^{\pm 2})}}\rme^{-\frac{A_{k,\pm}}{g_{\text{s}}}}+\cdots,\quad A_{k,\pm}=\frac{4\sqrt{2}b^{\pm1}(-1)^{k+1}\sin\left(\pi k b^{\pm 2}\right)}{1-b^{\pm 4}}.
\end{equation}
\noindent
This result indeed agrees with the one presented in \cite{cemr23}. For the asymptotic tests in section \ref{subsec:largeorderchecks} we will need the instanton contribution with the saddle closest to the origin ($k=1,+$). We have
\begin{equation}
\label{eq:oneinstclosest}
\mathcal{F}^{(1)}_{\text{V}, 1,+} \simeq \frac{\rmi \sqrt{g_{\text{s}}}}{4\cdot 2^{3/4} \pi^{3/2}\sqrt{b\sin(b^2\pi)}}\rme^{-\frac{1}{g_{\text{s}}}\frac{4\sqrt{2}b\sin(b^2\pi)}{1-b^4}}+\cdots\,.
\end{equation}

\begin{figure}
\centering
\includegraphics[scale=0.9]{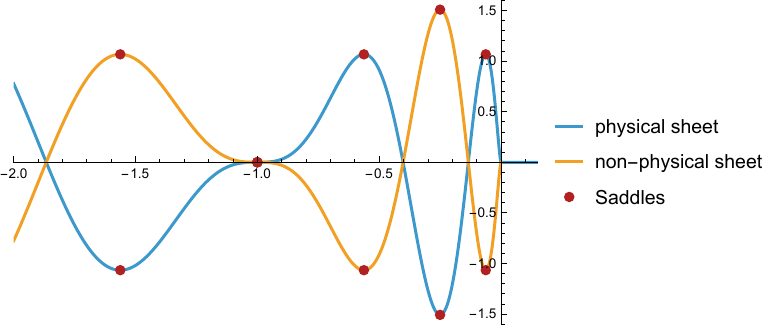}
\caption{Depiction of the effective potential $V_{\text{eff}}(E) = \int_0^{E}\text{d}E^{\prime} y(E^{\prime})$ as a function of $E$ at $b=1/2$. Same as the spectral curve \eqref{eq:spectralCurveVMS} it has two sheets: the physical sheet ({\color{LightBlue}light blue}) and the non-physical one ({\color{darktangerine}light orange}). Compare also to figure \ref{fig:MSspectralcurve}. The saddles ({\color{cornellred}red}) of the potential on the physical sheet are -- up to a factor of 2 -- exactly the resurgent instanton actions \eqref{eq:action} and we can indeed see that they oscillate between positive and negative values. In addition, taking into account the non-physical sheet we find additional saddle points ({\color{cornellred}red}) (On the uniformization cover each saddle point splits into two, see for example formula \eqref{eq:naiveminussignforaction}.), exactly with a relative minus sign in comparison with the physical sheet.}
\label{fig:Potentialplot}
\end{figure}

Having established the one-instanton correction it is natural to compute higher instanton contributions. The formula is most conveniently written as a ratio of partition functions and reads for a generic double scaled curve \cite{sst23}
\begin{equation}
\label{eq:generic-ell-instanton}
\frac{\mathcal{Z}^{(\ell)}(g_{\text{s}})}{\mathcal{Z}^{(0)}(g_{\text{s}})} \simeq \frac{G_{2}(\ell+1)}{\left(2\pi\right)^{\ell/2}}\, \rme^{-\frac{\ell}{g_{\text{s}}}2 \left(V_{\text{h;eff}}(E^{\star})-V_{\text{h;eff}}(a)\right)} \left(\frac{g_{\text{s}}}{32 M^{\prime}(E^{\star}) \left(a-E^{\star}\right)^{5/2}}\right)^{\frac{\ell^2}{2}} + \cdots,
\end{equation}
\noindent
where $G_2$ is the Barnes function and the holomorphic effective potential is the primitive of the spectral curve $V_{\text{h;eff}}^{\prime}(E)=y(E)$. This result means that to each saddle point we are led to associate an infinite tower of instantons -- integer multiples of the same action. In the matrix model such contributions are created by tunnelling multiple eigenvalues to the same saddle.  Notice furthermore that the starting genus is quadratic in the instanton number $\ell$ which is consistent with previous results \cite{msw08}. Specializing the above formula to the VMS we arrive at the multi-instanton prediction
\begin{equation}
\frac{\mathcal{Z}^{(\ell)}_{\text{V}, k,\pm}(g_{\text{s}})}{\mathcal{Z}^{(0)}_{\text{V}}(g_{\text{s}})} \simeq 2^{-\frac{1}{4}\ell(2+9\ell)}\pi^{-\frac{1}{2}\ell(1+2\ell)}G_{2}(1+\ell)\rme^{-\ell\frac{4\sqrt{2}b^{\pm1}(-1)^{k+1}\sin\left(\pi k b^{\pm 2}\right)}{g_{\text{s}}(1-b^{\pm 4})}}\left(\frac{g_{\text{s}}(-1)^k}{k^2 b^{\pm 1}\sin\left(k\pi b^{\pm 2}\right)}\right)^{\frac{\ell^2}{2}}\hspace{-0.2cm}+\cdots
\end{equation}
\noindent
Having computed those non-perturbative contributions we observe a peculiar fact. The instanton actions \eqref{eq:action} are not all positive (see also \eqref{eq:oneinst}) -- we do not only find falling exponentials from the physical sheet of the spectral curve. This is consistent with findings in minimal string theory \cite{gs21, sst23}. See figure \ref{fig:Potentialplot} for an illustration. This issue is exactly resolved through the discussion in \cite{mss22, sst23} that stipulates that one should also take into account contributions from the non-physical sheet. Indeed this gives additional resurgent instanton actions exactly with a relative minus sign -- the missing falling exponentials (see again figure \ref{fig:Potentialplot}). In fact, this means indeed that all instantons appear in resonant pairs (a falling and a growing exponential) -- exactly as expected from a string genus expansion (see also figure \ref{fig:innerworkingsofresurgence})\footnote{At this point one might wonder whether the original nomenclature of physical and non-physical sheet still makes sense for the VMS -- as there are falling (meaning physical) exponentials also on the non-physical sheet.} \cite{sst23}.

%%%%%%%%%%%%%%%%%%%%%%%%%%%%%%%%%%%%%%%%%%%%%%%%%%%%%%%%%%%%%%%%%
%%%%%%%%%%%%%%%%%%%%%%%%%%%%%%%%%%%%%%%%%%%%%%%%%%%%%%%%%%%%%%%%%

\paragraph{Instantons from the Non-Physical Sheet:}
After having computed known non-perturbative corrections, we now turn to uncover new resurgent contributions: as mentioned already in the introduction, the VMS is \emph{resonant}, meaning for each resurgent instanton action $A$ it will have a sibling instanton action $-A$ \cite{mss22}. This can be seen at the level of a generic spectral curve of the form \eqref{eq:genericCurve} as follows: the curve, as a function of $E$, has square-root double-sheeted structure. We introduce the uniformizing coordinate $z^2=-E$ with the involution map $\sigma$ given by\footnote{Details on the subsequent discussion can be found in \cite{mss22}.}
\begin{equation}
z\to\sigma(z) = -z,
\end{equation}
\noindent
such that 
\begin{equation}
\label{eq:involution-map}
E(\sigma(z)) = E(z),\quad \mathcal{Y}(\sigma(z)) = -\mathcal{Y}(z),
\end{equation}
\noindent
where on the principal sheet we have identified $\mathcal{Y}(z)=y(E(z))$ -- meaning $\mathcal{Y}(z)$ is the spectral curve in uniformized variables. Notice that in the uniformization formulation each saddle-point $E^{\star}$ now splits into two saddles that we will denote by 
\begin{equation}
z^{\star},\quad\bar{z}^{\star} = -z^{\star}.
\end{equation}
\noindent
Also, in our nomenclature we follow \cite{mss22} and by $\bar{\bullet}$ we mean the non-physical sheet and not complex conjugation. With the above terminology we are ready to reformulate our computation of the resurgent instanton action, but this time taking all sheets into account. On the principal sheet we still have exactly formula \eqref{eq:action} but on the involuted sheet we find \cite{mss22}
\begin{align}
\label{eq:naiveminussignforaction}
\bar{A} &= 2\int_{0}^{\bar{z}^{\star}} \text{d}z\, \frac{\partial E}{\partial z}\, \mathcal{Y}(z) = 2\int_{0}^{z^{\star}} \text{d}z\, \frac{\partial E}{\partial z}\, \mathcal{Y}(\sigma(z)) = - 2\int_{0}^{z^{\star}} \text{d}z\, \frac{\partial E}{\partial z}\, \mathcal{Y}(z) \equiv -A.
\end{align}
\noindent
This means that for each saddle-point $E^{\star}$ we should not expect to find just one instanton action but two -- one for each sheet. The integration contours on the spectral curve are explicitly visualized in figure \ref{fig:MSspectralcurve}. For this reason it is common to introduce the notation
\begin{equation}
\mathcal{F}^{(n|m)} \simeq \rme^{-(n-m)\frac{A}{g_{\text{s}}}} \left(\text{asymptotic series in }g_{\text{s}}\right),
\end{equation}
\noindent
where $n,m$ label multiples of the exponential contributions associated to the uniformized saddles $z^{\star}, \bar{z}^{\star}$ respectively. Such contributions have been computed extensively in \cite{mss22, sst23} and we want to apply the formulae derived therein to the VMS. The simplest such contribution is that of a single involuted sheet instanton, which for the generic curve \eqref{eq:genericCurve} reads \cite{sst23}
\begin{align}
\label{eq:backwardforward}
\mathcal{F}^{(0|1)} &\simeq \rme^{\frac{A}{g_{\text{s}}}}\frac{1}{4}\sqrt{\frac{g_{\text{s}}}{-4\pi M^{\prime}(E^{\star})(a-E^{\star})^{\frac{5}{2}}}}+\cdots=\nonumber\\
&\simeq-\rmi\rme^{\frac{A}{g_{\text{s}}}}\frac{1}{4}\sqrt{\frac{g_{\text{s}}}{4\pi M^{\prime}(E^{\star})(a-E^{\star})^{\frac{5}{2}}}}+\cdots, 
\end{align}
\noindent
where the correct sign choice in the second equality was determined in \cite{sst23}. For the VMS this contribution then reads
\begin{equation}
\label{eq:backwardinstanton}
\mathcal{F}^{(0|1)}_{\text{V}, k,\pm} \simeq \frac{-\rmi\sqrt{(-1)^k g_{\text{s}}}}{4\cdot 2^{3/4} \pi^{3/2}k\sqrt{b^{\pm 1}\sin(\pi k b^{\pm 2})}}\rme^{\frac{1}{g_{\text{s}}}\frac{4\sqrt{2}b^{\pm1}(-1)^{k+1}\sin\left(\pi k b^{\pm 2}\right)}{1-b^{\pm 4}}}+\cdots.
\end{equation}
\noindent
Let us comment on this expression: it is very close to its resonant sibling \eqref{eq:oneinst} -- the only difference being a global phase and the sign in the exponential. Though small, these differences are crucial for the asymptotic analysis which will be explored in detail in section \ref{subsec:largeorderchecks}. Furthermore, one might worry about the sign in the exponential contribution and the physical instabilities that it might carry: here it is important that in this work we never consider such contributions on their own but solely as non-perturbative corrections to the free energy -- which we can control using resurgence methods\footnote{In fact we will see in section \ref{sec:exactZ} how such growing instanton effects can contribute to a Zak transform formulation of the partition function. See also \cite{krsst25a}.}.

Having established the simplest resonant instanton let us turn to a rather non-trivial contribution: the $(1|1)$ sector. From \cite{ot06} one might get the impression that the annulus diagram between a brane and a negative tension brane should be trivial but that turns out not to be the case: indeed this contribution does not come with exponentials attached, but one still finds a non-trivial asymptotic series. The generic prediction for its lowest order reads \cite{mss22, sst23}
\begin{equation}
\label{eq:11-prediction}
\mathcal{F}^{(1|1)} \simeq \frac{\rmi}{2\pi g_{\text{s}}}A+O(g_{\text{s}}),
\end{equation}

\noindent
meaning for the VMS at the saddle point $E^{\star}_{k, \pm}$ we find
\begin{equation}
\mathcal{F}^{(1|1)}_{\text{V}, k,\pm} \simeq \frac{\rmi}{2\pi g_{\text{s}}}\frac{4\sqrt{2}b^{\pm1}(-1)^{k+1}\sin\left(\pi k b^{\pm 2}\right)}{1-b^{\pm 4}}+O(g_{\text{s}}).
\end{equation}

As a last non-trivial resonant computation we want to present a result for the $(2|1)$ contribution. For details on this contribution we refer the reader to \cite{mss22, sst23} and we will here merely present the result. For an arbitrary curve of the form \eqref{eq:genericCurve} it is most conveniently written as
\begin{align}
\label{eq:dsl-21-sector}
\frac{\mathcal{Z}^{(2|1)} (g_{\text{s}})}{\mathcal{Z}^{(0|0)} (g_{\text{s}})} - \frac{\mathcal{Z}^{(1|0)} (g_{\text{s}})}{\mathcal{Z}^{(0|0)} (g_{\text{s}})}\, \frac{\mathcal{Z}^{(1|1)} (g_{\text{s}})}{\mathcal{Z}^{(0|0)} (g_{\text{s}})} &\simeq \rme^{-\frac{1}{g_{\text{s}}} 2\left(V_{\text{h;eff}}(E^{\star}) - V_{\text{h;eff}}(a)\right)} \times \\
&
\hspace{-100pt}
\times \frac{\rmi}{8 \left(2\pi\right)^2}\, \sqrt{\frac{\pi g_{\text{s}}}{M^{\prime}(E^{\star}) \left(a-E^{\star}\right)^{5/2}}} \left\{ 2 \gamma_{\text{E}} + \log \left( \frac{2^{10}}{g_{\text{s}}^2}\, M^{\prime}(E^{\star})^2 \left(a-E^{\star}\right)^{5} \right) \right\} + \cdots, \nonumber
\end{align}
\noindent
where $\gamma_{\text{E}}$ is the Euler–Mascheroni constant. Plugging in the VMS spectral curve we arrive at
\begin{align}
\label{eq:dsl-21-sector-vms}
\frac{\mathcal{Z}^{(2|1)}_{\text{V}, k,\pm} (g_{\text{s}})}{\mathcal{Z}^{(0|0)}_{\text{V}} (g_{\text{s}})} - \frac{\mathcal{Z}^{(1|0)} _{\text{V}, k,\pm}(g_{\text{s}})}{\mathcal{Z}^{(0|0)}_{\text{V}}(g_{\text{s}})}\, \frac{\mathcal{Z}^{(1|1)}_{\text{V}, k,\pm} (g_{\text{s}})}{\mathcal{Z}^{(0|0)}_{\text{V}} (g_{\text{s}})} &\simeq \rme^{-\frac{1}{g_{\text{s}}}\frac{4\sqrt{2}b^{\pm1}(-1)^{k+1}\sin\left(\pi k b^{\pm 2}\right)}{1-b^{\pm 4}}} \times \\
&
\hspace{-100pt}
\times \frac{\rmi}{16\cdot 2^{3/4}\pi^{5/2}k}\sqrt{\frac{g_{\text{s}}(-1)^{k}}{b^{\pm 1}\sin(\pi b^{\pm 2} k)}} \left\{ 2 \gamma_{\text{E}} + \log \left( \frac{2^9 b^{\pm 2} k^4\pi^4\sin^2(b^{\pm 2} k\pi)}{g_{\text{s}}^2} \right) \right\} + \cdots. \nonumber
\end{align}
\noindent
To finish this section let us further comment that the above contributions can be conveniently assembled into a (schematic) transseries structure. Including the contributions from just one singularity it is of the form
\begin{equation}
F \simeq \sum\limits_{n,m} \sigma_{1}^n\sigma_{2}^m F^{(n|m)}(g_{\text{s}}).
\end{equation}
\noindent
We can generalize this to multiple -- say $\kappa$ -- saddles where we arrive at
\begin{equation}
\label{eq:free-energy-transseries}
F \simeq \sum\limits_{n,m\in\mathbb{N}^{\kappa}} \prod\limits_{i=1}^{\kappa}\sigma_{i,1}^{n_i}\sigma_{i,2}^{m_i} F^{(n_1|m_1), \dots, (n_{\kappa}, m_{\kappa})}(g_{\text{s}}),
\end{equation}
\noindent
which for the partition function $Z = \exp\left(F\right)$ reads analogously.
Notice that in all our above computations the transseries parameters $\sigma_i$ have been set to specific Stokes constants which is also why we have changed notation from curly to straight letters when writing the transseries \eqref{eq:free-energy-transseries}. Details on this can be found in \cite{gikm10, mss22, sst23} and we will elaborate on it further in the context of the partition function in section \ref{sec:exactZ}.

%%%%%%%%%%%%%%%%%%%%%%%%%%%%%%%%%%%%%%%%%%%%%%%%%%%%%%%%%%%%%%%%%
%%%%%%%%%%%%%%%%%%%%%%%%%%%%%%%%%%%%%%%%%%%%%%%%%%%%%%%%%%%%%%%%%
\subsection{On Negative Branes from Boundary Conformal Field Theory}
\label{subsec:bcft}
%%%%%%%%%%%%%%%%%%%%%%%%%%%%%%%%%%%%%%%%%%%%%%%%%%%%%%%%%%%%%%%%%
%%%%%%%%%%%%%%%%%%%%%%%%%%%%%%%%%%%%%%%%%%%%%%%%%%%%%%%%%%%%%%%%%

Having computed negative brane contributions using the matrix integral formalism in section \ref{subsec:instantonsVMM} we comment briefly on the analogous result from asymptotic boundaries. It was shown in \cite{cemr23} that the density for the double-scaled Virasoro matrix integral can be obtained from the disk partition function by inverse Laplace transform, which then by analytic continuation gives the spectral curve
\begin{equation}
\label{eq:scfromdisk}
\varrho_{\text{V}, 0}(E) = \int\limits_{-\rmi \infty+\gamma}^{\rmi \infty+\gamma}\frac{\text{d}\beta}{2\pi\rmi}\rme^{\beta E} Z_{\text{disk}}^{(b)}(\beta),\qquad y_{\text{V}}(E+\rmi 0) = \rmi\pi \varrho_{\text{V}, 0}(E),
\end{equation}
\noindent
where $\gamma\in\mathbb{R}_{+}$ such that we avoid the singularities of the disk partition function $Z_{\text{disk}}^{(b)}(\beta)$ to the right. Furthermore the parameter $\beta$ can be interpreted as a boundary length. For further details we refer the reader to the original paper \cite{cemr23}. Formula \eqref{eq:scfromdisk} implies that also $Z_{\text{disk}}^{(b)}(\beta)$ "sees" both sheets of the spectral curve. It therefore suggests that one should be able to reproduce the negative tension brane contribution also from a direct ZZ-disk computation. 

Let us quickly state the standard exponentiated ZZ-brane contribution. More specifically it was shown in \cite{cemr23} that the first exponential correction to the VMS free energy \eqref{eq:oneinst} can also be computed via
\begin{equation}
\mathcal{F}^{(1)}_{\text{V}, k,\pm} \simeq \exp\left(\DiscZZ{(k,\pm)} + \AnnulusZZ{\tiny (k,\pm)}{(k,\pm)} + \cdots\right),
\end{equation}
\noindent
where the unmarked disk and the annulus are computed with standard ZZ-boundary conditions in the $c$ theory and with "Half-" ZZ-boundary conditions in the $\hat{c}$ theory. Explicitly one considers the boundary state\footnote{We again follow exactly the notations in \cite{cemr23}.}
\begin{equation}
\label{eq:ZZboundarystate}
\ket{\text{ZZ}^{(b)}_{(1,1)}}\otimes\ket{\widehat{\text{ZZ}}^{(\rmi b)}_{k,\pm}}.
\end{equation}
\noindent
One should furthermore mention that the ZZ-annulus at face value diverges due to an improper use of the Siegel gauge condition and it is therefore subject to a regularization using string field theory \cite{emms22a, cemr23}. Here one divergent boson and two divergent fermion modes in the annulus integral 
\begin{equation}
\label{eq:annulusintegral}
\AnnulusZZ{\tiny (k,\pm)}{(k,\pm)} = \int\limits_{0}^{\infty}\frac{\text{d}t}{2t}\left(\sum\limits_{\text{b}}\rme^{-2\pi h_{\text{b}} t} - \sum\limits_{\text{f}}\rme^{-2\pi \hat{h}_{\text{f}} t}\right),
\end{equation}
\noindent
are cured via the assignment (for details we refer the reader to \cite{emms22a})
\begin{equation}
\label{eq:sftregularization}
\int\limits_{0}^{\infty}\frac{\text{d}t}{2t} \left(\rme^{2\pi t}+\rme^{-2\pi t} -2\right)\quad\longrightarrow\quad -\frac{1}{2}\log\left(-2^{5}\pi^{3}\, \DiscZZ{(k,\pm)}\right).
\end{equation}
\noindent

To compute negative tension contributions we are instructed to consider the ZZ-boundary state \eqref{eq:ZZboundarystate} with an overall minus sign\footnote{The attentive reader might wonder that this overall minus sign is just a normalization issue, but this does not seem to be the case as the effects are clearly visible in the asymptotics of the free energy (see section \ref{subsec:largeorderchecks}).
%In fact we will see that this minus sign fundamentally changes the annulus contribution between a brane and a negative tension brane.%%
}. This will produce a minus sign in front of the disk contribution but leave the annulus untouched -- up to the disk in the regularization prescription \eqref{eq:sftregularization}. This minus will contribute as a global prefactor of $1/\sqrt{-1}$. Upon the appropriate choice of sign for the square-root we find
\begin{equation}
\mathcal{F}^{(0|1)}_{\text{V}, k,\pm} \simeq -\rmi\exp\left(-\DiscZZ{(k,\pm)} + \AnnulusZZ{\tiny (k,\pm)}{(k,\pm)} + \cdots\right),
\end{equation}
\noindent
This is exactly the familiar pre-factor of $-\rmi$ between positive and negative tension instantons. Comparing with formula \eqref{eq:backwardinstanton} yields perfect agreement\footnote{One can furthermore try to reproduce the $(1|1)$ contribution in formula \eqref{eq:11-prediction} directly from conformal field theory. Such a computation has been performed in \cite{sst23} by integrating over the FZZT moduli space to arrive at ZZ-brane computations. A direct computation of \eqref{eq:11-prediction} from ZZ-branes along the lines of \cite{emms22a} would be very interesting.}.

%BCFT calculation (3a introduce CFT formulation of MVS (basically just the liouvilles are needed!); 3b-c-d-e repeat above following SST; 3f show BCFT matches matrix integrals)

%%%%%%%%%%%%%%%%%%%%%%%%%%%%%%%%%%%%%%%%%%%%%%%%%%%%%%%%%%%%%%%%%
%%%%%%%%%%%%%%%%%%%%%%%%%%%%%%%%%%%%%%%%%%%%%%%%%%%%%%%%%%%%%%%%%
\subsection{The JT Gravity Limit}
\label{subsec:JT}
%%%%%%%%%%%%%%%%%%%%%%%%%%%%%%%%%%%%%%%%%%%%%%%%%%%%%%%%%%%%%%%%%
%%%%%%%%%%%%%%%%%%%%%%%%%%%%%%%%%%%%%%%%%%%%%%%%%%%%%%%%%%%%%%%%%

It was established in the original work \cite{cemr23} that in the limit $b\to 0$ ($c\to\infty$) the VMS reduces to JT-gravity upon scaling $g_{\text{s}}$ by $(8\pi^2 b^2)^{3/2}$. Let us perform this map for the non-perturbative contributions computed above: This is a good check of our previous results because JT-gravity is known to be resonant -- to have both positive and negative exponential corrections \cite{gs21, sst23, eggls23}. Furthermore we focus our study on the saddle-point closest to the origin ($k=1, +$), as this is the one that has been checked numerically to great accuracy in the aforementioned publications. Let us start with the well-studied one-instanton contribution: we take formula \eqref{eq:oneinstclosest} and take the JT gravity limit. We then find
\begin{equation}
\mathcal{F}^{(1)}_{\text{V}, 1,+} \xrightarrow{\quad\text{JT}\quad} \mathcal{F}^{(1)}_{\text{JT}, 1} \simeq \rmi\sqrt{\frac{g_{\text{s}}}{2\pi}}\,\rme^{-\frac{1}{g_{\text{s}}}\frac{1}{4\pi^2}}+\cdots.
\end{equation}
\noindent
This agrees exactly with the results in \cite{sst23, eggls23}. Similarly we can proceed with the other non-perturbative contributions associated to higher instanton numbers
\begin{equation}
\frac{\mathcal{Z}^{(\ell)}_{\text{V}, 1,+}}{\mathcal{Z}^{(0)}_{\text{V}}} \xrightarrow{\quad\text{JT}\quad} \frac{\mathcal{Z}^{(\ell)}_{\text{JT}, 1}}{\mathcal{Z}^{(0)}_{\text{JT}}} \simeq \frac{G_2(1+\ell)}{(2\pi)^{\ell/2}}(-g_{\text{s}})^{\frac{\ell^2}{2}}\rme^{-\ell\frac{1}{g_{\text{s}}}\frac{1}{4\pi^2}}+\cdots .
\end{equation}
\noindent
We now come to the resonant contributions:
\begin{align}
\mathcal{F}^{(0|1)}_{\text{V}, 1,+} &\xrightarrow{\quad\text{JT}\quad} \mathcal{F}^{(0|1)}_{\text{JT}, 1} \simeq \sqrt{\frac{g_{\text{s}}}{2\pi}}\,\rme^{\frac{1}{g_{\text{s}}}\frac{1}{4\pi^2}}+\cdots,\\
\mathcal{F}^{(1|1)}_{\text{V}, 1,+} &\xrightarrow{\quad\text{JT}\quad} \mathcal{F}^{(1|1)}_{\text{JT}, 1} \simeq \frac{1}{g_{\text{s}}}\frac{\rmi}{8\pi^3} +\cdots.
\end{align}
\noindent
At this stage let us note that also for JT-gravity the $(1|1)$ contribution is not trivial. This was already established in \cite{sst23} by studying the JT limit of minimal string theory and our results agree with those presented therein. To finish this subsection let us also provide the JT limit of the $(2|1)$ contribution 
\begin{align}
\frac{\mathcal{Z}^{(2|1)}_{\text{V}, 1,+} (g_{\text{s}})}{\mathcal{Z}^{(0|0)}_{\text{V}} (g_{\text{s}})} &- \frac{\mathcal{Z}^{(1|0)} _{\text{V}, 1,+}(g_{\text{s}})}{\mathcal{Z}^{(0|0)}_{\text{V}}(g_{\text{s}})}\, \frac{\mathcal{Z}^{(1|1)}_{\text{V}, 1,+} (g_{\text{s}})}{\mathcal{Z}^{(0|0)}_{\text{V}} (g_{\text{s}})} \xrightarrow{\quad\text{JT}\quad}\frac{\mathcal{Z}^{(2|1)}_{\text{JT}, 1} (g_{\text{s}})}{\mathcal{Z}^{(0|0)}_{\text{JT}} (g_{\text{s}})} - \frac{\mathcal{Z}^{(1|0)} _{\text{JT}, 1}(g_{\text{s}})}{\mathcal{Z}^{(0|0)}_{\text{JT}}(g_{\text{s}})}\, \frac{\mathcal{Z}^{(1|1)}_{\text{JT}, 1} (g_{\text{s}})}{\mathcal{Z}^{(0|0)}_{\text{JT}} (g_{\text{s}})} \simeq\nonumber\\
&\simeq -\frac{1}{2\pi}\sqrt{\frac{g_{\text{s}}}{2\pi}}\left(\gamma_{\text{E}}-\log(g_{\text{s}})\right)\rme^{-\frac{1}{g_{\text{s}}}\frac{1}{4\pi^2}}+\cdots.
\end{align}
\noindent

%JT gravity limit (4a recall how MVS is a c-deformation of JT gravity; 4b-c-d-e take the limits of all above to JT gravity and match against GS, EGGLS)

%%%%%%%%%%%%%%%%%%%%%%%%%%%%%%%%%%%%%%%%%%%%%%%%%%%%%%%%%%%%%%%%%
%%%%%%%%%%%%%%%%%%%%%%%%%%%%%%%%%%%%%%%%%%%%%%%%%%%%%%%%%%%%%%%%%
\subsection{Large-Order Resurgent Checks}
\label{subsec:largeorderchecks}
%%%%%%%%%%%%%%%%%%%%%%%%%%%%%%%%%%%%%%%%%%%%%%%%%%%%%%%%%%%%%%%%%
%%%%%%%%%%%%%%%%%%%%%%%%%%%%%%%%%%%%%%%%%%%%%%%%%%%%%%%%%%%%%%%%%

%%%%%%%%%%%%%%%%%%%%%%%%%%%%%%%%%%%%%%%%%%%%%%%%%%%%%%%%%%%%%%%%%
\begin{figure}
\centering
\includegraphics[scale=0.8]{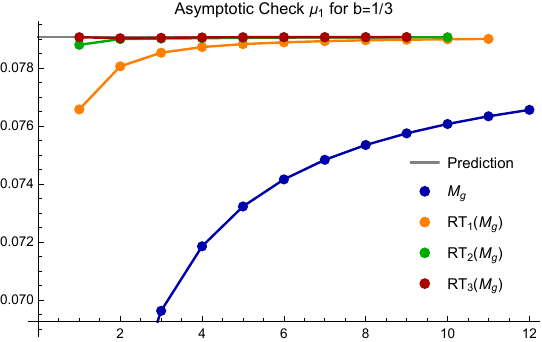}
\hspace{0.3cm}
\includegraphics[scale=0.8]{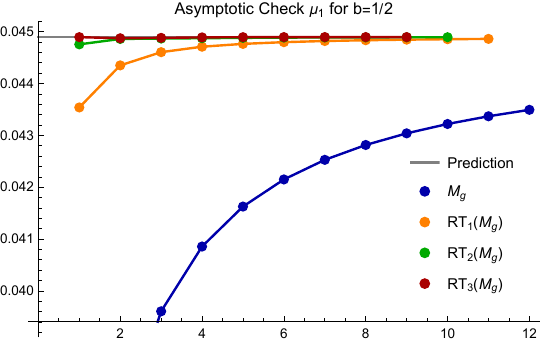}
\caption{Plot of the sequence $M_g(b)$ and its Richardson transforms of order $n$ $\text{RT}_n$ as a function of $g$ for the values $b=1/3$ (left) and $b=1/2$ (right). We also show the theoretical prediction in formula \eqref{eq:asymptoticscheck} in gray. Even though we do not have many terms in the asymptotic series we achieve an agreement of 5 digits with the predicted value.}
\label{fig:Asymptotics}
\end{figure}
%%%%%%%%%%%%%%%%%%%%%%%%%%%%%%%%%%%%%%%%%%%%%%%%%%%%%%%%%%%%%%%%%

%%%%%%%%%%%%%%%%%%%%%%%%%%%%%%%%%%%%%%%%%%%%%%%%%%%%%%%%%%%%%%%%%
\begin{figure}
\centering
\includegraphics[scale=1]{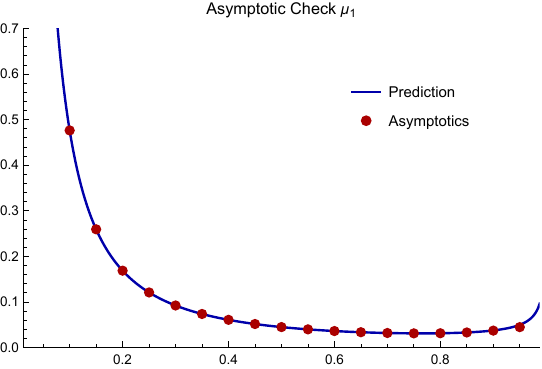}
\caption{Plot of the numerical extrapolation $M_{\infty}(b)$ as a function of $b$. We also show the theoretical prediction in formula \eqref{eq:asymptoticscheck} in blue. The precision is 4 to 5 digits for each point, but seems to drop a little bit close to $b=1$ exactly as expected in \cite{cemr23}. In fact a very similar plot has already appeared in \cite{cemr23}: The difference here is that here we got this plot from the corrected asymptotics \eqref{eq:asymptoticscheck} that take into account that the free energy is even.}
\label{fig:AsymptoticsVaryB}
\end{figure}
%%%%%%%%%%%%%%%%%%%%%%%%%%%%%%%%%%%%%%%%%%%%%%%%%%%%%%%%%%%%%%%%%

%%%%%%%%%%%%%%%%%%%%%%%%%%%%%%%%%%%%%%%%%%%%%%%%%%%%%%%%%%%%%%%%%
\begin{figure}
\centering
\includegraphics[scale=0.8]{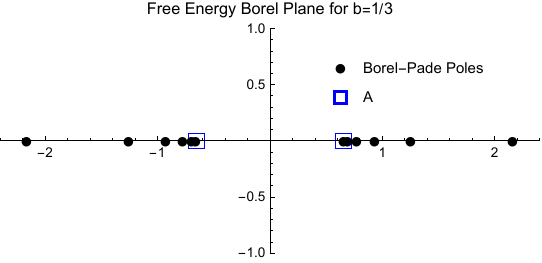}
\hspace{0.3cm}
\includegraphics[scale=0.8]{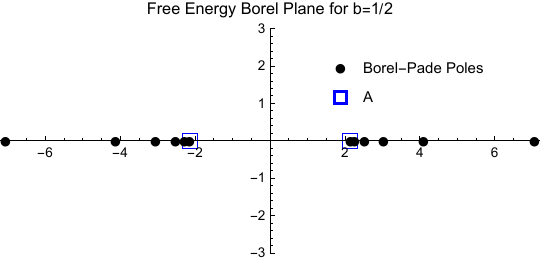}
\caption{Plots of the poles of the Borel Padé approximant of the free energy for the VMS in the Borel plane. We have again chosen the values $b=1/3$ (left) and $b=1/2$ (right). We clearly see the resonance phenomenon where the instanton actions appear in pairs $(A, -A)$ -- the Borel plane is indeed symmetric. Furthermore the locations of the Borel singularities agree exactly with the prediction for those said actions.}
\label{fig:borelfreeenergy}
\end{figure}
%%%%%%%%%%%%%%%%%%%%%%%%%%%%%%%%%%%%%%%%%%%%%%%%%%%%%%%%%%%%%%%%%

Having established non-perturbative contributions to the free energy of the VMS we want to check if they are seen by large-order resurgent asymptotic methods. The slight difference between the analysis in \cite{cemr23} and the one performed here is the inclusion of negative tension branes which completely explains the asymptotics of the genus expansion of the free energy. To this end let us here recall quickly how the asymptotics of generic \emph{even} asymptotic series are constructed \cite{msw07, gikm10, m12}: consider a generic asymptotic perturbative series
\begin{equation}
\label{eq:genericseries}
\sum\limits_{g\geq 0} a_g g_{\text{s}}^{g-2}.
\end{equation}
For an exponential contribution of the form
\begin{equation}
\rmi g_{\text{s}}^{\beta}\rme^{-\frac{A}{g_{\text{s}}}}\sum\limits_{g\geq 0} c_g g_{\text{s}}^g,
\end{equation}
\noindent
the asymptotic contribution to the series \eqref{eq:genericseries} reads
\begin{equation}
\label{eq:genericasymptotics}
a_k \sim \frac{1}{2\pi}\frac{\Gamma(k-2-\beta)}{A^{(k-2-\beta)}}\left(c_0 + c_1\frac{A}{(k-2-\beta-1)}+\cdots\right).
\end{equation}
\noindent
Notice how the perturbative series \eqref{eq:genericseries} in contrast to the free energy \eqref{eq:freeEnergyVMS} also contains odd terms in $g$ that equally appear in the asymptotic relation \eqref{eq:genericasymptotics}. This means that we are missing terms in the asymptotics that resolve this mismatch. This is exactly solved in our resonant setting: for each instanton action $A$ we are instructed to also consider its resonant sibling $-A$, which contributes to the asymptotics as
\begin{equation}
\label{eq:genericbackwardasymptotics}
a_k \sim \frac{1}{2\pi}\frac{\Gamma(k-2-\beta)}{(-A)^{(k-2-\beta)}}\left(\bar{c}_0 - \bar{c}_1\frac{A}{(k-2-\beta-1)}+\cdots\right).
\end{equation}
\noindent
Adding the two we find the total asymptotics
\begin{align}
a_k\sim \frac{1}{2\pi}\frac{\Gamma(k-2-\beta)}{A^{(k-2-\beta)}}
\begin{cases}
\left(c_0+\frac{\bar{c}_0}{(-1)^{\beta}}\right)+\left(c_1-\frac{\bar{c}_1}{(-1)^{\beta}}\right)\frac{A}{(k-2-\beta-1)}+\cdots, \quad & k\text{ even},\\
\left(c_0-\frac{\bar{c}_0}{(-1)^{\beta}}\right)+\left(c_1+\frac{\bar{c}_1}{(-1)^{\beta}}\right)\frac{A}{(k-2-\beta-1)}+\cdots, \quad & k\text{ odd},
\end{cases}
\end{align}
\noindent
where cancellation of the odd contributions implies
\begin{equation}
c_0-\frac{\bar{c}_0}{(-1)^{\beta}}=0,\quad c_1+\frac{\bar{c}_1}{(-1)^{\beta}}=0,\quad\cdots.
\end{equation}
\noindent
Notice how this is exactly consistent with the sign changes in our previous formulae between brane \eqref{eq:oneinstclosest} and negative brane contributions \eqref{eq:backwardforward}\footnote{This is where our asymptotic analysis differs from \cite{cemr23}: We take into account the fact that the perturbative series is even, which is why we find additional contributions.}.

Having reviewed the generic asymptotic formula for resonant problems let us apply it to the VMS. Here we focus on the saddle
\begin{equation}
\label{eq:extremalpoint}
E_{1,+}^{\star} = -b^2/4, \qquad 0< b < 1,
\end{equation}
\noindent
which is closest to the origin\footnote{Notice, that the saddle closest to the origin in the $E$ plane is the relevant one for the perturbative sector even though there might be other action whose absolute value is smaller. This is addressed in \cite{sst23}.}. Using formula \eqref{eq:oneinstclosest} and \eqref{eq:backwardforward} we find the leading large-order prediction
\begin{equation}
\label{eq:asymptoticscheck}
M_g(b) := \pi \frac{F_{\text{V},g}A^{2g-\frac{5}{2}}}{\Gamma\left(2g-\frac{5}{2}\right)}\quad \xrightarrow{g\to\infty}\quad\frac{1}{4\cdot 2^{3/4} \pi^{3/2}\sqrt{b\sin(b^2\pi)}},
\end{equation}
\noindent
where we remember 
\begin{equation}
\label{eq:ZZaction}
A = \frac{4\sqrt{2}b\sin(b^2\pi)}{1-b^4}
\end{equation}
\noindent
from formula \eqref{eq:oneinstclosest}. Let us now check the behavior of the auxiliary sequence $M_g(b)$ numerically. To this end we have produced explicitly the coefficients of the VMS free energy up to order $g=14$ using the methods provided in \cite{cemr23}. We plot the sequence $M_g(b)$ together for various values of $b$ in figure \ref{fig:Asymptotics}. Indeed in this way we verify formula \eqref{eq:asymptoticscheck} to 5 digits. Furthermore we can also plot the numerically extrapolated values for $M_{\infty}(b)$ against their analytical prediction. To this end we have taken the value of the third Richardson transform. This is visualized in figure \ref{fig:AsymptoticsVaryB}.

To finish the numerical exploration of the free energy let us give more evidence for the resonant behavior of the VMS in another way: For this we show a numerical approximation of the Borel plane singularities that is done via Padé approximants. The locations of those singularities correspond exactly to instanton actions\footnote{Notice that the poles of the Padé approximant are mimicking branch cuts, meaning that we have to identify the instanton actions with the locations where the pole accumulations start.}. The result is displayed in figure \ref{fig:borelfreeenergy}. Indeed we can clearly see the resonant behavior of the Borel plane, that predicts the instanton actions with both signs.

%%%%%%%%%%%%%%%%%%%%%%%%%%%%%%%%%%%%%%%%%%%%%%%%%%%%%%%%%%%%%%%%%
%%%%%%%%%%%%%%%%%%%%%%%%%%%%%%%%%%%%%%%%%%%%%%%%%%%%%%%%%%%%%%%%%
\subsection{Wall Crossing in the Virasoro Minimal String}
\label{subsec:wall-crossing}
%%%%%%%%%%%%%%%%%%%%%%%%%%%%%%%%%%%%%%%%%%%%%%%%%%%%%%%%%%%%%%%%%
%%%%%%%%%%%%%%%%%%%%%%%%%%%%%%%%%%%%%%%%%%%%%%%%%%%%%%%%%%%%%%%%%

%%%%%%%%%%%%%%%%%%%%%%%%%%%%%%%%%%%%%%%%%%%%%%%%%%%%%%%%%%%%%%%%%
\begin{figure}
\centering
\includegraphics[scale=0.8]{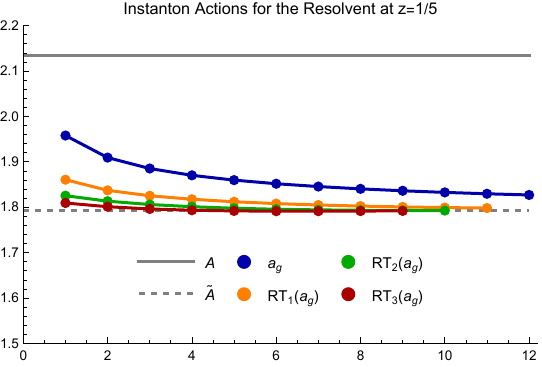}
\hspace{0.3cm}
\includegraphics[scale=0.8]{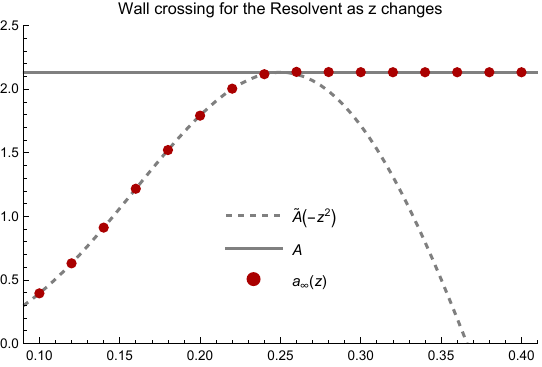}
\caption{Visualization of the wall crossing phenomenon for the resolvent. On the left we check the appearance of the instanton action $\tilde{A}(-z^2)$ while on the right we visualize how it can disappear from the asymptotics as we vary $z$. On the left we plot the sequence $a_g$ and its Richardson transforms of order $n$, namely $\text{RT}_n$, as a function of $g$. We show the theoretical prediction for the instanton action in formula \eqref{eq:actionresolvent} in gray for the case $z=1/5<1/4$ and $b=1/2$. Indeed the asymptotics follow the smaller instanton action $\tilde{A}(-z^2)$ in this case. On the right hand side we visualize the numerical extrapolation of $a_{\infty}$ ($y$-axis) as we vary $z$ ($x$ axis) together with the theoretical prediction of the instanton actions. We still choose $b=1/2$. Indeed we observe exactly the prediction \eqref{eq:actionresolvent} where the asymptotics sees the action $\tilde{A}(-z^2)$ until we reach $z=b/2$ and then the action $A$ takes over.}
\label{fig:tworesolventactions}
\end{figure}
%%%%%%%%%%%%%%%%%%%%%%%%%%%%%%%%%%%%%%%%%%%%%%%%%%%%%%%%%%%%%%%%%

%%%%%%%%%%%%%%%%%%%%%%%%%%%%%%%%%%%%%%%%%%%%%%%%%%%%%%%%%%%%%%%%%
\begin{figure}
\centering
\includegraphics[scale=0.85]{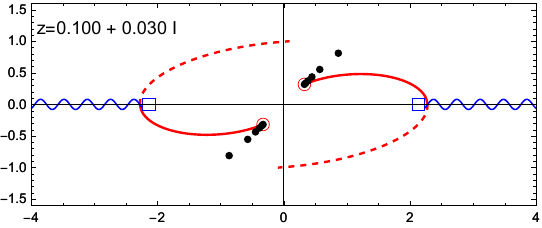}
\hspace{0.2cm}
\includegraphics[scale=0.85]{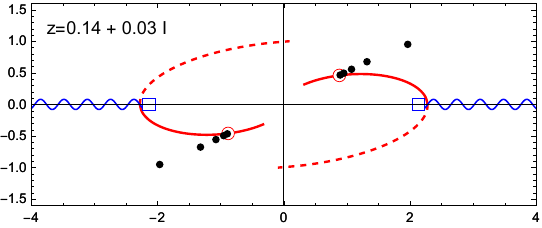}
\includegraphics[scale=0.85]{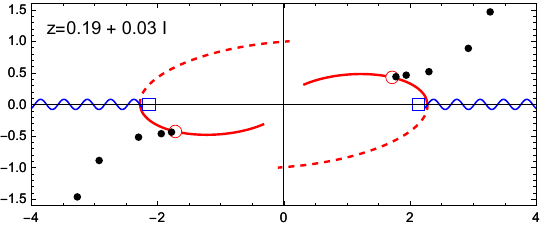}
\hspace{0.2cm}
\includegraphics[scale=0.85]{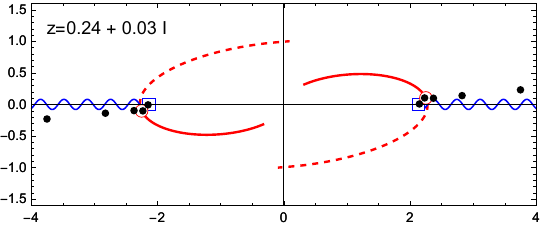}
\includegraphics[scale=0.85]{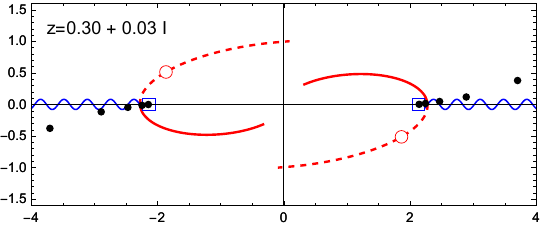}
\hspace{0.2cm}
\includegraphics[scale=0.85]{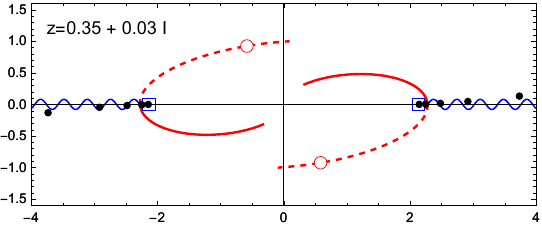}
\caption{Visualization of the wall crossing phenomenon for the resolvent. We plot the Borel plane for the resolvent \eqref{eq:resolvent}. The ZZ action $A$ is shown as blue squares where the branch cut of the associated logarithmic singularity at $A$ is visualized as a wiggly blue line. We want to study the movement of the singularity $\tilde{A}(-z^2)$ shown as a red path, where the small circles correspond to the value of $\tilde{A}(-z^2)$ for the various plots. Here we have taken $z$ slightly above the real line to move from $1/10+\rmi\epsilon$ to $7/20+\rmi\epsilon$ and $b=1/2$ (the plots are ordered from left to right, top to bottom). Each plot further shows in black the poles of the Borel-Padé approximant for the assigned value of $z$. Indeed we see how the singularity $\tilde{A}(-z^2)$ vanishes from the numerics as soon as it passes the branch cut of the stationary singularity $A$ even though $|\tilde{A}|<|A|$. The underlying reason is that this singularity left the principal sheet of the Borel transformed resolvent. This is exactly the phenomenon of wall crossing. The presented plot builds on similar plots from \cite{eggls23}.}
\label{fig:WCresolvent}
\end{figure}
%%%%%%%%%%%%%%%%%%%%%%%%%%%%%%%%%%%%%%%%%%%%%%%%%%%%%%%%%%%%%%%%%

Having studied the free energy it is natural to extend this discussion to correlation functions -- the simplest example being the resolvent. We want to study
\begin{equation}
\label{eq:resolvent}
R_1(E) = \left\langle\text{Tr}\frac{1}{E-M}\right\rangle \simeq \sum\limits_{g=0}^{\infty} g_{\text{s}}^{2g-1} R_{g,1}\left(E\right),
\end{equation}
\noindent
where we again follow the notation and conventions from \cite{cemr23} and the coefficients $R_{g,1}\left(-z^2\right)$ can be computed from the VMS curve \eqref{eq:spectralCurveVMS} using topological recursion\footnote{See for example formula (5.16) in \cite{cemr23}.}. Further in this subsection we will often use the uniformization variable $z$ given by $E=-z^2$. In the resolvent yet again new non-perturbative contributions appear: because of the explicit dependence of the resolvent on $z$ we expect FZZT-like\footnote{In this paper when referring to FZZT branes in the VMS we mean the hermitian matrix model analogue with instanton action \eqref{eq:atilde} -- which is expected to be consistent with putting FZZT boundary conditions on the spacelike Liouville sector and ZZ boundary conditions on the timelike one in the VMS \cite{cemr23}.} non-perturbative effects to play a role \cite{eggls23}. They are expected to have the instanton action 
\begin{align}
\label{eq:atilde}
\tilde{A}(E) &= 2\int\limits_{0}^{E} y_{\text{V}}(E) \text{d}E = \\
&= \frac{2\sqrt{2}b}{b^4-1}\left(\left(1+b^2\right)\sin\left[2\pi\left(b-\frac{1}{b}\right)\sqrt{-E}\right]+\left(1-b^2\right)\sin\left[2\pi\left(b+\frac{1}{b}\right)\sqrt{-E}\right]\right),\nonumber
\end{align}
\noindent
which, in contrast to the ZZ instanton \eqref{eq:ZZaction}, is $E$ dependent. Of those two actions ($\tilde{A}(E)$ for FZZT and $A$ for ZZ contributions) the asymptotics will see the leading one -- but only if it is turned on by resurgent relations. In other words it is natural to expect wall crossing phenomena to appear similarly to the minimal string discussion in \cite{eggls23}\footnote{Notice that we are here referring to the resurgent wall crossing phenomenon, which just means that some Borel plane singularities switch sheet. This is a more general phenomenon than the well known discussion from supersymmetric gauge theories in \cite{gmn09}.}. The natural expectation is that the wall (the crossing location) is located where the two actions agree: exactly at $z = \sqrt{-E_{1,+}^{\star}} = b/2$ as in formula \eqref{eq:extremalpoint}. For positive real $z$ this translates to the following prediction for the asymptotics
\begin{equation}
\label{eq:actionresolvent}
a_g(z) := \sqrt{\frac{4 g^2 R_{g,1}(-z^2)}{R_{g+1,1}(-z^2)}} \quad \xrightarrow{g\to\infty}\quad 
\begin{cases}
\tilde{A}(-z^2),\quad & z < \frac{b}{2},\\
A,\quad & z\geq \frac{b}{2}.
\end{cases}
\end{equation}
\noindent
To check this let us first see that the instanton action \eqref{eq:atilde} can really contribute to the asymptotics: fix the values $b=1/2, z=1/5$ which implies $\tilde{A} = 2/3\cdot\sqrt{5+\sqrt{5}} < A = 32/15$. Then again using the terms up to $g=14$ we can confirm the relation \eqref{eq:actionresolvent} with 3 digits after employing 3 Richardson transforms. The check is visualized on the left of figure \ref{fig:tworesolventactions}. 

Furthermore we can track the value of $a_{g\to\infty}$ as we vary $z$, using the approximation that we get from the third Richardson transform. This is shown on the right in figure \ref{fig:tworesolventactions}. We clearly see how we first follow the action $\tilde{A}(-z^2)$ until we reach $z=b/2$ where $A$ takes over. 

Let us understand this wall crossing phenomenon further and give a detailed account of the whereabouts of the $\tilde{A}(-z^2)$ action for $z>b/2$: as shown in figure \ref{fig:tworesolventactions} clearly $\tilde{A}(-z^2) < A$ is still true for $z>b/2$, yet the action no longer contributes to the asymptotics. The reason for this is that the Stokes constant associated to this action is turned off by a wall crossing phenomenon -- this means that the action $\tilde{A}(-z^2)$ leaves the principal sheet of the Borel plane. This can be checked explicitly using Padé-approximants and the analysis is presented in figure \ref{fig:WCresolvent}: We vary the value of $z$ on the real line with a slight imaginary offset such that the FZZT action \eqref{eq:atilde} passes the branch cut created by the ZZ-brane \eqref{eq:ZZaction}. Then we visualize the Borel plane of the resolvent for various values of $z$ along that path to check when the FZZT action is turned on or off. Indeed as shown in figure \ref{fig:WCresolvent} once the FZZT action passes the branch cut, it leaves the principal sheet of the Borel plane and is no longer visible -- it is turned off by a wall crossing phenomenon. We conclude that there are indeed FZZT like, non-perturbative contributions $\tilde{A}(-z^2)$ in the VMS, that are moreover subject to wall crossing phenomena. It is natural to expect that the same will hold for higher resolvents.

%%%%%%%%%%%%%%%%%%%%%%%%%%%%%%%%%%%%%%%%%%%%%%%%%%%%%%%%%%%%%%%%%
%%%%%%%%%%%%%%%%%%%%%%%%%%%%%%%%%%%%%%%%%%%%%%%%%%%%%%%%%%%%%%%%%
\section{Non-Perturbative Partition Function for the Virasoro Minimal String}
\label{sec:exactZ}
%%%%%%%%%%%%%%%%%%%%%%%%%%%%%%%%%%%%%%%%%%%%%%%%%%%%%%%%%%%%%%%%%
%%%%%%%%%%%%%%%%%%%%%%%%%%%%%%%%%%%%%%%%%%%%%%%%%%%%%%%%%%%%%%%%%

In this section we present a non-perturbative partition function for the VMS, meaning the complete resurgent family of non-perturbative completions indexed by 2 transseries parameters for each saddle point. The construction is based on \cite{krsst25a} and is supported by our previous computations of non-perturbative VMS effects in section \ref{sec:VMS}. We present the non-perturbative partition function in subsection \ref{subsec:generic-exactZ}. Finally we briefly discuss a sensible choice of non-perturbative completion in \ref{subsec:non-pert-com}.

%%%%%%%%%%%%%%%%%%%%%%%%%%%%%%%%%%%%%%%%%%%%%%%%%%%%%%%%%%%%%%%%%
%%%%%%%%%%%%%%%%%%%%%%%%%%%%%%%%%%%%%%%%%%%%%%%%%%%%%%%%%%%%%%%%%
\subsection{Generic Non-Perturbative Completion of the Partition Function}
\label{subsec:generic-exactZ}
%%%%%%%%%%%%%%%%%%%%%%%%%%%%%%%%%%%%%%%%%%%%%%%%%%%%%%%%%%%%%%%%%
%%%%%%%%%%%%%%%%%%%%%%%%%%%%%%%%%%%%%%%%%%%%%%%%%%%%%%%%%%%%%%%%%

We can assemble the non-perturbative contributions computed above into a closed form transseries expression for the partition function given in terms of a Zak transform. 

For this, let us first review a recent result for hermitian matrix models (HMM). Here, such a partition function was conjectured in  \cite{krsst25a}. We study a hermitian matrix model with a polynomial potential $V(\lambda)$ with $\kappa+1$ saddles\footnote{Meaning we have $\kappa$ non-perturbative saddles and one perturbative one.} and $N$ eigenvalues. Here we will be interested in the solution at large $N$, expanded around the one-cut configuration in the perturbative saddle with total 't Hooft parameter $t=g_{\text{s}}N$. The generic proposal comes from intuition in matrix integrals: Non-perturbative contributions are associated to eigenvalues that are tunnelling to other saddles of the spectral curve -- where those eigenvalues accumulate and open new cuts \cite{msw08}. The final result is the following, non-perturbative partition function for a hermitian matrix model\footnote{Here the convention of $\upmu_i$ has been chosen such that we have maximal symmetry between the backward and the forward transseries sectors. For details see \cite{krsst25a}.}\,\footnote{It is interesting to compare the partition function \eqref{eq:HMM-partition-function-conjecture} with a similar result in \cite{em08}. In fact \eqref{eq:HMM-partition-function-conjecture} can be interpreted as the one cut limit of the result in \cite{em08}.} \cite{krsst25a}
\begin{align}
\label{eq:HMM-partition-function-conjecture}
Z^{\text{HMM}} &\simeq \sum\limits_{\ell\in\mathbb{Z}^{\kappa}}\prod\limits_{i=1}^{\kappa}\uprho_{i}^{\ell_i}\frac{\mathcal{Z}_{\text{G}}\left(\ell_i+\frac{\upmu_i}{2\pi\rmi}\right)}{\mathcal{Z}_{\text{G}}\left(\frac{\upmu_i}{2\pi\rmi}\right)}\times\\
&\times\exp\left[\sum\limits_{g=0}^{\infty}\hat{F}_{g}^{\text{HMM}}\left(t-g_{\text{s}}\left(\sum\limits_{i=1}^{\kappa}\ell_i+\frac{\upmu_i}{2\pi\rmi}\right),g_{\text{s}}\left(\ell_1+\frac{\upmu_1}{2\pi\rmi}\right), \dots, g_{\text{s}}\left(\ell_{\kappa}+\frac{\upmu_{\kappa}}{2\pi\rmi}\right)\right) g_{\text{s}}^{2g-2}\right],\nonumber
\end{align}
\noindent
where the gaussian partition function is computed as
\begin{equation}
Z_{\text{G}}(N) = \frac{g_{\text{s}}^{\frac{N^2}{2}}}{(2\pi)^{\frac{N}{2}}}G_{2}(1+N),
\end{equation}
\noindent
and $\hat{F}_{g}^{\text{HMM}}(t_1, \dots, t_{\kappa+1})$ are the regularized multi-cut free energies of genus $g$, where each partial 't Hooft parameter $t_i$ controls the size of the cut opened around the $i$-th saddle $E_i^{\star}$. It is related to the standard multi-cut free energy $F_g^{\text{HMM}}\left(t_1,\dots,t_{\kappa+1}\right)$ by subtracting the gaussian terms at each genus for each non-perturbative saddle \cite{msw08, ms24}
\begin{equation}
\hat{F}_{g}^{\text{HMM}}\left(t_1,\dots,t_{\kappa+1}\right) = F_g^{\text{HMM}}\left(t_1,\dots,t_{\kappa+1}\right) - \sum\limits_{i=2}^{\kappa+1} \mathcal{F}_{g}^{\text{G}}(t_i),
\end{equation}
\noindent
where the asymptotic expansion of the gaussian reads
\begin{equation}
\mathcal{F}_0^{\text{G}}(t) = \frac{t^2}{2}\left(\log(t)-\frac{3}{2}\right),\quad \mathcal{F}_1^{\text{G}}(t) = -\frac{1}{12}\log(t),\quad \mathcal{F}_g^{\text{G}}(t) = \frac{B_{2g}}{2g(2g-2)}t^{2-2g};\,\,g\geq 2.
\end{equation}
\noindent
Last but not least we have $2\kappa$ transseries parameters $\uprho_i,\upmu_i; i=1,\dots,\kappa$, whose choice constitutes a non-perturbative completion of the partition function (see subsection \ref{subsec:non-pert-com}). Notice also that this is an asymptotic series, which still needs to be resummed. For further details we refer the reader to \cite{em08, krsst25a}. Let us also mention for completeness that there are 2 ways of computing the multi-cut free energy $F_g^{\text{HMM}}\left(t_1,\dots,t_{\kappa+1}\right)$:
\begin{itemize}
\item It can be constructed directly from the original HMM potential via multi-cut spectral geometry and topological recursion -- which, having the VMS in mind, is not useful as we do not have such an underlying potential\footnote{A proposal for constructing a multi-cut curve underlying the double scaled analogue of $F^{\text{HMM}}$ will be made in \cite{ss25}.}.
\item It can be constructed as a series around the one-cut setting $\hat{F}_{g}^{\text{HMM}}\left(t,0,\dots,0\right)$ where each term can be computed from the original one-cut curve \cite{msw08}. Notice that this type of series shows up naturally when expanding \eqref{eq:HMM-partition-function-conjecture} in $g_{\text{s}}$. The simplest such example
\begin{equation}
\label{eq:der-free-energy-for-action}
\partial_{\nu_i} \hat{F}_{0}^{\text{HMM}}\left.\left(t-\sum\limits_{j=1}^{\kappa}\nu_j,\nu_1,\dots, \nu_{\kappa}\right)\right|_{\nu_1=\dots=\nu_{\kappa}=0} = -\int\limits_{a}^{E_i^{\star}}y^{\text{HMM}}(E)\text{d}E
\end{equation}
\noindent
gives exactly the instanton action \eqref{eq:action} up to a minus sign\footnote{In comparison to the $s$ variable commonly introduced in matrix models \cite{msw08, krsst25a} there is exactly a minus sign in comparison with the $\nu_i$ variable used here.} and the familiar factor of $2$ that is due to the half-cycle integration in the matrix model as opposed to the full-cycle in the minimal string \cite{sst23}.
\end{itemize}
\noindent
We now make the case that a similar construction as in \cite{em08, krsst25a} also works for the VMS: The perturbative content of the VMS can be computed via topological recursion -- a method born from hermitian matrix models. Moreover we have demonstrated in subsections \ref{subsec:instantonsVMM}, \ref{subsec:largeorderchecks} and \ref{subsec:wall-crossing} above, how the non-perturbative contributions of the VMS can be computed with the non-perturbative hermitian matrix model formalism developed in \cite{msw07, msw08, mss22, sst23}. In fact we had emphasized how all the formulae we used are valid for generic one cut spectral curves. This suggests that the non-perturbative partition function for the VMS should be an analogue of \eqref{eq:HMM-partition-function-conjecture}, where the VMS version of the regularized multi-cut free energy can be computed as a series in terms of the VMS spectral curve as in \eqref{eq:der-free-energy-for-action}. To write the explicit partition function for the VMS however we need to address the fact that the spectral curve \eqref{eq:spectralCurveVMS} has infinitely many zeros:  therefore let us momentarily assume that only $\kappa$ saddles of the VMS are populated with eigenvalues\footnote{Later on we can still take $\kappa$ to infinity to get the full result for the VMS. In practice considering only finitely many contributions for each computation suffices.}. Furthermore the VMS spectral curve is of double scaled type \eqref{eq:genericCurve}. In the following we will write formulae based on the generic curve \eqref{eq:genericCurve} but we keep in mind that we want to apply them to the VMS. We can then predict \cite{krsst25a} the VMS partition function
\begin{align}
\label{eq:VMS-partition-function}
Z^{\text{V}}_{\kappa} &\simeq \sum\limits_{\ell\in\mathbb{Z}^{\kappa}}\prod\limits_{i=1}^{\kappa}\uprho_{i}^{\ell_i}\frac{\mathcal{Z}_{\text{G}}\left(\ell_i+\frac{\upmu_i}{2\pi\rmi}\right)}{\mathcal{Z}_{\text{G}}\left(\frac{\upmu_i}{2\pi\rmi}\right)}\exp\left[\sum\limits_{g=0}^{\infty}f_{g}\left(g_{\text{s}}\left(\ell_1+\frac{\upmu_1}{2\pi\rmi}\right), \dots, g_{\text{s}}\left(\ell_{\kappa}+\frac{\upmu_{\kappa}}{2\pi\rmi}\right)\right) g_{\text{s}}^{2g-2}\right],
\end{align}
\noindent
where $f_g(\nu_1,\dots,\nu_{\kappa})$ is the double scaled analogue of the multi-cut free energy in formula \eqref{eq:HMM-partition-function-conjecture}. Furthermore each $f_g(\nu_1,\dots,\nu_{\kappa}):\mathbb{C}^{\kappa}\to\mathbb{C}$ is analytic at the origin and we can give explicit expressions for its derivatives in terms of the curve \eqref{eq:genericCurve} (or \eqref{eq:spectralCurveVMS} for the VMS). If no eigenvalues have tunnelled it reduces to the perturbative free energy, which can be computed from topological recursion. In the case of the VMS we then find
\begin{equation}
f_g(0,\dots,0) = F_{\text{V}, g},
\end{equation}
\noindent
which was already discussed in \eqref{eq:freeEnergyVMS}. In the same spirit, its derivatives can be computed for any saddle $E^{\star}_i, i=1,\dots,\kappa$. The first few read \cite{krsst25a}
\begin{align}
\label{eq:dsl-b-cycle}
\partial_{\nu_i}f_0(\nu_1,\dots,\nu_{\kappa})\Big|_{\nu_1=\dots=\nu_{\kappa}=0} &= -2\int\limits_{a}^{E^{\star}_i}y(E)\text{d}E = -\oint\limits_{B_{i}} y(E) \text{d}E,\\
\partial_{\nu_i}^2f_0(\nu_1,\dots,\nu_{\kappa})\Big|_{\nu_1=\dots=\nu_{\kappa}=0} &= -\log\left(32\left(a-E^{\star}_i\right)^{5/2} M^{\prime}\left(E^{\star}_i\right)\right), \\
\partial_{\nu_i}^3f_0(\nu_1,\dots,\nu_{\kappa})\Big|_{\nu_1=\dots=\nu_{\kappa}=0} &=\frac{8\left(a-E^{\star}_i\right)^2 M^{\prime\prime}\left(E^{\star}_i\right)^2+77 M^{\prime}\left(E^{\star}_i\right)^2}{16\left(a-E^{\star}_i\right)^{5/2} M^{\prime}\left(E^{\star}_i\right)^3}-\\
&-\frac{2\left(a-E^{\star}_i\right)M^{\prime\prime\prime}\left(E^{\star}_i\right)+17 M^{\prime\prime}\left(E^{\star}_i\right)}{8\left(a-E^{\star}_i\right)^{3/2} M^{\prime}\left(E^{\star}_i\right)^2}+\frac{4}{\left(a-E_i^{\star}\right)^{3/2} M(a)},\nonumber\\
\label{eq:dsl-d1-f1}
\partial_{\nu_i}f_1(\nu_1,\dots,\nu_{\kappa})\Big|_{\nu_1=\dots=\nu_{\kappa}=0} &=\frac{4\left(a-E^{\star}_i\right)^2 M^{\prime\prime}\left(E^{\star}_i\right)^2+19 M^{\prime}\left(E^{\star}_i\right)^2}{192\left(a-E^{\star}_i\right)^{5/2} M^{\prime}\left(E^{\star}_i\right)^3}-\\
&-\frac{2\left(a-E^{\star}_i\right)M^{\prime\prime\prime}\left(E^{\star}_i\right)+ M^{\prime\prime}\left(E^{\star}_i\right)}{96\left(a-E^{\star}_i\right)^{3/2} M^{\prime}\left(E^{\star}_i\right)^2}+\frac{M(a)+3\left(a-E_i^{\star}\right)M^{\prime}(a)}{24\left(a-E_i^{\star}\right)^{3/2}M(a)^2},\nonumber
\end{align}
\noindent
where in accordance with our generic curve \eqref{eq:genericCurve} we have to set the endpoint of the cut $a=0$. Notice also, that in contrast to the half cycles \eqref{eq:der-free-energy-for-action} in matrix models, in the double scaling limit people often express things in terms of full cycles on the spectral curve as in \eqref{eq:dsl-b-cycle} \cite{ss03,ss04}. We are here following this convention and have therefore included all the appropriate factors of 2 regarding the full cycles of the double scaling limit in the expressions \eqref{eq:dsl-b-cycle}-\eqref{eq:dsl-d1-f1}\footnote{The question of using half or full cycles on spectral curves is purely a matter of convention. See \cite{sst23, krsst25a} for details.}. The more involved case with 2 saddles $E^{\star}_i,E^{\star}_j, i\neq j$ goes as follows \cite{krsst25a}
\begin{align}
\partial_{\nu_j}\partial_{\nu_i}f_0(\nu_1,\dots,\nu_{\kappa})\Big|_{\nu_1=\dots=\nu_{\kappa}=0} &= -4\text{arctanh}\left[ \sqrt{\frac{a-E_i^{\star}}{a-E_j^{\star}}}\,\right],
\end{align}
\noindent
where again $a=0$. In principle the production of these derivative formulae is algorithmic following \cite{msw08, sst23}, but computations quickly become very involved. Nevertheless the above expressions can be immediately applied to the VMS spectral curve \eqref{eq:spectralCurveVMS} to compute the full non-perturbative partition function as an asymptotic series in $g_{\text{s}}$\footnote{The coefficients $F^{\text{HMM}}_g$ fulfil topological recursion on a $\kappa+1$-cut spectral curve and $f_g$ are their double scaled analogue. This means, that the expression \eqref{eq:VMS-partition-function} grows factorially \cite{beg25}. One might then hope that it is 1-summable in the appropriate directions -- In this sense one can make sense of \eqref{eq:VMS-partition-function} as a function for example by Borel-Laplace summation \cite{krsst25a}.}. In addition it fully includes all non-perturbative ZZ-brane contributions in the VMS (See formulae \eqref{eq:one-inst-check}, \eqref{eq:neg-one-inst-check} and \eqref{eq:one-one-check}.). Having established the computation of $f_g(\nu_1,\dots,\nu_{\kappa})$ let us study the non-perturbative proposal \eqref{eq:VMS-partition-function} in more detail: We have $2\kappa$ transseries parameters $\uprho_i,\upmu_i; i=1,\dots,\kappa$, where the map to the rectangular transseries parameters in \eqref{eq:free-energy-transseries} has the form \cite{krsst25a}
\begin{equation}
\label{eq:transseries-map}
\upmu_i=\sigma_{i,1}\sigma_{i,2},\qquad \uprho_i = \frac{\sigma_{i,1}}{\Gamma\left(1+\frac{\upmu_i}{2\pi\rmi}\right)}\text{e}^{\alpha_i\upmu_i}.
\end{equation}
\noindent
Here, the $\alpha_i\in\mathbb{C}$ are constants that depend on the specific choice of rectangular transseries conventions. For details we refer the interested reader to \cite{krsst25a}. A specific choice for the parameters $\uprho_i$ and $\upmu_i$ constitutes a specific choice of non-perturbative completion and we have picked the normalization in formula \eqref{eq:VMS-partition-function} to be consistent with the matrix integral results in section \ref{sec:VMS}, meaning such that the trivial Stokes constants in the forward and backward direction are $1$. Furthermore the partition function in formula \eqref{eq:VMS-partition-function} is given in terms of a Zak transform\footnote{Partition functions as Zak transforms have some history: See for example \cite{em08, ghm16, blmst16}.}: All the instanton dependence has been conveniently rewritten in terms of a single sum for each involved saddle. Comments on the background independence of the expression \eqref{eq:HMM-partition-function-conjecture} can be found in \cite{ss25}.

Let us now give some explicit checks on the above proposal by comparing to the non-perturbative contributions computed in section \ref{subsec:instantonsVMM}. We will study 3 cases, namely standard instanton contributions, negative instantons and finally their pairs:
\begin{itemize}
\item \underline{\textbf{Physical Sheet Instantons:}} We compute the $\ell\geq 1$ instanton associated to a saddle $E^{\star}_i$ by picking up the $\ell$ mode of the Zak transform in \eqref{eq:VMS-partition-function} and normalizing by the perturbative partition function. Furthermore we are only interested in instantons, meaning for now take $\upmu_i=0$ (This implies $\uprho_i\to\sigma_{i,1}$). We will restrict ourselves here to the leading order $g_{\text{s}}$ contribution. Then we find
\begin{align}
\label{eq:one-inst-check}
\sigma_{i,1}^{\ell} &\mathcal{Z}_{\text{G}}(\ell)\exp\left[\frac{\ell}{g_{\text{s}}}\partial_{\nu_i} f_0(\nu_1,\dots,\nu_{\kappa})\Big|_{\nu_1=\dots=\nu_{\kappa}=0}\right]\exp\left[\frac{\ell^2}{2}\partial_{\nu_i}^2 f_0(\nu_1,\dots,\nu_{\kappa})\Big|_{\nu_1=\dots=\nu_{\kappa}=0}\right] = \nonumber\\
&=\sigma_{i,1}^{\ell}\mathcal{Z}_{\text{G}}(\ell)\text{e}^{-\ell\frac{A_i}{g_{\text{s}}}}\left(\frac{1}{32 M^{\prime}(E^{\star}) \left(a-E^{\star}\right)^{5/2}}\right)^{\frac{\ell^2}{2}},
\end{align}
\noindent
where we have introduced the instanton action $A_i$ exactly as in \eqref{eq:action}. Then formula \eqref{eq:one-inst-check} agrees exactly with formula \eqref{eq:generic-ell-instanton}. In fact up to the double scaling limit this is analogous to the computation presented in \cite{msw08} and therefore the match is expected. 
\item \underline{\textbf{Non-Physical Sheet Instantons:}}
The negative instantons contained in \eqref{eq:VMS-partition-function} are new contributions so let us check them carefully: They emerge in the limit $\sigma_{i,1}\to 0$, meaning $\uprho_i\to 0, \upmu_i\to 0$ while $\upmu_i/\uprho_i \to\sigma_{i,2}$, which can be seen by studying the map \eqref{eq:transseries-map}. We pick the $\ell=-1$ mode in the Zak transform \eqref{eq:VMS-partition-function} and expand in $\upmu_i$ at lowest order to find 
\begin{align}
\label{eq:neg-one-inst-check}
&\qquad\left.\frac{\mathcal{Z}_{\text{G}}\left(-1+\frac{\upmu_i}{2\pi\rmi}\right)}{\uprho_i}\right|_{\sigma_{i,1}\to 0} \exp\left[-\frac{1}{g_{\text{s}}}\partial_{\nu_i} f_0(\nu_1,\dots,\nu_{\kappa})\Big|_{\nu_1=\dots=\nu_{\kappa}=0}\right]\times\\&\times\exp\left[\frac{1}{2}\partial_{\nu_i}^2 f_0(\nu_1,\dots,\nu_{\kappa})\Big|_{\nu_1=\dots=\nu_{\kappa}=0}\right] = -\rmi\sqrt{\frac{g_{\text{s}}}{2\pi}}\text{e}^{\frac{A_i}{g_{\text{s}}}}\frac{1}{\sqrt{32 M^{\prime}(E^{\star}) \left(a-E^{\star}\right)^{5/2}}}\sigma_{i,2}.\nonumber
\end{align}
\noindent
Again this agrees perfectly with formula \eqref{eq:backwardforward}. Higher order negative instantons can be computed in the same way and we checked them up to $\ell=-3$. 
\item \underline{\textbf{Instanton -- Non-Physical Sheet Instanton Pairs:}} Last but not least we compute the mixed $(1|1)$ sector by picking up the zero mode of the Zak transform \eqref{eq:VMS-partition-function} and expanding in $\upmu$ to first order. We find
\begin{align}
\label{eq:one-one-check}
\frac{\upmu_i}{2\pi\rmi g_{\text{s}}}\partial_{\nu_i} f_0(\nu_1,\dots,\nu_{\kappa})\Big|_{\nu_1=\dots=\nu_{\kappa}=0} = \upmu_i\frac{\rmi}{2\pi}\frac{A_{i}}{g_{\text{s}}},
\end{align}
\noindent
which exactly agrees with \eqref{eq:11-prediction}. Further checks of this type of formula are provided in \cite{krsst25a} and we refer the interested reader to those references for details.
\end{itemize}
\noindent
With this we conclude our checks on the partition function \eqref{eq:VMS-partition-function}.

%%%%%%%%%%%%%%%%%%%%%%%%%%%%%%%%%%%%%%%%%%%%%%%%%%%%%%%%%%%%%%%%%
%%%%%%%%%%%%%%%%%%%%%%%%%%%%%%%%%%%%%%%%%%%%%%%%%%%%%%%%%%%%%%%%%
\subsection{Choosing a Non-Perturbative Completion}
\label{subsec:non-pert-com}
%%%%%%%%%%%%%%%%%%%%%%%%%%%%%%%%%%%%%%%%%%%%%%%%%%%%%%%%%%%%%%%%%
%%%%%%%%%%%%%%%%%%%%%%%%%%%%%%%%%%%%%%%%%%%%%%%%%%%%%%%%%%%%%%%%%

To finish this discussion, let us understand the non-perturbative partition function in a more explicit setting and construct a specific non-perturbative completion: First -- assuming the problem is sufficiently well behaved -- resum the transseries following the methods in \cite{krsst25a}. Let us for the moment assume that such a resummation also makes sense here. The missing piece is now a choice of initial transseries parameters. 

Let us start in the phase of the VMS where the free energy \eqref{eq:freeEnergyVMS} is dominant. This is the analogue of the one cut phase of hermitian matrix models and the tritronquée solution of minimal strings \cite{krsst25a}, where the non-perturbative completion is given by median summation -- meaning the falling instantons are turned on with the transseries parameter being half the value of their Stokes constants \cite{e81}. But in the VMS a subtlety appears: As mentioned before, the falling VMS instantons do not all live on the physical sheet of the spectral curve (see also figures \ref{fig:MSspectralcurve} and \ref{fig:Potentialplot}). This means for each saddle $i$ one has to decide if the effective potential is positive or negative and then pick the corresponding falling action: Let $A_i$ be the action \eqref{eq:action} associated to the $i$-th saddle point on the physical sheet. If\footnote{In this discussion we are ignoring some saddles that have $0$ action. Those will not give non-perturbative effects, but in other phases of the VMS they might modify what is meant with the perturbative sector. See also \cite{ss25}. It would be interesting to investigate this further. Let us also mention that the endpoint value $b=1$ needs extra consideration.}
\begin{itemize}
\item \underline{\textbf{$A_i>0$:}} choose $\sigma_{i,1}=1/2, \sigma_{i,2}=0$, which translates to
\begin{equation}
\uprho_i=\frac{1}{2},\qquad\upmu_i=0.
\end{equation}
\item \underline{\textbf{$A_i<0$:}} choose $\sigma_{i,1}=0, \sigma_{i,2}=1/2$, which translates to
\begin{equation}
\frac{\upmu_i}{\uprho_i}=\frac{1}{2},\qquad\uprho_i\to0,\quad\upmu_i\to 0.
\end{equation}
\end{itemize}
\noindent
To see that this choice of initial conditions indeed only turns on falling instanton effects the reader might compare with formula \eqref{eq:one-inst-check} and \eqref{eq:neg-one-inst-check}. Interestingly, even this simple choice of non-perturbative completion already includes exponential contributions from both sheets of the spectral curve.

At this stage we should mention that the above choice of non-perturbative completion via median summation is in no way the unique possible choice\footnote{The choice of non-perturbative completion in a matrix model is predicated on choosing a special eigenvalue integration contour -- often the contour that minimizes eigenvalue energy. But starting from this setting matrix models still exhibit various phases, where the non-perturbative completion changes through Stokes phenomena: For example study the $(2,3)$ - minimal string and its related cubic matrix model with potential
\begin{equation}
V(\lambda) = \lambda^2-\frac{1}{6}\lambda^3,\nonumber
\end{equation}
\noindent
which has two real saddle points at $\lambda^{\star}_1=0,\lambda^{\star}_2=4$ (but none of the associated steepest descent contours are strictly real). As shown in \cite{krsst25a} the cubic matrix model has phases where it behaves quasi-periodically -- where its expansion is dominated by exponentials. In these phases the transseries parameters have values different from median summation, because the system experiences the aforementioned Stokes phenomena.}: One might therefore wonder also about other choices of transseries parameters for the VMS where new behavior can emerge -- driven by new dominant exponential contributions \cite{krsst25a}. Let us quickly outline what to expect from the partition function \eqref{eq:VMS-partition-function} in this case: Focus on the $i$-th non-perturbative saddle and study a generic configuration $\uprho_i\neq 0,\upmu_i\neq 0$. At leading order in $g_{\text{s}}$ the VMS partition function then reads
\begin{align}
&\exp\left[\frac{1}{g_{\text{s}}^2}F_{\text{V}, 0}+F_{\text{V}, 1}+\dots\right]\sum\limits_{\ell\in\mathbb{Z}}\uprho_i^{\ell} \frac{g_{\text{s}}^{\left(\ell+\frac{\upmu_i}{2\pi\rmi}\right)^2/2}}{(2\pi)^{\left(\ell+\frac{\upmu_i}{2\pi\rmi}\right)/2}}\frac{G_2\left(1+\ell+\frac{\upmu_i}{2\pi\rmi}\right)}{\mathcal{Z}_{\text{G}}\left(\frac{\upmu_i}{2\pi\rmi}\right)}\exp\left[\frac{1}{g_{\text{s}}}\left(\ell+\frac{\upmu_i}{2\pi\rmi}\right)\partial_{\nu_i}f_0\right]\times\\&\times\exp\left[\frac{1}{2}\left(\ell+\frac{\upmu_i}{2\pi\rmi}\right)^2\partial_{\nu_i}^2 f_0\right]\left(1+g_{\text{s}}\left(\left(\ell+\frac{\upmu_i}{2\pi\rmi}\right)\partial_{\nu_i}f_1+\frac{1}{6}\left(\ell+\frac{\upmu_i}{2\pi\rmi}\right)^3\partial_{\nu_i}^3 f_0\right)+\mathcal{O}\left(g_{\text{s}}^2\right)\right).\nonumber
\end{align}
\noindent
This is a modified theta function: It is a theta function sum with the extra insertion of a Barnes function $G_2(\nu)$\footnote{The attentive reader might be worried about the convergence properties of the instanton sum: This growth originates from the genus sum and it is discussed in detail in \cite{krsst25a}.}. This means we should no longer expect clean exponential asymptotics \cite{bde00} -- but rather quasi-periodic, Zak transform like behaviour \cite{em08, msw08, krsst25a} when both parameters $\uprho_i$ and $\upmu_i$ are turned on.

%%%%%%%%%%%%%%%%%%%%%%%%%%%%%%%%%%%%%%%%%%%%%%%%%%%%%%%%%%%%%%%%%
%%%%%%%%%%%%%%%%%%%%%%%%%%%%%%%%%%%%%%%%%%%%%%%%%%%%%%%%%%%%%%%%%
\section{Non-Perturbative Effects in 3d Gravity}
\label{sec:3d-gravity}
%%%%%%%%%%%%%%%%%%%%%%%%%%%%%%%%%%%%%%%%%%%%%%%%%%%%%%%%%%%%%%%%%
%%%%%%%%%%%%%%%%%%%%%%%%%%%%%%%%%%%%%%%%%%%%%%%%%%%%%%%%%%%%%%%%%

%%%%%%%%%%%%%%%%%%%%%%%%%%%%%%%%%%%%%%%%%%%%%%%%%%%%%%%%%%%%%%%%%
\begin{figure}
\centering
\includegraphics[scale=0.8]{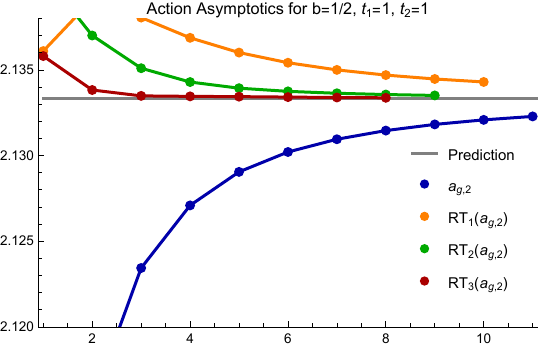}
\hspace{0.2cm}
\includegraphics[scale=0.8]{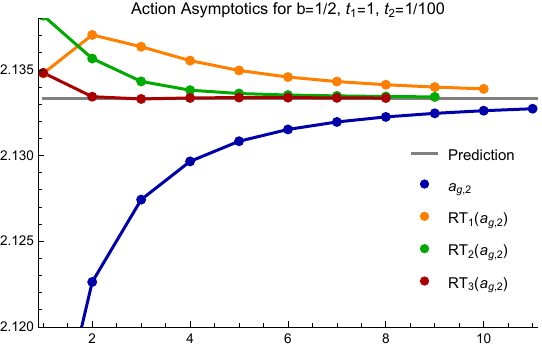}

\vspace{0.6cm}
\includegraphics[scale=0.8]{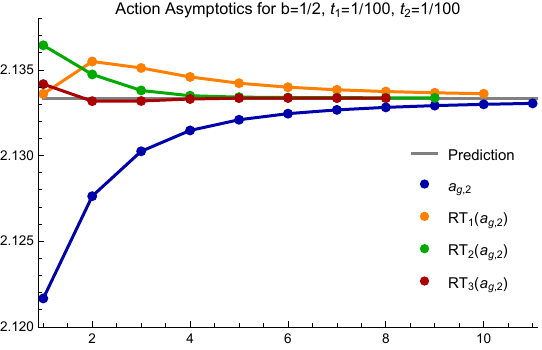}
\hspace{0.2cm}
\includegraphics[scale=0.8]{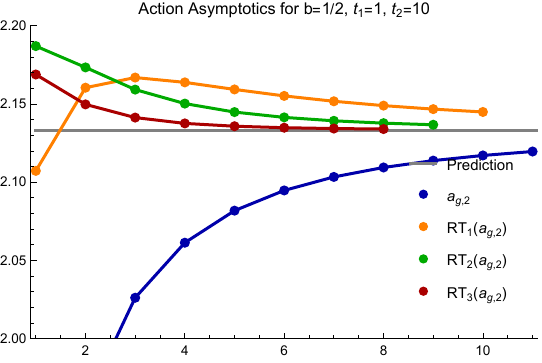}

\vspace{0.6cm}
\includegraphics[scale=0.8]{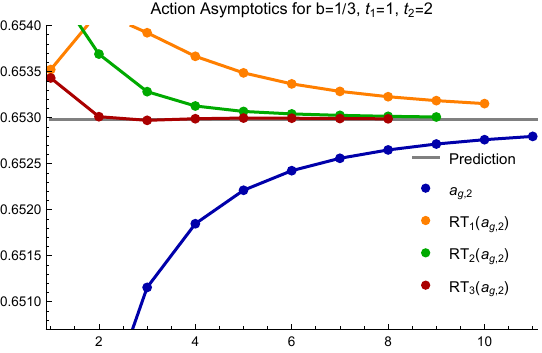}
\hspace{0.2cm}
\includegraphics[scale=0.8]{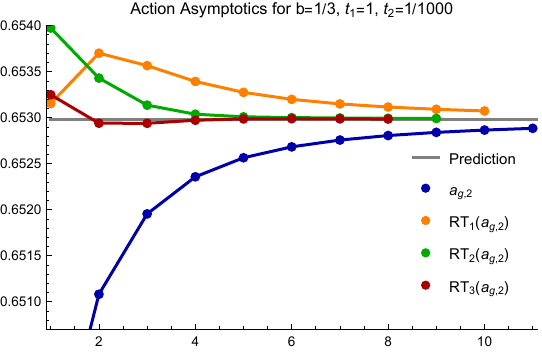}
\caption{Asymptotic checks for the leading instanton action of the series \eqref{eq:grav-2-boundary-series} for various values of $t_1,t_2$. Explicitly we plot the sequence $a_{g,2}(t_1,t_2)$ from formula \eqref{eq:actionresolvent-2} together with its first 3 Richardson transforms. We find good agreement with the prediction $A_{1,+}$ with a precision of around 3 digits.}
\label{fig:ads3-action}
\end{figure}
%%%%%%%%%%%%%%%%%%%%%%%%%%%%%%%%%%%%%%%%%%%%%%%%%%%%%%%%%%%%%%%%%

%%%%%%%%%%%%%%%%%%%%%%%%%%%%%%%%%%%%%%%%%%%%%%%%%%%%%%%%%%%%%%%%%
\begin{figure}
\centering
\includegraphics[scale=0.8]{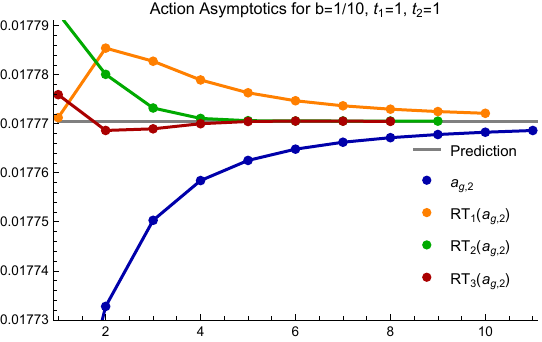}
\hspace{0.2cm}
\includegraphics[scale=0.8]{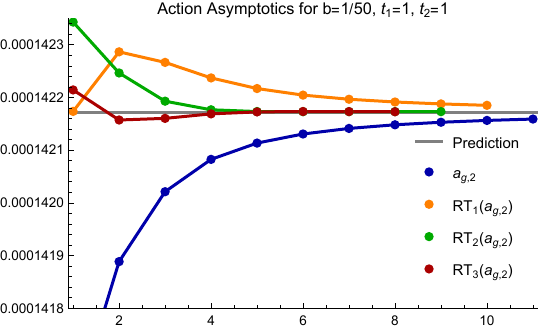}

\vspace{0.6cm}
\includegraphics[scale=0.8]{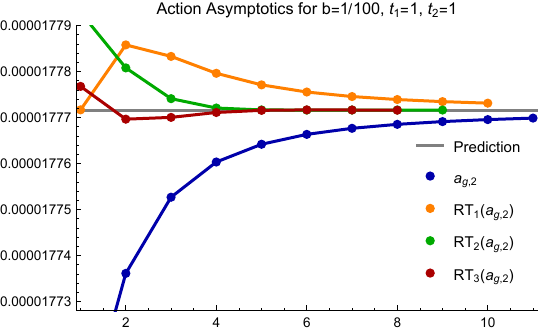}
\hspace{0.2cm}
\includegraphics[scale=0.8]{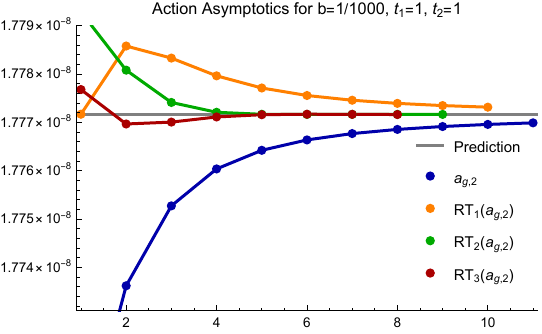}

\caption{Asymptotic checks for the leading instanton action of the series \eqref{eq:grav-2-boundary-series} for increasing values of $c$. As in figure \ref{fig:ads3-action} we plot the sequence $a_{g,2}(t_1,t_2)$ from formula \eqref{eq:actionresolvent-2} together with its first 3 Richardson transforms. We find good agreement with the prediction $A_{1,+}$ with a precision of 3 relevant digits even as $c$ increases.}
\label{fig:ads3-action-large-c}
\end{figure}
%%%%%%%%%%%%%%%%%%%%%%%%%%%%%%%%%%%%%%%%%%%%%%%%%%%%%%%%%%%%%%%%%

Having understood non-perturbative contributions in the VMS we want to understand their relation to 3d gravity. To be more specific, we focus on the relation between the VMS and 3d gravity on manifolds that are topologically a thickened Riemann surface $\Sigma_{g,n}\times I$ with end-of-the-world (EOW) branes at the ends of the interval: A relation that was conjectured in \cite{jrw25} and recently proven in \cite{jrsw25}. For notations and conventions on 3d gravity we will follow \cite{jrsw25}. Explicitly, the following correspondence is established\footnote{A comment on notation: In 3d gravity we will follow the conventions from \cite{jrsw25} where the dependence of $1/g_{\text{s}}=Z_{\text{disk}}^{(a)}Z_{\text{disk}}^{(b)}$ is absorbed in the partition function. On the other hand in the VMS for the resolvent we will keep $g_{\text{s}}$ explicit to be consistent with section \ref{sec:VMS}: For $\mathcal{Z}^{\text{primary}}_{\Sigma_{g,n}\times I}, \mathsf{Z}_{g,n}$ we keep the dependence implicit, while for the resolvent $R_i$ or the free energy the factors of $g_{\text{s}}$ are spelled out.}
\begin{align}
\label{eq:grav-z-map}
\mathcal{Z}^{\text{primary}}_{\Sigma_{g,n}\times I}\left(\rmi t_1,\dots,\rmi t_n\right) = \rme^{-\frac{\beta_1+\dots +\beta_n}{24}}\mathsf{Z}_{g,n}\left(\beta_1,\dots,\beta_n\right),
\end{align}
\noindent
where on the left hand side $\mathcal{Z}^{\text{primary}}_{\Sigma_{g,n}\times I}$ denotes the gravitational path integral on $\Sigma_{g,n}\times I$ with EOW branes and with descendants removed (see formula (2.8) of \cite{jrsw25}). Furthermore $\rmi \cdot t_j$ are annulus moduli for each of the $n$ boundaries of $\Sigma_{g,n}$. On the right hand side of equation \eqref{eq:grav-z-map}, $\mathsf{Z}_{g,n}$ is the VMS partition function on $\Sigma_{g,n}$. Lastly $\beta_i = 2\pi t_i$ are inverse temperatures and the expression still contains the $g_{\text{s}}=\exp(-S_0)$ dependence implicitly. This dependence is encoded on the gravity side via $1/g_{\text{s}}=Z_{\text{disk}}^{(a)}Z_{\text{disk}}^{(b)}$, where $Z_{\text{disk}}^{(a)}$ denotes the empty disk partition function with boundary condition $a$ at one end of the interval $I$ (and $b$ on the other respectively). Conveniently $\mathsf{Z}_{g,n}$ can be computed from the VMS matrix model correlators via inverse Laplace transform
\begin{equation}
\label{eq:z-laplace-res}
\mathsf{Z}_{g,n}\left(\beta_1,\dots,\beta_n\right) = \int\limits_{\Gamma^{\downarrow}}\left(\prod\limits_{j=1}^{n}\frac{\text{d}u_j}{2\pi\rmi}\rme^{\beta_j u_j}\right)g_{\text{s}}^{2g-2+n}R_{g, n}\left(-u_1,\dots,-u_n\right),
\end{equation}
\noindent
where $\Gamma^{\downarrow}$ is a downward Bromwich contour. Furthermore, in accordance with the notation in \cite{cemr23} (and our previous section \ref{sec:VMS}) we have made the $g_{\text{s}}$ dependence for the resolvent explicit on the right hand side.

We want to understand the consequences of the sum over genus in 3d gravity with EOW branes. For this endeavour let us focus on $n=2$ boundaries. We will start on the VMS side, where the genus sum for the resolvent reads
\begin{equation}
\label{eq:vms-resolvent-2}
R_2(-u_1, -u_2) \simeq \sum\limits_{g=0}^{\infty}g_{\text{s}}^{2g}R_{g, 2}\left(-u_1,-u_2\right),
\end{equation}
\noindent
and then map the results to 3d gravity via
\begin{equation}
\label{eq:grav-inverse-laplace-2}
\sum\limits_{g=0}^{\infty}\mathcal{Z}^{\text{primary}}_{\Sigma_{g,n=2}\times I}\left(\rmi t_1,\rmi t_2\right)\simeq\int\limits_{\Gamma^{\downarrow}}\left(\prod\limits_{j=1}^{2}\frac{\text{d}u_j}{2\pi\rmi}\rme^{\beta_j \left(u_j-\frac{1}{24}\right)}\right) R_{2}\left(-u_1,-u_2\right).
\end{equation}
\noindent
Formula \eqref{eq:grav-inverse-laplace-2} can be obtained by combining \eqref{eq:grav-z-map} with \eqref{eq:z-laplace-res} and switching the inverse Laplace transform with the formal genus sum. Notice again that the expansion parameter on the gravity side is given by $1/g_{\text{s}}=Z_{\text{disk}}^{(a)}Z_{\text{disk}}^{(b)}$ \cite{jrsw25} and that -- in the 3d gravity language -- those factors are absorbed in the partition function.  
Consequently we need to study the non-perturbative contributions to the VMS resolvent \eqref{eq:vms-resolvent-2} and understand their behaviour under the inverse Laplace transform \eqref{eq:grav-inverse-laplace-2}: This is an asymptotic study in $g_{\text{s}}$ and we expect the VMS resolvent transseries to include the following non-perturbative effects \cite{sss19, os20, eggls23, jr25}:
\begin{itemize}
\item \underline{\textbf{ZZ-type Instantons and Their Negative Siblings:}} These are the standard ZZ-type instanton contributions computed in \cite{cemr23} and their negative siblings from section \ref{sec:VMS}. For $n$ eigenvalues and $m$ anti-eigenvalues at each saddle point $E^{\star}_{k,\pm}$ (see formula \eqref{eq:saddlesVMS}) they contribute to the transseries as
\begin{equation}
\label{eq:zz-resolvent-2}
\rme^{-(n-m)\frac{A_{k,\pm}}{g_{\text{s}}}}\tilde{R}^{(n|m)}_{E^{\star}_{k,\pm}}(E_1, E_2)
\end{equation}
\noindent
where $A_{k,\pm}$ is given in \eqref{eq:oneinst} and $\tilde{R}^{(n|m)}_{E^{\star}_{k,\pm}}(E_1, E_2)$ is an asymptotic series attached to the instanton. We want to understand what those contributions correspond to for 3d gravity -- meaning after performing the inverse Laplace transform \eqref{eq:grav-inverse-laplace-2}.  Notice how the instanton actions in \eqref{eq:zz-resolvent-2} are independent of $u_1, u_2$. This means that when performing the inverse Laplace transform on \eqref{eq:zz-resolvent-2} the non-perturbative exponential can be treated as just a prefactor and taken out of the integral: The instanton action does not change. Consequently we expect exactly the ZZ exponentials 
\begin{equation}
\label{eq:grav-zz}
\rme^{-(n-m)\frac{A_{k,\pm}}{g_{\text{s}}}}
\end{equation}
\noindent
to appear in the non-perturbative completion of the series \eqref{eq:grav-inverse-laplace-2}. Exactly as in the case of the VMS they can be turned on and off by making a choice of non-perturbative completion or by crossing Stokes lines. We also emphasize again that some of the \emph{falling} instanton contributions above come originally from the non-physical sheet of the VMS spectral curve as not all the $A_{k,\pm}$ are positive.
\item \underline{\textbf{FZZT-type Instantons:}}
As shown in subsection \ref{subsec:wall-crossing} the VMS resolvents receive also non-perturbative contributions from FZZT-type exponentials (see formula \eqref{eq:atilde})
\begin{equation}
\label{eq:VMS-fzzt-action}
\rme^{-\frac{\tilde{A}(E)}{g_{\text{s}}}},\qquad \tilde{A}(E) = 2\int\limits_{0}^{E} y_{\text{V}}(E^{\prime}) \text{d}E^{\prime}.
\end{equation}
\noindent
In this context we have shown the existence of wall crossing in subsection \ref{subsec:wall-crossing}. One might then wonder how these effects translate to the genus sum over primary partition functions in 3d gravity. For this endeavour we need to understand the effect of the inverse Laplace transform \eqref{eq:grav-inverse-laplace-2} on the action \eqref{eq:VMS-fzzt-action}. It can be evaluated using saddle point approximation for small $g_{\text{s}}$ on integrals of the type
\begin{equation}
\int_{\Gamma}\frac{\text{d}u}{2\pi\rmi}\rme^{\beta u}\rme^{-\frac{\tilde{A}(-u)}{g_{\text{s}}}}\left(\dots\right).
\end{equation}
\noindent
Because $\beta u$ is $g_{\text{s}}$ subleading in comparison to the action $\tilde{A}(-u)$, the inverse Laplace transform \eqref{eq:grav-inverse-laplace-2} localizes exactly around the saddle points of the action $\tilde{A}(-u)$ in \eqref{eq:VMS-fzzt-action} -- which correspond to the zeros of the spectral curve \eqref{eq:spectralCurveVMS} given by the saddle points $E^{\star}_{k,\pm}$\footnote{Exactly which saddle is turned on depends on the specific steepest descent contour decomposition of the contour $\Gamma$.}. In this way the ZZ-type instanton contributions\footnote{This is analogous to the FZZT-branes in minimal string theory that also localize on ZZ-branes upon integration over their moduli-space \cite{ss03, ss04, sst23}.} reappear. Therefore for 3d gravity the FZZT-like VMS effects produce exactly the actions $A_{k,\pm}$ in \eqref{eq:grav-zz}. In this sense they mix with the original ZZ-type non-perturbative effects and are no longer individually detectable. It remains to be investigated however if the wall crossing survives the transformation \eqref{eq:grav-inverse-laplace-2} by shifting around Stokes constants -- meaning by jumps of the prefactors of the non-perturbative exponentials \eqref{eq:grav-zz}. This result is exactly consistent with similar studies for JT gravity in \cite{eggls23}.
\end{itemize}
Summing up the above discussion we expect to find the non-perturbative exponentials \eqref{eq:grav-zz} for the genus sum \eqref{eq:grav-inverse-laplace-2}. To confirm this we can study the asymptotics of 
\begin{equation}
\label{eq:grav-2-boundary-series}
\sum\limits_{g=0}^{\infty}\mathcal{Z}^{\text{primary}}_{\Sigma_{g,n=2}\times I}\left(\rmi t_1,\rmi t_2\right).
\end{equation}
\noindent
This is an asymptotic series in $1/g_{\text{s}}=Z_{\text{disk}}^{(a)}Z_{\text{disk}}^{(b)}$ and we should see the ZZ instanton contribution $A_{1,+}$ leading the asymptotics, meaning
\begin{equation}
\label{eq:actionresolvent-2}
a_{g, 2}\left(t_1,t_2\right) := \sqrt{\frac{4 g^2}{\left(Z_{\text{disk}}^{(a)}Z_{\text{disk}}^{(b)}\right)^2}\frac{\mathcal{Z}^{\text{primary}}_{\Sigma_{g,n=2}\times I}\left(\rmi t_1,\rmi t_2\right)}{\mathcal{Z}^{\text{primary}}_{\Sigma_{g+1,n=2}\times I}\left(\rmi t_1,\rmi t_2\right)}} \quad \xrightarrow{g\to\infty}\quad 
A_{1,+}
\end{equation}
\noindent
Notice the disk factors that appear in \eqref{eq:actionresolvent-2}, to cancel the implicit disk dependence of the coefficients in the series \eqref{eq:grav-2-boundary-series}. Moreover we should find the same action $A_{1,+}$ for any $t_1,t_2$ as any FZZT-type effects should be absent. This is indeed the case: We have computed the first 12 orders of the series \eqref{eq:grav-2-boundary-series} from the VMS data by using the map \eqref{eq:grav-z-map} and we have tested the asymptotics \eqref{eq:actionresolvent-2} for various values $t_1,t_2$. The checks are visualized in figure \ref{fig:ads3-action} and they confirm our prediction with a precision of 3 digits.

Let us now understand the above computation in the natural language for 3d gravity, namely as an expansion in large $c$. Firstly we need to understand if the above asymptotic study in $1/g_{\text{s}}=Z_{\text{disk}}^{(a)}Z_{\text{disk}}^{(b)}$ can be carried over to the large $c$ analysis. This is not immediate because in analogy to \cite{sss19} we might expect exponential $c$ dependence in the disk contributions \cite{jrw25-gen}
\begin{equation}
\log\left(Z_{\text{disk}}^{(a)}\right)\sim c.
\end{equation}
\noindent
and in addition for fixed saddle label $k$, the instanton action scales as $|A_{k,+}|\sim c^{-3/2}$ at large $c$. Lastly, $\mathcal{Z}^{\text{primary}}_{\Sigma_{g,n=2}\times I}\left(\rmi t_1,\rmi t_2\right)$ is a polynomial in $c$ at each genus $g$ (with the disk contributions taken out). In other words the asymptotic study \eqref{eq:actionresolvent-2} is not strictly a large $c$ analysis. Nevertheless, given the above scales at large $c$ it does imply doubly exponential behaviour. Furthermore, there is a strict separation of scales between the exponential $c$ dependence of the disk and the polynomial $c$ dependence of the coefficients -- meaning the asymptotics \eqref{eq:actionresolvent-2} should hold. In fact, we can test the asymptotics \eqref{eq:actionresolvent-2} numerically for large $c$ (assuming the separation of scales between $g_{\text{s}}$ and $c$\footnote{Of course this is not a proper large $c$ analysis, but it gives us confidence that the limit $c\to\infty$ does not break the asymptotics \eqref{eq:actionresolvent-2}.}) which we have visualized in figure \ref{fig:ads3-action-large-c}. We find perfect agreement of 3 relevant digits with the prediction. Concluding, we have made a strong case for the existence of the following doubly exponential contributions in $c$ for the genus sum \eqref{eq:grav-2-boundary-series}
\begin{equation}
\exp\left[-(n-m)\frac{A_{k,\pm}}{g_{\text{s}}}\right],
\end{equation}
\noindent
where $(n|m)$ counts instantons (and negative instantons) of the type $A_{k,\pm}$ given in \eqref{eq:grav-zz} with the specific formula for $A_{k,\pm}$ given in \eqref{eq:oneinst}. This type of doubly exponential contribution is in line with the expectation in \cite{jrw25-gen} at least for the falling exponentials. One might further wonder, if these contributions correspond to non-perturbative saddles of the gravity path integral. Lastly we should mention that a generic closed form construction of the VMS resolvents along the lines of the non-perturbative partition function in section \ref{sec:exactZ} might be very useful to extend the above study past the leading instanton actions.

%%%%%%%%%%%%%%%%%%%%%%%%%%%%%%%%%%%%%%%%%%%%%%%%%%%%%%%%%%%%%%%%%
%%%%%%%%%%%%%%%%%%%%%%%%%%%%%%%%%%%%%%%%%%%%%%%%%%%%%%%%%%%%%%%%%
\section{Black Hole Threshold as a Stokes Transition}
\label{sec:black-hole-stokes}
%%%%%%%%%%%%%%%%%%%%%%%%%%%%%%%%%%%%%%%%%%%%%%%%%%%%%%%%%%%%%%%%%
%%%%%%%%%%%%%%%%%%%%%%%%%%%%%%%%%%%%%%%%%%%%%%%%%%%%%%%%%%%%%%%%%

Above, we have studied the resurgent properties of partition functions (and correlators) for the VMS and 3d gravity. What we have not discussed yet is the eigenvalue density. More specifically we want to understand how non-perturbative effects behave as we cross the edge of the eigenvalue distribution in matrix models. For the setting with a generic double scaled spectral curve as given in formula \eqref{eq:genericCurve} compute the resolvent
\begin{equation}
\label{eq:generic-resolvent}
R(E) = \left\langle\text{Tr}\frac{1}{E-M}\right\rangle,
\end{equation}
\noindent
and study the density of states on the real line \cite{m04, sss19}
\begin{equation}
\label{eq:definition-generic-density}
\langle\varrho(E)\rangle = -\frac{1}{2\pi\rmi}\text{Disc}\left[ R(E)\right] = -\frac{1}{2\pi\rmi}\lim\limits_{\epsilon\to 0}\left(R(E+\rmi\epsilon)-R(E-\rmi\epsilon) \right), \quad E\in\mathbb{R}.
\end{equation}
\noindent
We want to investigate the behavior of the density as the threshold $E=0$ is crossed. In this context the following leading behavior was proposed in \cite{sss19} 
\begin{align}
\label{eq:triple-s-formula}
\left\langle\varrho(E)\right\rangle \simeq 
\begin{cases}
\frac{1}{g_{\text{s}}}\varrho_0(E)-\frac{1}{4\pi E}\cos\left(\frac{2\pi}{g_{\text{s}}}\int\limits_{0}^{E}\text{d}E^{\prime}\varrho_0\left(E^{\prime}\right)\right),\qquad & E>0, \\
-\frac{1}{8\pi E}\exp\left(-\frac{2}{g_{\text{s}}}\int\limits_{0}^{E}\text{d}E^{\prime}\, y\left(E^{\prime}\right)\right),& E<0,
\end{cases}
\end{align}
\noindent
where $\varrho_0(E)$ is the classical part of the density \eqref{eq:definition-generic-density}. In the following we will generalize this result from a resurgent viewpoint and understand it in terms of Stokes and anti-Stokes transitions. We will further show that \eqref{eq:triple-s-formula} is the universal piece of a more generic expression that depends on a choice of non-perturbative completion. We will first present the generic construction for any hermitian matrix model followed by two examples: firstly we discuss implications for the VMS and for 3d gravity and secondly we comment on JT gravity.

%%%%%%%%%%%%%%%%%%%%%%%%%%%%%%%%%%%%%%%%%%%%%%%%%%%%%%%%%%%%%%%%%
\subsection{Generic Construction}
\label{subsec:gen-construction}
%%%%%%%%%%%%%%%%%%%%%%%%%%%%%%%%%%%%%%%%%%%%%%%%%%%%%%%%%%%%%%%%%

We want to understand how non-perturbative effects correct the eigenvalue density \eqref{eq:definition-generic-density}, which is obtained from the resolvent by taking the discontinuity along the real line \cite{sss19}. Consequently let us start with the generic matrix model resolvent \eqref{eq:resolvent} and construct the resurgent non-perturbative completion for the asymptotic series in $g_{\text{s}}$. Such effects have been computed in \cite{sss19, os20, jr25} and we have already seen them in the VMS explicitly in subsection \ref{subsec:wall-crossing}. Famously there are two types of non-perturbative effects: 
\begin{itemize}
\item \underline{\textbf{FZZT Contributions}} whose leading order for the action on the physical sheet was computed in \cite{sss19, os20, jr25} and reads\footnote{Notice that the result for the FZZT effects is normalized to match matrix integral computations \cite{jr25}.}
\begin{align}
\label{eq:falling-FZZT}
R^{\text{FZZT}_{-}}(E) \simeq \left(\frac{\rmi}{4 E}+\mathcal{O}(g_{\text{s}})\right)\text{e}^{-\frac{\tilde{A}(E)}{g_{\text{s}}}}
\end{align}
\noindent
where (see equation \eqref{eq:atilde} for the VMS analogue) 
\begin{equation}
\label{eq:generic-FZZT-action}
\tilde{A}(E) = 2\int_{0}^{E} y(E^{\prime}) \text{d}E^{\prime}.
\end{equation}
\noindent
On the other hand, on the non-physical sheet we have
\begin{align}
\label{eq:growing-FZZT}
R^{\text{FZZT}_{+}}(E) \simeq \left(-\frac{\rmi}{4 E}+\mathcal{O}(g_{\text{s}})\right)\text{e}^{\frac{\tilde{A}(E)}{g_{\text{s}}}}
\end{align}
\noindent
Here there are no towers of multiples of the FZZT action.
\item \underline{\textbf{ZZ Contributions}} that are associated to saddle points of the effective action. For a saddle $E^{\star}$ we have the one (anti-) instanton contribution at leading order \cite{eggls23} (compare with \eqref{eq:generic-one-inst})
\begin{align}
\label{eq:resolvent-ZZ-p}
R^{(1|0)_{\text{ZZ}}}(E) &\simeq\rme^{-\frac{\tilde{A}\left(E^{\star}\right)}{g_{\text{s}}}}\left(\frac{1}{4}\sqrt{\frac{g_{\text{s}}}{4\pi M^{\prime}(E^{\star})(a-E^{\star})^{\frac{5}{2}}}}\sqrt{\frac{-E^{\star}}{-E}}\frac{1}{E-E^{\star}}+\mathcal{O}\left(g_{\text{s}}^{3/2}\right)\right),\\
\label{eq:resolvent-ZZ-m}
R^{(0|1)_{\text{ZZ}}}(E) &\simeq\rmi\rme^{\frac{\tilde{A}\left(E^{\star}\right)}{g_{\text{s}}}}\left(\frac{1}{4}\sqrt{\frac{g_{\text{s}}}{4\pi M^{\prime}(E^{\star})(a-E^{\star})^{\frac{5}{2}}}}\sqrt{\frac{-E^{\star}}{-E}}\frac{1}{E-E^{\star}}+\mathcal{O}\left(g_{\text{s}}^{3/2}\right)\right),
\end{align}
\noindent
where $\tilde{A}(E)$ is given in formula \eqref{eq:generic-FZZT-action}. Each ZZ contribution of course comes with an infinite tower of siblings associated with the number of $n$ eigenvalues and $m$ anti eigenvalues that tunnel to $E^{\star}$. Generically we then denote the ZZ contributions by $R^{(n|m)_{\text{ZZ}}}(E)$.
\end{itemize}
%
%%%%%%%%%%%%%%%%%%%%%%%%%%%%%%%%%%%%%%%%%%%%%%%%%%%%%%%%%%%%%%%%%
\begin{figure}
\centering
	\begin{tikzpicture}
\begin{scope}[scale=1,  shift={({0},{0})}]	
\draw[gray, ->, line width=1.5pt] (-7,0) -- (7,0);
\draw[gray, ->, line width=1.5pt] (0,-1.3) -- (0,2.5);
\node[black] at (-6.75, 2.25) {$E$};
\draw[gray, line width=1.5pt] (-7, 2) -- (-6.5, 2) -- (-6.5, 2.5);
\draw[blue, line width=2pt, dashed] (180:3) -- (180:4);
%\draw[blue, line width=2pt, dashed] (60:2) -- (60:3);
%\draw[blue, line width=2pt, dashed] (300:2) -- (300:3);
\draw[orange, line width=2pt, dashed] (0:3) -- (0:4);
\draw[orange, line width=2pt, dashed] (120:2) -- (120:3);
\draw[orange, line width=2pt, dashed] (240:1) -- (240:1.5);
\node[blue] at (-2.2, 0.35) {Stokes};
\node[orange] at (-1.7, -0.7) {anti-Stokes};
\draw[blue, line width=2pt] (0,0) -- (180:3);
%\draw[blue, line width=2pt] (0,0) -- (60:2);
%\draw[blue, line width=2pt] (0,0) -- (300:2);
\draw[orange, line width=2pt] (0,0) -- (0:3);
\draw[orange, line width=2pt] (0,0) -- (120:2);
\draw[orange, line width=2pt] (0,0) -- (240:1);
\draw[purple, line width=2pt] (-4, 0.7) to[out=-10, in=165] (-0.8, 1.05);
\draw[purple, line width=2pt] (-0.57,1) circle (6pt);
\draw[purple, line width=2pt] (-0.34, 0.97) to[out=-10, in=175] (0.6, 1);
\draw[purple, line width=2pt, ->] (0.6, 1) to[out=-5, in=190] (3.5,0.7);	
\node[purple] at (-4.9, 0.9) {\Large $\tau_1\frac{\rmi}{4E}\text{e}^{\frac{-\tilde{A}(E)}{g_{\text{s}}}}$};
\node[purple] at (5, 0.9) {\Large $\tau_1\frac{\rmi}{4E}\text{e}^{-\frac{\rmi\tilde{\mathsf{A}}(E)}{g_{\text{s}}}}$};
%\node[rectangle, line width=1.5pt, draw=blue,align=center,rounded corners,minimum height=2em,fill=white, rotate=-30] at (1.3, 1.9) {{\color{purple}$\tau\to\tau+S$}};
%\node[rectangle, line width=1.5pt, draw=orange,align=center,rounded corners,minimum height=2em,fill=white, rotate=0] at (-1.4, 1.9) {{\color{purple}$\tilde{A}(z)\to\rmi\tilde{\mathsf{A}}(z)$}};
\end{scope}
\begin{scope}[scale=1,  shift={({0},{-4.5})}]	
\draw[gray, ->, line width=1.5pt] (-7,0) -- (7,0);
\draw[gray, ->, line width=1.5pt] (0,-1.3) -- (0,2.5);
%\draw[snake it, draw=red, line width=0.8pt] (0, 0) -- (5, 0);
\node[black] at (-6.75, 2.25) {$E$};
\draw[gray, line width=1.5pt] (-7, 2) -- (-6.5, 2) -- (-6.5, 2.5);
%%
%\draw[blue, line width=2pt, dashed] (180:2) -- (180:3);
\draw[blue, line width=2pt, dashed] (60:2) -- (60:3);
\draw[blue, line width=2pt, dashed] (300:1) -- (300:1.5);
\draw[orange, line width=2pt, dashed] (0:3) -- (0:4);
\draw[orange, line width=2pt, dashed] (120:2) -- (120:3);
\draw[orange, line width=2pt, dashed] (240:1) -- (240:1.5);
\node[blue] at (1.2, -0.7) {Stokes};
\node[orange] at (-1.7, -0.7) {anti-Stokes};
%%
%\draw[blue, line width=2pt] (0,0) -- (180:2);
\draw[blue, line width=2pt] (0,0) -- (60:2);
\draw[blue, line width=2pt] (0,0) -- (300:1);
\draw[orange, line width=2pt] (0,0) -- (0:3);
\draw[orange, line width=2pt] (0,0) -- (120:2);
\draw[orange, line width=2pt] (0,0) -- (240:1);
\draw[purple, line width=2pt] (-4, 0.7) to[out=-10, in=165] (-0.8, 1.05);
\draw[purple, line width=2pt] (-0.57,1) circle (6pt);
\draw[purple, line width=2pt] (-0.34, 0.97) to[out=-10, in=175] (0.34, 1.03);
\draw[purple, line width=2pt] (0.57,1) circle (6pt);
\draw[purple, line width=2pt, ->] (0.8, 0.95) to[out=-10, in=190] (3.5,0.7);	
\node[purple] at (-4.9, 0.9) {\Large $\tau_2\frac{-\rmi}{4E}\text{e}^{\frac{\tilde{A}(E)}{g_{\text{s}}}}$};
\node[purple] at (5.3, 0.9) {\Large $\left(\tau_2-S_2\right)\frac{-\rmi}{4E}\text{e}^{\frac{\rmi\tilde{\mathsf{A}}(E)}{g_{\text{s}}}}$};
\node[rectangle, line width=1.5pt, draw=blue,align=center,rounded corners,minimum height=2em,fill=white, rotate=0] at (1.4, 1.9) {{\color{purple}$\tau_2\to\tau_2-S_2$}};
%\node[rectangle, line width=1.5pt, draw=orange,align=center,rounded corners,minimum height=2em,fill=white, rotate=0] at (-1.5, 1.9) {{\color{purple}$\tilde{A}(z)\to\rmi\tilde{\mathsf{A}}(z)$}};
\end{scope}
\end{tikzpicture}
\caption{We show the analytic continuation of FZZT-branes from the region $E<0$ to $E>0$ in a neighbourhood of the origin. We focus on the transition above the real line \eqref{eq:transition-above}. Start in the upper graphic with the falling FZZT contribution shown: We have a Stokes line ({\color{blue}blue}) on the negative real line and anti-Stokes lines ({\color{orange}orange}) emanating from the origin in the directions $0$, $\frac{2}{3}\pi$ and $\frac{4}{3}\pi$. Then when moving from left to right the instanton action will experience the effects of the anti-Stokes lines and change from exponentially damped to oscillating.
Compare this picture with the graphic below that depicts the growing exponential: Now there are Stokes lines ({\color{blue}blue}) in the directions $\frac{1}{3}\pi$ and $\frac{5}{3}\pi$, and the anti-Stokes lines ({\color{orange}orange}) remain unchanged. Now when moving from left to right the action still changes dominance same as before, but on top of it we are passing a Stokes line, which shifts $\tau_2$ to $\tau_2-S_2$.}
\label{fig:Generic-Treshold-Stokes}
\end{figure}
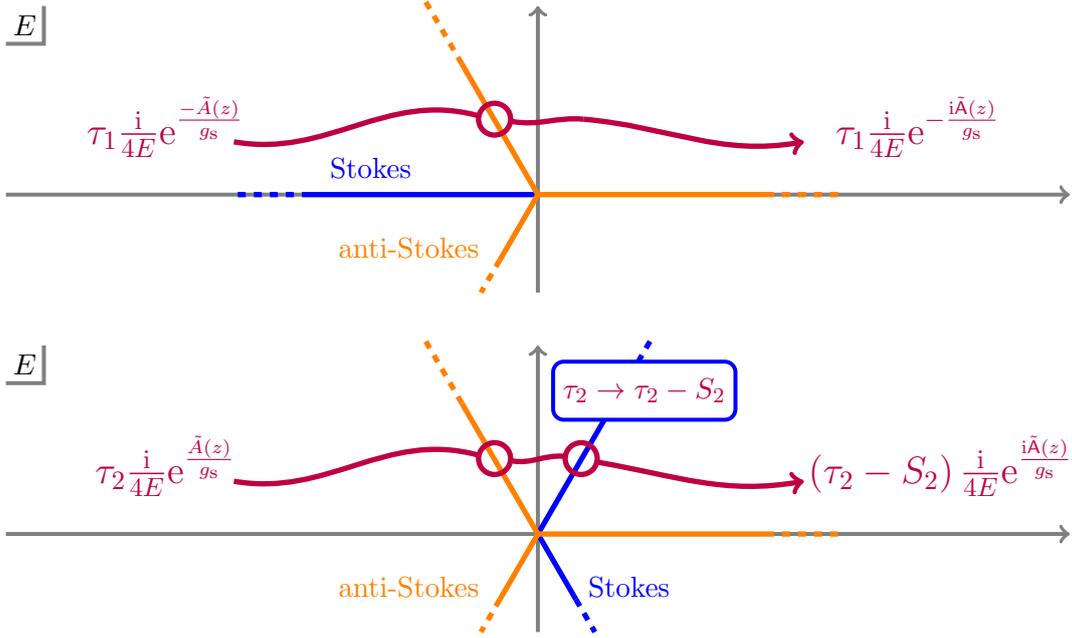
%%%%%%%%%%%%%%%%%%%%%%%%%%%%%%%%%%%%%%%%%%%%%%%%%%%%%%%%%%%%%%%%%
%
\noindent
Putting the above contributions together we can write the generic transseries for the resolvent. In general it will contain all the possible combinations of ZZ- and FZZT-branes, as shown in \cite{jr25}. However we are interested in the resurgent behaviour as the threshold $E=0$ is crossed, meaning we want to study the instanton actions that depend on $E$: the FZZT-branes \eqref{eq:falling-FZZT} and \eqref{eq:growing-FZZT}. In other words we focus on the pure FZZT subsector of the transseries that reads\footnote{We comment on the inclusion of ZZ-branes towards the end of this subsection.}
\begin{align}
\label{eq:resolvent-transseries}
R(E, \tau_1, \tau_2) &\simeq \sum\limits_{g=0}^{\infty} R_g(E) g_{\text{s}}^{2g-1} + \tau_1 R^{\text{FZZT}_{-}}(E)+\tau_2 R^{\text{FZZT}_{+}}(E),
\end{align}
\noindent
where $\tau_1, \tau_2\in\mathbb{C}$ are the FZZT transseries parameters whose choice fixes a non-perturbative completion\footnote{Of course the ZZ exponentials come attached with their own transseries parameters so for a complete non-perturbative completion one also has to fix those on top of the FZZT parameters mentioned here.}. Let us now understand the resurgent behaviour of the transseries \eqref{eq:resolvent-transseries} around $E=0$.  Specifically we want to understand (anti-) Stokes transitions. For this endeavour, assume $M(0)\neq 0$ and focus on a small neighbourhood $\mathsf{U}$ of the origin $E=0$, where the action \eqref{eq:generic-FZZT-action} behaves as\footnote{Notice that in order to compute the density \eqref{eq:definition-generic-density} we need to understand the resolvent in a neighbourhood of the real line, because we have to take a discontinuity: Even if we are only interested in the density strictly on the real line we need to understand the behaviour of the resolvent also slightly away from it.}
\begin{equation}
\tilde{A}(E) \sim (-E)^{3/2},\qquad E\in\mathsf{U}.
\end{equation}
\noindent 
For $\mathsf{U}$ sufficiently small this is a good approximation. Then the Stokes and anti-Stokes lines in a neighbourhood of the origin are\footnote{Often the nomenclature for (anti-) Stokes lines in the mathematical literature differs from the one in physics: What is an anti-Stokes line here is a Stokes line in the mathematics literature.}
\begin{itemize}
\item \underline{\textbf{Stokes Lines}} given by
\begin{equation}
\text{Im}\left(\tilde{A}(E)\right) = 0,
\end{equation} 
\noindent
with $\text{Re}\left(\tilde{A}(E)\right)>0$. Study first the physical sheet FZZT action $\exp\left(-\tilde{A}(E)/g_{\text{s}}\right)$: It has only one Stokes line emanating from $E=0$ in the $\pi$ direction. Take $\mathsf{E}\in\mathbb{R}_{<0}$ such that $\mathsf{E}\pm\rmi\epsilon\in\mathsf{U}$ with $\epsilon>0$ and small: The Stokes line is then crossed as
\begin{equation}
\label{eq:disc-pi}
R(\mathsf{E}+\rmi\epsilon, \tau_1, \tau_2) \longrightarrow R(\mathsf{E}-\rmi\epsilon, \tau_1+S_1, \tau_2),
\end{equation}
\noindent
where we have introduced the Stokes constant $S_1=1$. Notice how it is normalized to the identity. This is because the transseries sectors are normalized to agree exactly with eigenvalue tunnelling -- which corresponds in turn exactly to Stokes transitions (for details see \cite{msw07, msw08, mss22}). Now turn to the action on the non-physical sheet $\exp\left(\tilde{A}(E)/g_{\text{s}}\right)$: Here we find two Stokes lines emanating from $E=0$ in the directions $\pi/3$ and $-\pi/3$. See figure \ref{fig:Generic-Treshold-Stokes} for a visualization. Consider now $\mathsf{E}_{+} = \mathsf{E}\exp\left(\rmi\frac{\pi}{3}\right), \mathsf{E}_{-} = \mathsf{E}\exp\left(\rmi\frac{5\pi}{3}\right), \mathsf{E}\in\mathbb{R}_{> 0}$ such that $\mathsf{E}_{+}\exp\left(\pm\rmi\epsilon\right)\in\mathsf{U}$ and $\mathsf{E}_{-}\exp\left(\pm\rmi\epsilon\right)\in\mathsf{U}$ for small and real positive $\epsilon$. Then we can cross the Stokes lines via
\begin{align}
\label{eq:disc-u}
R(\mathsf{E}_{+}\rme^{+\rmi\epsilon}, \tau_1, \tau_2) &\longrightarrow R(\mathsf{E}_{+}\rme^{-\rmi\epsilon}, \tau_1, \tau_2-S_2),\\
\label{eq:disc-b}
R(\mathsf{E}_{-}\rme^{-\rmi\epsilon}, \tau_1, \tau_2) &\longrightarrow R(\mathsf{E}_{-}\rme^{+\rmi\epsilon}, \tau_1, \tau_2+S_2),
\end{align}
\noindent
where the Stokes constant reads $S_2=1$ as expected by now and the extra minus sign of the shift in the first line comes from the direction of travel through the Stokes line. 
\item \underline{\textbf{Anti-Stokes Lines}} given by
\begin{equation}
\text{Re}\left(\tilde{A}(E)\right) = 0
\end{equation} 
\noindent
which are the lines where exponentials that were previously suppressed become dominant or vice versa.  From $E=0$ they emanate to the directions $0, 2\pi/3, 4\pi/3$ for both FZZT actions. See again figure \ref{fig:Generic-Treshold-Stokes} for a visualization. 
\end{itemize}
\noindent
We can assemble the above Stokes and anti-Stokes transitions into the local picture close to the threshold $E=0$ depicted in figure \ref{fig:Generic-Treshold-Stokes}. Armed with this discussion we are now ready to compute non-perturbative contributions to the density by analysing the discontinuity \eqref{eq:definition-generic-density}. We first focus on $E<0$: Here there is a discontinuity that is created by the Stokes line on the negative real axis, which we can compute from formula \eqref{eq:disc-pi} to read
\begin{equation}
\frac{1}{2\pi\rmi}\text{Disc}\left[R(E, \tau_1, \tau_2)\right] \simeq \underbrace{S_1}_{=1}\left(\frac{1}{8\pi E}+ \mathcal{O}(g_{\text{s}})\right)\text{e}^{-\frac{\tilde{A}(E)}{g_{\text{s}}}},\qquad E<0.
\end{equation}
\noindent
Notice that this result does not depend on the non-perturbative completion (the transseries parameters). Notice also how this discontinuity comes purely from the Stokes transition on the negative real line: It is expected to be smoothed out by resumming higher $g_{\text{s}}$ contributions. It is also worthwhile to mention that only the damped exponential appears here because this is the one that experiences a Stokes line on the negative real axis -- in contrast to the growing exponential, which does not: This is intrinsic to the construction of the density \eqref{eq:definition-generic-density} and not an external choice that we had to make.  There are no further contributions to the discontinuity in the region $E<0$. Consequently move to $E>0$. We will proceed by considering continuous paths starting from $E<0$ and keep track of the changes as we move to $E>0$ (see figure \ref{fig:Generic-Treshold-Stokes}). The FZZT contributions will experience an anti Stokes line as they cross close to $E=0$ and for this reason they will change dominance. Moreover, in the case considered here where the spectral curve is strictly real or imaginary on the real line, the FZZT action becomes purely imaginary close to the positive real line. This anti-Stokes phenomenon is exactly what drives the change of behaviour from exponential for $E<0$ to oscillating for $E>0$. Above the real line let us use the offset $E=\mathsf{E}+\rmi\epsilon$ and imply $\epsilon$ to be small. We will then write
\begin{equation}
\label{eq:FZZT-anti-Stokes-action}
\rmi\tilde{\mathsf{A}}(E) = \lim\limits_{\epsilon\to 0} \tilde{A}(\mathsf{E}+\rmi\epsilon),\qquad \mathsf{E}\in\mathbb{R}_{>0}.
\end{equation}
\noindent
Following the path drawn in figure \ref{fig:Generic-Treshold-Stokes} -- above the real line -- then amounts to
\begin{align}
\label{eq:transition-above}
\begin{tikzpicture}[baseline=(current  bounding  box.center)]
\node[darkgray] at (-5, 1) {\underline{$E=\mathsf{E}+\rmi\epsilon,\,\,\mathsf{E}<0:$}};
\node at (-5,0) {$\tau_1 \left(\frac{\rmi}{4 E}+\mathcal{O}\left(g_{\text{s}}\right)\right) \text{e}^{-\frac{\tilde{A}(E)}{g_{\text{s}}}}$};
\node at (-5,-1) {$\tau_2 \left(-\frac{\rmi}{4 E}+\mathcal{O}\left(g_{\text{s}}\right)\right) \text{e}^{+\frac{\tilde{A}(E)}{g_{\text{s}}}}$};
\node[darkgray] at (4, 1) {\underline{$E=\mathsf{E}+\rmi\epsilon,\,\,\mathsf{E}>0:$}};
\node at (4,0) {$\tau_1 \left(\frac{\rmi}{4 E}+\mathcal{O}\left(g_{\text{s}}\right)\right) \text{e}^{-\frac{\rmi\tilde{\mathsf{A}}(E)}{g_{\text{s}}}},$};
\node at (4,-1) {$\left(\tau_2 - S_2\right) \left(-\frac{\rmi}{4 E}+\mathcal{O}\left(g_{\text{s}}\right)\right) \text{e}^{+\frac{\rmi\tilde{\mathsf{A}}(E)}{g_{\text{s}}}},$};
\draw[purple, line width=2pt, ->] (-2, -0.5) to[out=10, in=165] (-0.8,-0.6) to[out=345, in=170] (1, -0.5);
\node[purple] at (-0.5, -0.1) {(Figure \ref{fig:Generic-Treshold-Stokes})};
\end{tikzpicture}
\end{align}
\noindent
where the first line corresponds to the physical sheet FZZT contribution \eqref{eq:falling-FZZT}, whereas the second line shows the change of the non-physical sheet one \eqref{eq:growing-FZZT}. Further we remember that $S_2=1$. In the same spirit we can follow the path below the real line, which yields
\begin{align}
\begin{tikzpicture}[baseline=(current  bounding  box.center)]
\node[darkgray] at (-5, 1) {\underline{$E=\mathsf{E}-\rmi\epsilon,\,\,\mathsf{E}<0:$}};
\node at (-5,0) {$\tau_1 \left(\frac{\rmi}{4 E}+\mathcal{O}\left(g_{\text{s}}\right)\right) \text{e}^{-\frac{\tilde{A}(E)}{g_{\text{s}}}}$};
\node at (-5,-1) {$\tau_2 \left(-\frac{\rmi}{4 E}+\mathcal{O}\left(g_{\text{s}}\right)\right) \text{e}^{+\frac{\tilde{A}(E)}{g_{\text{s}}}}$};
\node[darkgray] at (4, 1) {\underline{$E=\mathsf{E}-\rmi\epsilon,\,\,\mathsf{E}>0:$}};
\node at (4,0) {$\tau_1 \left(\frac{\rmi}{4 E}+\mathcal{O}\left(g_{\text{s}}\right)\right) \text{e}^{+\frac{\rmi\tilde{\mathsf{A}}(E)}{g_{\text{s}}}},$};
\node at (4,-1) {$\left(\tau_2 + S_2\right) \left(-\frac{\rmi}{4 E}+\mathcal{O}\left(g_{\text{s}}\right)\right) \text{e}^{-\frac{\rmi\tilde{\mathsf{A}}(E)}{g_{\text{s}}}},$};
\draw[purple, line width=2pt, ->] (-2, -0.5) to[out=10, in=165] (-0.8,-0.6) to[out=345, in=170] (1, -0.5);
%\node[purple] at (-0.5, -0.1) {(Figure \ref{fig:Generic-Treshold-Stokes})};
\end{tikzpicture}
\end{align}
\noindent
where again the first line corresponds to the physical sheet FZZT and the second line to its sibling and again $S_2=1$. Then the FZZT contribution to the discontinuity for $E>0$ reads
\begin{align}
\underbrace{S_2}_{=1}\left(\frac{\rmi}{2 E}+\mathcal{O}\left(g_{\text{s}}\right)\right)\cos\left(\frac{\tilde{\mathsf{A}}\left(E\right)}{g_{\text{s}}}\right) + \left(\tau_2+\tau_1\right)\left(\frac{1}{2 E}+\mathcal{O}\left(g_{\text{s}}\right)\right)\sin\left(\frac{\tilde{\mathsf{A}}\left(E\right)}{g_{\text{s}}}\right).
\end{align}
\noindent
Interestingly the cosine part only depends on Stokes constants, while the sine part depends on transseries parameters! We are now in a position to put together all of the above contributions to find the non-perturbative density at leading order in $g_{\text{s}}$ 
\begin{align}
\label{eq:non-pert-rho}
\left\langle\varrho\left(E, \tau_1, \tau_2\right)\right\rangle \simeq
\begin{cases}
\frac{1}{g_{\text{s}}}\varrho_0(E)-\frac{1}{4\pi E}\cos\left(\frac{\tilde{\mathsf{A}}\left(E\right)}{g_{\text{s}}}\right)+ \left(\tau_2+\tau_1\right)\frac{\rmi}{4\pi E}\sin\left(\frac{\tilde{\mathsf{A}}\left(E\right)}{g_{\text{s}}}\right),\quad & E>0,\\[8pt]
-\frac{1}{8\pi E}\exp\left(-\frac{\tilde{A}\left(E\right)}{g_{\text{s}}}\right), & E<0.
\end{cases}
\end{align}
\noindent
Some comments are in order: Firstly, the expression \eqref{eq:non-pert-rho} is just the leading term of a full transseries that is still asymptotic and should be resummed to find the non-perturbative density as a function. Secondly, the oscillatory behaviour of the non-perturbative density \eqref{eq:non-pert-rho} for $E>0$ comes from FZZT-branes that have undergone Stokes and anti-Stokes transitions\footnote{Anti-Stokes transitions are sometimes related to phase transitions \cite{m08, krsst25a, brr26}. It could be interesting to explore this connection further.} as computed in \eqref{eq:transition-above}. We identify\footnote{This is interesting to study because of the following analogy: In JT gravity, oscillatory terms in the matrix model density might indicate that any exact microscopic realization (like a single hamiltonian system drawn from the ensemble) should have a discrete spectrum \cite{sss-semi, sss19, sw20}.}
\begin{center}
\begin{tikzpicture}[
	grayframe/.style={
		rectangle,
		draw=gray,
		text width=7em,
		align=center,
		rounded corners,
		minimum height=2em
	}, line width=1pt]
\node[grayframe] at (0, 0) {oscillatory density contribution};
\node[grayframe] at (5,0) {resurgent anti-Stokes behavior of FZZT branes};
\draw[gray, <->, line width=1.9pt] (2, 0) -- (3,0);
\node at (6.7, -0.5) {.};
\end{tikzpicture}
\end{center}
\noindent
Thirdly, the universal piece -- meaning the $\tau_1,\tau_2$ independent piece -- of formula \eqref{eq:non-pert-rho} corresponds exactly to the result \eqref{eq:triple-s-formula} from \cite{sss19}. Indeed starting from the explicit form for the spectral curve \eqref{eq:genericCurve} we can use the definition of the density \eqref{eq:definition-generic-density} to find
\begin{align}
\tilde{\mathsf{A}}(E) = 2\pi \int\limits_{0}^{E} \varrho_{0}(E^{\prime})\text{d}E^{\prime}.
\end{align}
\noindent
Notice furthermore, that in contrast to the computation in \cite{sss19} no particular choice of matrix integral contour was needed to obtain the aforementioned universal piece -- in fact it is \emph{universally} there for any choice of non-perturbative completion. On top of this, there are further contributions that do depend on the transseries parameters $\tau_1, \tau_2$ -- and it is only here where the choice of non-perturbative completion enters. These non-universal terms are still oscillatory -- in fact they are of order $\mathcal{O}(1)$ and in this sense cannot be ignored when turned on. Last but not least, let us again comment on the fact that the discontinuity of the resolvent on the negative real line is mitigated by a Stokes transition, which implies the expectation that it should be smoothed out by $g_{\text{s}}$ corrections upon resummation of the asymptotic perturbative series. In this sense the full non-perturbative eigenvalue density is still expected to be exponentially small in the region $E<0$. Let us finish this subsection with two comments:

\paragraph{Universality of Oscillatory Behaviour:} We want to elaborate on the fact, that formula \eqref{eq:non-pert-rho} has a universal -- meaning $\tau_i$ independent -- piece that emerges purely from Stokes transitions. It comes about as follows: even when initially turning off all non-perturbative FZZT contributions (choosing $\tau_1=\tau_2=0$ for the resolvent), the Stokes transitions at $E=0$ will turn them back on and generate oscillatory behaviour for the density. In other words such oscillations should appear in the density for any hermitian matrix model \emph{independent of the choice of non-perturbative completion}. No boundary value matching is needed. The resurgence machinery above constructs them automatically: They do not need to be put in by hand, be it by choice of eigenvalue integration contour or by choice of transseries parameters. In this context let us further remark that the minimal choice (where all FZZT instanton effects are initially turned off, meaning $\tau_1=\tau_2=0$) exactly produces the same result as in \cite{sss19} where it was obtained by choosing a real eigenvalue integration contour. As a final comment, formula \eqref{eq:non-pert-rho} was derived for generic hermitian matrix models: It is therefore natural to conjecture that it holds for any problem that fulfils topological recursion as long as it has the appropriate differential and Bergman kernel. In fact, our construction above does not rely on an explicit matrix integral formulation -- it is based purely on the non-perturbative structure of the resolvent.

\paragraph{Inclusion of ZZ-Branes:} Finally, let us comment on ZZ-brane contributions. For simplicity we focus our attention on the one instanton and anti-instanton contributions \eqref{eq:resolvent-ZZ-p} and \eqref{eq:resolvent-ZZ-m} for a saddle point $E^{\star}$. They contribute additively to the transseries \eqref{eq:resolvent-transseries} with
\begin{equation}
\sigma_1 \, R^{(1|0)_{\text{ZZ}}}(E)+\sigma_2 \, R^{(0|1)_{\text{ZZ}}}(E),
\end{equation}
\noindent
where $\sigma_1,\sigma_2$ are the ZZ transseries parameters. Notice that the instanton actions $A_{k,\pm}$ are $E$ independent, meaning in contrast to the FZZT analysis above there are no Stokes transitions when varying $E$. Nevertheless, the explicit expressions for the ZZ instantons \eqref{eq:resolvent-ZZ-m} and \eqref{eq:resolvent-ZZ-p} have square root branch cuts on the positive real line -- where they will contribute to the discontinuity of the resolvent \eqref{eq:definition-generic-density} (meaning to the density). Those square root branch cuts give\footnote{Notice that there is also the special point $E^{\star}$ where one might expect delta like spikes to appear from the ZZ-brane discontinuity.}
\begin{align}
\frac{1}{2\pi\rmi}\text{Disc}\left[\sigma_1 R^{(1|0)_{\text{ZZ}}}(E)\right] \simeq
\begin{cases}
0,& E< 0, E\neq E^{\star},\\
\sigma_1 \,\frac{1}{\pi\rmi} R^{(1|0)_{\text{ZZ}}}(E),& E>0,
\end{cases}\\
\frac{1}{2\pi\rmi}\text{Disc}\left[\sigma_2 R^{(0|1)_{\text{ZZ}}}(E)\right] \simeq 
\begin{cases}
0,& E < 0, E\neq E^{\star}\\
\sigma_2 \,\frac{1}{\pi\rmi} R^{(0|1)_{\text{ZZ}}}(E),& E>0.
\end{cases}
\end{align}
\noindent
We can see that there is no contribution in the region $E<0$ which is expected. After crossing the threshold $E=0$ the ZZ-branes are turned back on\footnote{They are turned on by the discontinuity of the square-root, not by a Stokes phenomenon.} with two times the transseries parameters as compared to the resolvent transseries. We expect similar results for higher $(n|m)$ contributions, but there the switching of square root sheets might not have the simple effect of the factor of 2 above. Nevertheless, because of the absence of square root branch cuts on the negative real line, we do not expect any ZZ contribution to appear for $E<0$ even for higher $(n|m)$ ZZ contributions -- up to the subtleties around $E^{\star}$. It would be interesting to study the contributions of all ZZ-branes in more detail, especially those that appear together with the FZZT contributions (see also \cite{jr25}).

%%%%%%%%%%%%%%%%%%%%%%%%%%%%%%%%%%%%%%%%%%%%%%%%%%%%%%%%%%%%%%%%%
\subsection{Example: The Virasoro Minimal String and 3d Gravity}
\label{subsec:vms-stokes}
%%%%%%%%%%%%%%%%%%%%%%%%%%%%%%%%%%%%%%%%%%%%%%%%%%%%%%%%%%%%%%%%%

%%%%%%%%%%%%%%%%%%%%%%%%%%%%%%%%%%%%%%%%%%%%%%%%%%%%%%%%%%%%%%%%%
\begin{figure}
\centering
\begin{tikzpicture}
\node at (-5.2,-0.1) {\includegraphics[width=0.3\textwidth]{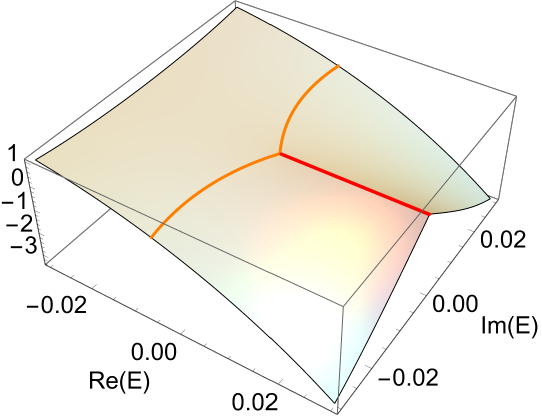}};
\node at (0,-0.1) {\includegraphics[width=0.3\textwidth]{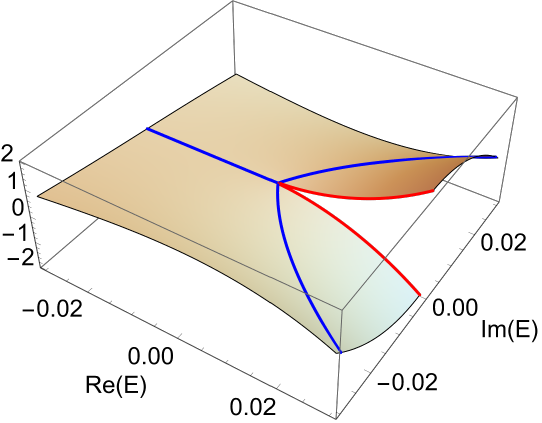}};
\node at (5.2,-0.3) {\includegraphics[width=0.285\textwidth]{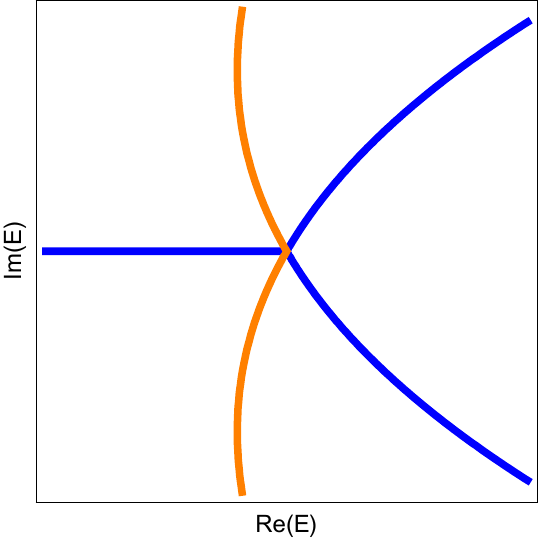}};
\draw[snake it, draw=red, line width=0.8pt] (5.37, -0.13) -- (7.4, -0.13);
\draw[orange, line width=1.9pt] (5.3, -0.13) -- (7.4, -0.13);
\end{tikzpicture}
\caption{We plot the effective potential of the VMS as a function of $E$. On the left we show the real part with the anti-Stokes lines in {\color{orange}orange}, in the middle we show the imaginary part together with Stokes lines in {\color{blue}blue}, and on the right we show the complex $E$ plane with both types of Stokes transitions drawn. In all figures we denote the branch cut on the positive real line in {\color{red}red}. Close to $E=0$ we indeed find the exact Stokes behaviour as predicted in the generic case of subsection \ref{subsec:gen-construction} as shown in figure \ref{fig:Generic-Treshold-Stokes}.}
\label{fig:veff-VMS}
\end{figure}
%%%%%%%%%%%%%%%%%%%%%%%%%%%%%%%%%%%%%%%%%%%%%%%%%%%%%%%%%%%%%%%%%

%%%%%%%%%%%%%%%%%%%%%%%%%%%%%%%%%%%%%%%%%%%%%%%%%%%%%%%%%%%%%%%%%
\begin{figure}
\centering
\includegraphics[width=0.45\textwidth]{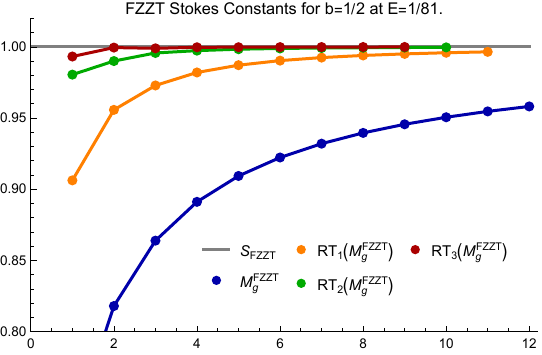}
\hspace{0.05\textwidth}
\includegraphics[width=0.45\textwidth]{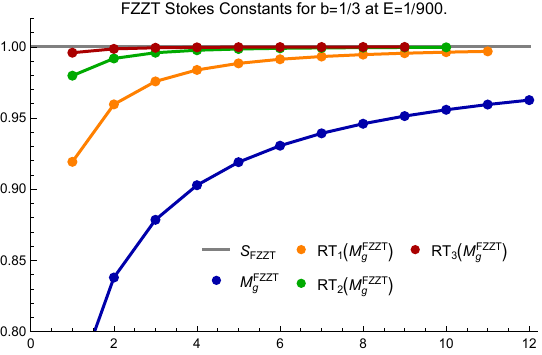}
\caption{Asymptotic checks for the Stokes constants of the FZZT contributions. Here we plot the sequence \eqref{eq:asymptoticscheck-m} and check its asymptotic behaviour. We depict the sequence itself together with its first 3 Richardson transforms. We find perfect agreement with the prediction.}
\label{fig:Asym-FZZT-action}
\end{figure}
%%%%%%%%%%%%%%%%%%%%%%%%%%%%%%%%%%%%%%%%%%%%%%%%%%%%%%%%%%%%%%%%%

%%%%%%%%%%%%%%%%%%%%%%%%%%%%%%%%%%%%%%%%%%%%%%%%%%%%%%%%%%%%%%%%%
\begin{figure}
\centering
\includegraphics[width=0.45\textwidth]{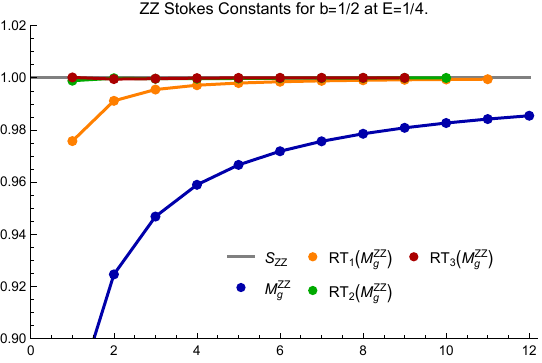}
\caption{Asymptotic checks for the Stokes constants of the ZZ contributions. We plot the sequence \eqref{eq:asymptoticscheck-m} and check its asymptotic behaviour. We depict the sequence itself together with its first 3 Richardson transforms and find perfect agreement with the prediction. Compare this asymptotic check with the FZZT check visualised in figure \ref{fig:Asym-FZZT-action}.}
\label{fig:Asym-ZZ-action}
\end{figure}
%%%%%%%%%%%%%%%%%%%%%%%%%%%%%%%%%%%%%%%%%%%%%%%%%%%%%%%%%%%%%%%%%

We want to check the above derivation on the example of the VMS and understand possible implications for 3d gravity as introduced in section \ref{sec:3d-gravity}. Let us first focus on the VMS and understand the discussion in subsection \ref{subsec:gen-construction} in that context. To give the explicit formula for the non-perturbative completion of the VMS density we need the FZZT instanton action $\tilde{A}(E)$ given in formula \eqref{eq:atilde} and we further compute the action $\tilde{\mathsf{A}}(E)$ given in formula \eqref{eq:FZZT-anti-Stokes-action} to be
\begin{equation}
\label{eq:VMS-FZZT-action}
\tilde{\mathsf{A}}(E) = \frac{2\sqrt{2}b}{1-b^4}\left(\left(1+b^2\right)\sinh\left[2\pi\left(b-\frac{1}{b}\right)\sqrt{E}\right]+\left(1-b^2\right)\sinh\left[2\pi\left(b+\frac{1}{b}\right)\sqrt{E}\right]\right).
\end{equation}
\noindent
Let us start by studying the FZZT instanton action \eqref{eq:atilde} close to the origin $E=0$. To this end we have visualized the effective potential of the VMS in figure \ref{fig:veff-VMS}. Indeed close to the origin the Stokes graph matches exactly to the discussion in subsection \ref{subsec:gen-construction} with expected deviation as we move further out into the $E$ plane (see also figure \ref{fig:Generic-Treshold-Stokes}).

We can furthermore check the Stokes constants $S_1=1$ in formula \eqref{eq:disc-pi} and $S_2=1$ in formulae \eqref{eq:disc-b} and \eqref{eq:disc-u} explicitly with large order methods. We are interested in the region $E>0$, where we have the prediction (see also subsection \ref{subsec:wall-crossing} for the $E<0$ discussion)
\begin{equation}
\label{eq:asymptoticscheck-m}
\begin{cases}
M^{\text{FZZT}}_g(E, b) := -\pi \frac{R_g(E+\rmi 0)\tilde{A}(E+\rmi 0)^{2g-1}}{\Gamma\left(2g-1\right)}\Big\lbrace 4 E \Big\rbrace\qquad\qquad\qquad\qquad\,\,\,\,\,\, \xrightarrow{g\to\infty}\quad 1,\quad & \tilde{\mathsf{A}}(E) < A,\\
M_g^{\text{ZZ}}(E, b) := \rmi\pi \frac{R_g(E+\rmi 0)A^{2g-\frac{3}{2}}}{\Gamma\left(2g-\frac{3}{2}\right)}\Big\lbrace  2^{7/4} \pi^{3/2}\sqrt{b\sin(b^2\pi)}\frac{\sqrt{E}(b^2+4E)}{b} \Big\rbrace \quad\xrightarrow{g\to\infty}\quad 1,\quad & \tilde{\mathsf{A}}(E) > A.
\end{cases}
\end{equation}
\noindent
Here the first line corresponds to the FZZT contribution while the second line is the standard ZZ prediction with leading instanton action $A$ (compare also to \eqref{eq:actionresolvent} in subsection \ref{subsec:wall-crossing}). We have checked these asymptotics with the available data (up to genus 14 as in subsection \ref{subsec:wall-crossing}) with 3 digits of precision. The checks are visualized in figures \ref{fig:Asym-FZZT-action} and \ref{fig:Asym-ZZ-action}. In addition we can compute the Stokes constants by Borel-Laplace summing and checking the discontinuity across the Stokes line. Explicitly, we will compute the Borel transform, approximate it with a Padé approximant and compute the residues of poles when crossing each of the Stokes lines (For similar techniques see \cite{bssv22, gm22, ms24, h25}). We choose $b=\frac{1}{2}$. Further we have chosen small values $z=-0.1, z=0.0560049\mp 0.114615\rmi$ that lie on each Stokes line respectively\footnote{Notice that the Stokes lines for the VMS deform slightly when moving away from the origin and they no longer lie cleanly on the rays $\frac{1+2k}{3}\pi$. Therefore we have numerically computed values of $z$ that hit the Stokes lines exactly.}. In this way we find 
\begin{align}
\text{Stokes line at arg}(E)=\pi: \qquad &S_1=0.9139\dots,\\[5pt]
\text{Stokes line at arg}(E)=\frac{\pi}{3}: \qquad &S_2=0.9577\dots\,-0.0061\dots\rmi,\\
\text{Stokes line at arg}(E)=\frac{5\pi}{3}: \qquad &S_2=0.9577\dots\,+0.0061\dots\rmi.
\end{align}
\noindent
Given that we have only 14 non zero terms in the resolvent series (which allows us to take $(14,13)$ and $(14,14)$ almost diagonal Padé approximants), this shows qualitative support for the predictions of $S_1=S_2=1$ in subsection \ref{subsec:gen-construction}, but as expected the precision is considerably lower than in the direct asymptotic tests. Based on the Stokes network in figure \ref{fig:veff-VMS} and the numerical checks for the Stokes constants we conclude that the analysis in subsection \ref{subsec:gen-construction} applies to the VMS verbatim and we find exactly formula \eqref{eq:non-pert-rho} which we specialize to the VMS case below
\begin{align}
\label{eq:non-pert-rho-VMS}
\left\langle\varrho_{\text{V}}\left(E, \tau_1^{\text{V}}, \tau_2^{\text{V}}\right)\right\rangle \simeq 
\begin{cases}
\frac{1}{g_{\text{s}}}\varrho_{\text{V},0}(E)-\frac{1}{4\pi E}\cos\left(\frac{\tilde{\mathsf{A}}\left(E\right)}{g_{\text{s}}}\right)+ \left(\tau_2^{\text{V}}+\tau_1^{\text{V}}\right)\frac{\rmi}{4\pi E}\sin\left(\frac{\tilde{\mathsf{A}}\left(E\right)}{g_{\text{s}}}\right),\quad & E>0,\\[8pt]
-\frac{1}{8\pi E}\exp\left(-\frac{\tilde{A}\left(E\right)}{g_{\text{s}}}\right), & E<0.
\end{cases}
\end{align}
\noindent
Let us briefly discuss the structure of that formula: This is the leading order of a transseries -- meaning the involved series are still asymptotic. $\varrho_{\text{V},0}$ is the leading order of the perturbative VMS density and $\tilde{\mathsf{A}}(E)$ is the FZZT contribution after anti-Stokes transition given in \eqref{eq:VMS-FZZT-action}. In the non-perturbative VMS density \eqref{eq:non-pert-rho-VMS} there is a falling exponential term on the negative real line. On the other hand oscillatory terms are turned on when crossing the Stokes and anti-Stokes lines at the origin $E=0$. Then, for $E>0$ we find a universal contribution coming from Stokes constants and an additional piece that depends on the non-perturbative completion (choosing $\tau_1^{\text{V}},\tau_2^{\text{V}}$). To keep the density real we should choose $\text{Re}\left(\tau_1^{\text{V}}+\tau_2^{\text{V}}\right)=0$. As expected this is exactly analogous to the generic result \eqref{eq:non-pert-rho}.

Let us now understand the connection to 3d gravity: We will again employ the setting of \cite{jrsw25} as already described in section \ref{sec:3d-gravity}, where the gravitational path integral on $\Sigma_{g,n}\times I$ with EOW branes at the ends of the interval $I$ is identified with correlators of the VMS. We can compute the 3d gravity primary density $\rho_{\text{grav}}^{(ab)}(h)$ as an integral of the disk partition function \cite{jrsw25}
\begin{equation}
\label{eq:rho-grav-inverse-laplace}
\left\langle\rho_{\text{grav}}^{(ab)}(h)\right\rangle_{g=0} = \frac{1}{g_{\text{s}}}\int\limits_{\Gamma^{\uparrow}}\frac{\text{d}\beta}{2\pi\rmi}\rme^{\beta\left(h-\frac{c-1}{24}\right)} \mathsf{Z}_{\text{disc}}^{(b)}(\beta) = \frac{1}{g_{\text{s}}}\left.\frac{2\sqrt{2}}{P}\sinh\left(2\pi b P\right)\sinh\left(2\pi b^{-1} P\right)\right|_{P^2=h-\frac{c-1}{24}},
\end{equation}
\noindent
where the upward contour $\Gamma^{\uparrow} = \gamma+\rmi\mathbb{R},\gamma\in\mathbb{R}_{>0}$ and $\mathsf Z_{\mathrm{disc}}^{(b)}(\beta)$ denotes the disk amplitude with its overall factor $g_{\text{s}}^{-1}$ removed. Here $a, b$ correspond to the boundary conditions of the EOW branes at the end of the interval $I$ and $h$ is a primary conformal weight. The brackets $\langle\,\cdots\rangle$ on the gravity side denote ensemble averaging and the subscript selects a genus. As in \cite{jrsw25} (and as reviewed in section \ref{sec:3d-gravity}) the disk factors and the string coupling at each genus are implicit in this formulation and we have the identification $1/g_{\text{s}}=Z_{\text{disk}}^{(a)}Z_{\text{disk}}^{(b)}$. 
The density shows Cardy behavior for $h>\frac{c-1}{24}$\footnote{We can expand \eqref{eq:rho-grav-inverse-laplace} at large $P$ to find
\begin{equation}
\exp\left(2\pi\sqrt{\frac{c-1}{6}}\sqrt{h-\frac{c-1}{24}}\right).\nonumber
\end{equation}}. We now use the map from \cite{jrsw25}
\begin{equation}
\label{eq:3d-grav-density-genus}
\left\langle\rho_{\text{grav}}^{(ab)}(h)\right\rangle_g = \left\langle\varrho_{\text{V}}(E)\right\rangle_g\Big|_{E=h-\frac{c-1}{24}},
\end{equation}
\noindent
where for the VMS quantities $\langle\,\cdots\rangle$ denotes the matrix model expectation value and the subscript $g$ selects a genus in the perturbative expansion. It is then straightforward to identify the onset of the Cardy behaviour with the threshold $E=0$ in the VMS matrix model\footnote{Under the identification $E=h-(c-1)/24$, the spectral edge corresponds to the threshold $h=(c-1)/24$ for Cardy growth in the restricted ensemble of primary weights described by this correspondence. The Stokes transition discussed in subsection 5.1 occurs at this edge. This identification concerns the ensemble-averaged observable and does not by itself establish a universal black-hole transition in unrestricted three-dimensional gravity.} -- exactly where the FZZT-branes undergo Stokes and anti-Stokes transitions (see also figure \ref{fig:veff-VMS}). With this machinery in place, we can now try to understand the non-perturbative effects that arise if one sums the 3d gravity density over the genus. Consequently, sum equation \eqref{eq:3d-grav-density-genus} over $g$ to form the following asymptotic equality
\begin{equation}
\label{eq:VMS-to-gravity}
\begin{tikzpicture}[baseline={([yshift=0 ex]current bounding box.center)}]
\node at (0,0) {$\sum\limits_{g=0}^{\infty}\left\langle\varrho_{\text{V}}(E)\right\rangle_g\Big|_{E=h-\frac{c-1}{24}} \quad = \quad  \sum\limits_{g=0}^{\infty}\left\langle\rho_{\text{grav}}^{(ab)}(h)\right\rangle_g ,$};
\node (a) at (-9, 0){$\left\langle\varrho_{\text{V}}(E)\right\rangle\Big|_{E=h-\frac{c-1}{24}}$};
\draw[cornellred, ->, line width=1pt] (-7.3, 0) -- (-4.2, 0);
\node[align=center, cornellred, text width=5em] at (-5.7, 0) {\footnotesize asymptotic expansion};
\end{tikzpicture}
\end{equation} 
\noindent
where we have written a true equality between the two asymptotic series, because they are really equal term by term. One can now wonder about their non-perturbative completion: First choose $1/g_{\text{s}}=Z_{\text{disk}}^{(a)}Z_{\text{disk}}^{(b)}$ as organizing parameter for the asymptotic study\footnote{Similar to the discussion in section \ref{sec:3d-gravity} it is not straightforward to disentangle the large $c$ growth of $1/g_{\text{s}}=Z_{\text{disk}}^{(a)}Z_{\text{disk}}^{(b)}$ and the $c$ dependence of the density itself. For the current discussion and based on our explicit checks in section \ref{sec:3d-gravity} let us assume that this procedure is valid. It would be interesting to check this further.}. Secondly, because the VMS and the gravity sum in formula \eqref{eq:VMS-to-gravity} are really exactly identical, they will have the exact same asymptotic behaviour\footnote{In fact, as shown in \cite{jrsw25} the map between the VMS and 3d gravity works term by term also for partition functions, and with this also for their Laplace transforms (the resolvent analogue). In this sense the discussion in subsection \ref{subsec:gen-construction} applies here too.} and therefore they will also generate the exact same transseries -- the only free choice being the transseries parameters. This choice is where the VMS non-perturbative completion (with parameters $\tau_1^{\text{V}},\tau_2^{\text{V}}$) might differ from the completion of the gravity sum in \eqref{eq:VMS-to-gravity}, where we will therefore introduce new transseries parameters $\tau_1^{\text{grav}},\tau_2^{\text{grav}}$\footnote{Of course one has to also include ZZ-branes in the full 'gravity' transseries with their own new transseries parameters. Here for brevity we omit them as the inclusion is straightforward. See also \cite{jr25}.}. Consequently we employ formula \eqref{eq:non-pert-rho-VMS} for $E>0$ to find at leading order
\begin{align}
\label{eq:leading-grav-density}
\left\langle\rho_{\text{grav}}^{(ab)}(h, \tau_1^{\text{grav}}, \tau_2^{\text{grav}})\right\rangle &\simeq \left\langle\rho_{\text{grav}}^{(ab)}(h)\right\rangle_0 +\dots \quad -\frac{1}{4\pi \left(h-\frac{c-1}{24}\right)}\left(1+\mathcal{O}(g_{\text{s}})\right)\cos\left(\frac{\tilde{\mathsf{A}}\left(h-\frac{c-1}{24}\right)}{g_{\text{s}}}\right)+\nonumber\\
&+ \left(\tau_2^{\text{grav}}+\tau_1^{\text{grav}}\right)\frac{\rmi}{4\pi \left(h-\frac{c-1}{24}\right)}\left(1+\mathcal{O}\left(g_{\text{s}}\right)\right)\sin\left(\frac{\tilde{\mathsf{A}}\left(h-\frac{c-1}{24}\right)}{g_{\text{s}}}\right),
\end{align}
\noindent
where $\tilde{\mathsf{A}}(E)$ is given in formula \eqref{eq:VMS-FZZT-action}. Interestingly formula \eqref{eq:leading-grav-density} still contains a universal, oscillating piece, that is inherited from the VMS description. On top of this we also have oscillatory terms that depend on the gravity transseries parameters. 

To summarize, the non-perturbative VMS density receives oscillatory contributions of order $\mathcal{O}(1)$ (universal and $\tau_{1}^{\text{V}},\tau_2^{\text{V}}$ dependent) from FZZT-branes that have experienced an anti-Stokes transition at the threshold $E=0$. On the 3d gravity (with EOW branes) side we can say that if one sums the primary density over the genus one finds similar oscillatory contributions. Here again there is a universal part and one that depends on the parameters $\tau_1^{\text{grav}},\tau_2^{\text{grav}}$. A priori it is not clear if one is supposed to set $\tau_1^{\text{grav}}=\tau_1^{\text{V}}, \tau_2^{\text{grav}}=\tau_2^{\text{V}}$. Nevertheless, it would be interesting to construct a full non-perturbative completion of the density along the lines of section \ref{sec:exactZ} and \cite{krsst25a} to exactly quantify the non-perturbative "freedom" inherent to the construction above\footnote{Perhaps also the approach in the recent work \cite{bcelp26} can shed some light on this matter.}.

%%%%%%%%%%%%%%%%%%%%%%%%%%%%%%%%%%%%%%%%%%%%%%%%%%%%%%%%%%%%%%%%%
\subsection{Example: JT Gravity}
\label{subsec:jt-stokes}
%%%%%%%%%%%%%%%%%%%%%%%%%%%%%%%%%%%%%%%%%%%%%%%%%%%%%%%%%%%%%%%%%

We want to further comment on JT gravity. As a limit of the VMS the above discussion clearly also applies here. We will follow the conventions of \cite{eggls23}, which is exactly consistent with double scaling the VMS\footnote{In the scaling limit to JT gravity the VMS energy $E$ scales with $b^2$. This follows from the scaling for $P$ in \cite{cemr23}.} results computed above in the conventions of \cite{cemr23}. At the same time, non-perturbative JT gravity has been discussed at length in the literature. Therefore some $g_{\text{s}}$ corrections to formula \eqref{eq:falling-FZZT} and \eqref{eq:growing-FZZT} are known explicitly for this example and we can compute higher order corrections to our generic formula \eqref{eq:non-pert-rho}. Namely we have the FZZT prediction \cite{os20, eggls23, jr25}
\begin{align}
\label{eq:JT-FZZT-prediction-p}
R^{\text{FZZT}_{-}}(E) &\simeq \frac{\rmi}{4E}\left(1-g_{\text{s}}\frac{17-2\pi^2 E}{12\left(-E\right)^{3/2}} - g_{\text{s}}^2\frac{1225-644\pi^2 E+148\pi^4 E^2}{288 E^3}+\dots\right)\text{e}^{-\frac{\tilde{A}(E)}{g_{\text{s}}}},\\
\label{eq:JT-FZZT-prediction-m}
R^{\text{FZZT}_{+}}(E) &\simeq -\frac{\rmi}{4E}\left(1+g_{\text{s}}\frac{17-2\pi^2 E}{12\left(-E\right)^{3/2}} - g_{\text{s}}^2\frac{1225-644\pi^2 E+148\pi^4 E^2}{288 E^3}+\dots\right)\text{e}^{+\frac{\tilde{A}(E)}{g_{\text{s}}}},
\end{align}
\noindent
where the FZZT action reads \cite{eggls23}
\begin{equation}
\label{eq:JT-FZZT-action}
\tilde{A}(E) = \frac{1}{4\pi^3}\left(\sin\left(2\pi\sqrt{-E}\right)-2\pi\sqrt{-E}\cos\left(2\pi\sqrt{-E}\right)\right).
\end{equation}
\noindent
After anti-Stokes transition -- and as given in formula \eqref{eq:FZZT-anti-Stokes-action} -- this action becomes
\begin{equation}
\label{eq:JT-FZZT-action-anti}
\tilde{\mathsf{A}}(E) = -\frac{1}{4\pi^3}\left(\sinh\left(2\pi\sqrt{E}\right)-2\pi\sqrt{E}\cosh\left(2\pi\sqrt{E}\right)\right).
\end{equation}
With this we can spell out the non-perturbative FZZT corrections to the JT gravity density to higher $g_{\text{s}}$ orders. We then find
\begin{align}
\label{eq:non-pert-rho-JT-b}
\left\langle\varrho_{\text{JT}}(E)\right\rangle \simeq -\frac{1}{8\pi E}\left(1-g_{\text{s}}\frac{17-2\pi^2 E}{12\left(-E\right)^{3/2}} - g_{\text{s}}^2\frac{1225-644\pi^2 E+148\pi^4 E^2}{288 E^3}+\dots\right)\text{e}^{-\frac{\tilde{A}\left(E\right)}{g_{\text{s}}}},
\end{align}
\noindent
for the density on the negative real line ($E<0$). Taking into account the extra square root branch cuts in \eqref{eq:JT-FZZT-prediction-p} and \eqref{eq:JT-FZZT-prediction-m} we find on the positive real line 
\begin{align}
\label{eq:non-pert-rho-JT-a}
\left\langle\varrho_{\text{JT}}(E)\right\rangle &\simeq 
\sum\limits_{g=0}^{\infty}g_{\text{s}}^{2g-1}\varrho_g(E)-\nonumber\\
&-\frac{1}{4\pi E}\left(1+ g_{\text{s}}\cdot 0 - g_{\text{s}}^2\frac{1225-644\pi^2 E+148\pi^4 E^2}{288 E^3} +\mathcal{O}\left(g_{\text{s}}^3\right)\right)\cos\left(\frac{\tilde{\mathsf{A}}\left(E\right)}{g_{\text{s}}}\right)-\nonumber\\
&-\left(g_{\text{s}}\frac{17-2\pi^2 E}{48\pi E^{5/2}}+\mathcal{O}\left(g_{\text{s}}^2\right)\right)\sin\left(\frac{\tilde{\mathsf{A}}\left(E\right)}{g_{\text{s}}}\right) +\\
&+\left(\tau_2+\tau_1\right)\frac{\rmi}{4\pi E}\left(1+ g_{\text{s}}\cdot 0 - g_{\text{s}}^2\frac{1225-644\pi^2 E+148\pi^4 E^2}{288 E^3} +\mathcal{O}\left(g_{\text{s}}^3\right)\right)\sin\left(\frac{\tilde{\mathsf{A}}\left(E\right)}{g_{\text{s}}}\right)-\nonumber\\
&-\rmi\left(\tau_2+\tau_1\right)\left(g_{\text{s}}\frac{17-2\pi^2 E}{48\pi E^{5/2}}+\mathcal{O}\left(g_{\text{s}}^2\right)\right)\cos\left(\frac{\tilde{\mathsf{A}}\left(E\right)}{g_{\text{s}}}\right) ,\qquad E>0.\nonumber
\end{align}
\noindent
We can conclude that the structure of formula \eqref{eq:triple-s-formula} gets slightly more intricate when adding higher $g_{\text{s}}$ orders. Nevertheless the main features stay the same: There is a universal oscillating piece dictated by Stokes constants and another also oscillating part that depends on the transseries parameters. To make contact with an actual function the transseries above still needs to be resummed. Last but not least we can connect the anti-Stokes behavior of the FZZT-branes with the idea that the oscillations in the eigenvalue density might originate from discreteness of microscopic hamiltonians drawn from the ensemble \cite{sss-semi, sss19, sw20}. We can then make the connection
\begin{center}
\begin{tikzpicture}[
	grayframe/.style={
		rectangle,
		draw=gray,
		text width=7em,
		align=center,
		rounded corners,
		minimum height=2em
	}, line width=1pt]
\node[grayframe] at (-5, 0) {microscopic discreteness};
\node[grayframe] at (0, 0) {oscillatory density contribution};
\node[grayframe] at (5,0) {resurgent anti-Stokes behavior in the JT MM};
\draw[gray, ->, line width=1.9pt, dashed] (-3, 0) -- (-2,0);
\draw[gray, <->, line width=1.9pt] (2, 0) -- (3,0);
\node at (6.7, -0.5) {.};
\end{tikzpicture}
\end{center}
\noindent
Given an ensemble description, the above construction constitutes a direct connection between the statistical properties of black holes and resurgent methods for FZZT-branes. This demonstrates that these methods can be used to study gravity via the matrix model approach.

%%%%%%%%%%%%%%%%%%%%%%%%%%%%%%%%%%%%%%%%%%%%%%%%%%%%%%%%%%%%%%%%%
\acknowledgments
It is a pleasure to thank 
Marcos~Mari\~no, 
Ricardo~Schiappa and Jasper Kager
for a careful reading of the manuscript. We would furthermore like to thank Ricardo Schiappa for pointing us to the connection between the VMS and 3d gravity. Lastly we would like to thank
Marcos~Mari\~no,
Ricardo~Schiappa,
Raghu Mahajan,
Noam~Tamarin,
Jo\~ao~Rodrigues and 
Jasper~Kager
for useful discussions, comments and/or correspondence. MS was partially supported by the Swiss National Science Foundation under grant number 185723 and grant number 200021$\_$219267. This paper is partly a result of the ERC-SyG project, Recursive and Exact New Quantum Theory (ReNewQuantum) funded by the European Research Council (ERC) under the European Union's Horizon 2020 research and innovation programme, grant agreement 810573.
%%%%%%%%%%%%%%%%%%%%%%%%%%%%%%%%%%%%%%%%%%%%%%%%%%%%%%%%%%%%%%%%%

%%%%%%%%%%%%%%%%%%%%%%%%%%%%%%%%%%%%%%%%%%%%%%%%%%%%%%%%%%%%%%%%%
%%%%%%%%%%%%%%%%%%%%%%%%%%%%%%%%%%%%%%%%%%%%%%%%%%%%%%%%%%%%%%%%%
\appendix
%%%%%%%%%%%%%%%%%%%%%%%%%%%%%%%%%%%%%%%%%%%%%%%%%%%%%%%%%%%%%%%%%
%%%%%%%%%%%%%%%%%%%%%%%%%%%%%%%%%%%%%%%%%%%%%%%%%%%%%%%%%%%%%%%%%

%%%%%%%%%%%%%%%%%%%%%%%%%%%%%%%%%%%%%%%%%%%%%%%%%%%%%%%%%%%%%%%%%
%%%%%%%%%%%%%%%%%%%%%%%%%%%%%%%%%%%%%%%%%%%%%%%%%%%%%%%%%%%%%%%%%
\section{Data for the Virasoro Minimal String}
\label{app:data}
%%%%%%%%%%%%%%%%%%%%%%%%%%%%%%%%%%%%%%%%%%%%%%%%%%%%%%%%%%%%%%%%%
%%%%%%%%%%%%%%%%%%%%%%%%%%%%%%%%%%%%%%%%%%%%%%%%%%%%%%%%%%%%%%%%%

Here we present some of the data points used in the main text. These data were produced with a slightly modified version of the \textsc{Mathematica} file provided in \cite{cemr23}. Let us start by giving some coefficients for the free energy \eqref{eq:freeEnergyVMS}. We have
\begin{align}
F_{\text{V}, 2} &= \frac{43 c^3}{238878720}-\frac{559 c^2}{79626240}+\frac{6691 c}{79626240}-\frac{72007}{238878720},\\
F_{\text{V}, 3} &= \frac{176557 c^6}{14794122225254400}-\frac{2295241 c^5}{2465687037542400}+\frac{4089107
   c^4}{140896402145280}-\\
   &-\frac{340534753 c^3}{739706111262720}+\frac{2740875239 c^2}{704482010726400}-\frac{8112433381
   c}{493137407508480}+\frac{396807114421}{14794122225254400},\nonumber\\
F_{\text{V}, 4} &= \frac{1959225867017 c^9}{667926185350848999063552000}-\frac{25469936271221 c^8}{74214020594538777673728000}+\nonumber\\
   &+\frac{9203210865259
   c^7}{530100147103848411955200}-\frac{3953544371405837 c^6}{7951502206557726179328000}+\nonumber\\
   &+\frac{9392348683949221
   c^5}{1060200294207696823910400}-\frac{76863561439881203 c^4}{757285924434069159936000}+\nonumber\\
   &+\frac{41450007951172042679
   c^3}{55660515445904083255296000}-\frac{62278985594273187389 c^2}{18553505148634694418432000}+\nonumber\\
   &+\frac{24874755570604002257
   c}{2968560823781551106949120}-\frac{5843451389177575283669}{667926185350848999063552000}.
\end{align}
\noindent
Furthermore we used the resolvent \eqref{eq:resolvent}, where the coefficients grow quickly in size. Let us nevertheless list one here
\begin{align}
R_{2,1}^{(b)}(-z^2) &= \frac{1}{20384317440 \sqrt{2} \pi ^9 z^{11}}\Big (\pi^8((c-26) c (145 (c-26) c+40722)+2782913) z^8+\nonumber\\
&+12 \pi ^6 (c-13) (169 (c-26) c+23713) z^6+\nonumber\\
&+180 \pi ^4 (139 (c-26) c+22099)
   z^4+219240 \pi ^2 (c-13) z^2+1020600\Big).
\end{align}
\noindent
We have computed them up to order 14. And finally the genus $1$ coefficient for a partition function with two boundaries that we used for the 3d gravity discussion reads
\begin{align}
&\mathsf{Z}_{1,2}(\beta_1,\beta_2) = \\&=g_{\text{s}}^2\frac{\sqrt{\beta _1} \sqrt{\beta _2} \left(24 \beta _1^2+24 \beta _2^2+\pi ^4 \left(c^2-26 c+153\right)+8 \beta _1 \left(3 \beta
   _2+\pi ^2 (c-13)\right)+8 \pi ^2 \beta _2 (c-13)\right)}{73728 \pi ^7}.\nonumber
\end{align}
\noindent
We have computed the series $\mathsf{Z}_{g,2}$ up to $g=12$.

%%%%%%%%%%%%%%%%%%%%%%%%%%%%%%%%%%%%%%%%%%%%%%%%%%%%%%%%%%%%%%%%%
%%%%%%%%%%%%%%%%%%%%%%%%%%%%%%%%%%%%%%%%%%%%%%%%%%%%%%%%%%%%%%%%%
\newpage
%%%%%%%%%%%%%%%%%%%%%%%%%%%%%%%%%%%%%%%%%%%%%%%%%%%%%%%%%%%%%%%%%
%%%%%%%%%%%%%%%%%%%%%%%%%%%%%%%%%%%%%%%%%%%%%%%%%%%%%%%%%%%%%%%%%

\bibliographystyle{plain}
%\bibliography{papers}

\end{document}